\documentstyle[11pt]{article}
\def\bbbone{{\mathchoice {\rm 1\mskip-4mu l} {\rm 1\mskip-4mu l}
{\rm 1\mskip-4.5mu l} {\rm 1\mskip-5mu l}}}
\newcommand{\vide} {\emptyset}

\newfam\gotfam
\font\twlgot=eufm10 at 12pt
\font\tengot=eufm10

\font\sevengot=eufm7
\textfont\gotfam=\twlgot
\scriptfont\gotfam=\tengot 
\scriptscriptfont\gotfam=\sevengot
\def\got{\fam\gotfam\twlgot}

\newtheorem{prop}  {Proposition}
\newtheorem{lem}  {Lemma}
\newtheorem{th}   {Theorem}

\newcommand{\be}  {\begin{equation}}
\newcommand{\ee}  {\end{equation}}
\newcommand{\bea} {\begin{eqnarray}}
\newcommand{\eea} {\end{eqnarray}}
\newcommand{\lp}  {\left(}
\newcommand{\rp}  {\right)}

\newcommand{\Br}  {\overline}

\newcommand{\cL}  {{\cal L}}
\newcommand{\cC}  {{\cal C}}
\newcommand{\cD}  {{\cal D}}
\newcommand{\cX}  {{\cal X}}
\newcommand{\cY}  {{\cal Y}}
\newcommand{\cO}  {{\cal O}}
\newcommand{\cM}  {{\cal M}}
\newcommand{\cS}  {{\cal S}}
\newcommand{\cP}  {{\cal P}}

\newcommand{\cT}  {{\cal T}}
\newcommand{\cI}  {{\cal I}}
\newcommand{\cR}  {{\cal R}}
\newcommand{\cA}  {{\cal A}}
\newcommand{\cE}  {{\cal E}}
\newcommand{\cU}  {{\cal U}}
\newcommand{\cV}  {{\cal V}}
\newcommand{\cG}  {{\cal G}}

\newcommand{\Om}  {\Omega}
\newcommand{\vph}  {\varphi}
\newcommand{\om}  {\omega}

\newcommand{\ep}  {\epsilon}
\newcommand{\si}  {\sigma}
\newcommand{\ga}  {\gamma}
\newcommand{\Ga}  {\Gamma}

\newcommand{\al}  {\alpha}
\newcommand{\la}  {\lambda}
\newcommand{\rh}  {\rho}
\newcommand{\et}  {\eta}
\newcommand{\ze}  {\zeta}
\newcommand{\La}  {\Lambda}
\newcommand{\de}  {\delta}
\newcommand{\De}  {\Delta}
\newcommand{\ch}  {\chi}
\newcommand{\Ph}  {\Phi}
\newcommand{\ph}  {\phi}

\newcommand{\Gt}  {{\got t}}
\newcommand{\Gc}  {{\got c}}
\newcommand{\GV}  {{\got V}}
\newcommand{\Gg}  {{\got g}}

\newcommand{\GC}  {{\got C}}
\newcommand{\GP}  {{\got P}}
\newcommand{\Gp}  {{\got p}}
\newcommand{\Gx}  {{\got x}}
\newcommand{\und}  {\underline}
\newcommand{\til}  {\tilde}

\def\half{\frac{1}{2}}
\def\prf{\noindent{\bf Proof:\ \ }}

\def\NN{\hbox to 9.3111pt{\rm I\hskip-1.8pt N}}
\def\RR{\hbox to 9.1722pt{\rm I\hskip-1.8pt R}}

\newcommand{\eqdef} {\stackrel{\rm def}{=}}
\def\Endproof
{\hfill\vrule height .6em width .6em depth 0pt\goodbreak\vskip.25in}

\begin{document}

\title{An Explicit Large Versus Small Field
Multiscale Cluster Expansion}
\author{
A. Abdesselam \& V. Rivasseau\\
Centre de Physique Th\'eorique, Ecole Polytechnique\\
99128 Palaiseau Cedex, FRANCE
}

\maketitle
\begin{abstract}
We introduce a new type of cluster expansion which generalizes
a previous formula of Brydges and Kennedy. The method is
especially suited for performing a phase-space multiscale
expansion in a just renormalizable theory, and allows
the writing of explicit non-perturbative formulas
for the Schwinger functions. The procedure is quite model independent,
but for simplicity we chose the infrared $\ph^4_4$ model as a
testing ground. We used also a large field versus small field
expansion.
The polymer amplitudes, corresponding to graphs without almost local
two and for point functions, are shown to satisfy
the polymer bound.
\end{abstract}

\section{Introduction}
\markboth{INTRODUCTION}{INTRODUCTION}

Many of the important results in constructive field theory, like
for instance the work done on the Yang-Mills theory in 4 dimensions
[Bal, MRS], make use of a mathematically rigorous
implementation of renormalization group techniques.
Albeit well understood at the perturbative level, their application
in a constructive framework is still a difficult enterprise.

After a period of first successes [GJ2, MS, FO, GK, FMRS, Bal],
progress in this branch of mathematical physics was
somewhat slowed down. It was partly because new conceptual tools
were needed, like for instance renormalization group around a surface
singularity, involving dynamical $\frac{1}{N}$ expansions and
random matrix methods [FT, FMRT, Poi].
But another reason is that the heavy technical apparatus
needed in the proofs, is reaching a critical mass
preventing potential readers from being 
fully convinced by their mathematical rigor, 
and discouraging actual authors
from venturing into the painful task of writing them
in detail.

Therefore many efforts were devoted since then to the clarification
and the improvement of these techniques.
One can already observe the growth of two complementary approaches.
In the first, a single step of renormalization group applied to algebras
of polymer activities is emphasized. One has then to define
the proper Banach spaces to support the iteration
of the procedure, and to study the corresponding dynamical
system.
Two versions of this method were developed mainly around
Brydges [Br2, Br3, BrY, BrDH]
on the one hand and Pordt [MP, Por] on the other.

The other approach to rigorous renormalization group methods,
is through phase-space expansions. It was initiated by Glimm
and Jaffe in [GJ1], and became the main tool of our
group at the Ecole Polytechnique [MS, FMRS, MRS].
In this approach, the iteration of many renormalization
group steps is treated as a whole.
The method looks closer to perturbation theory of which it is a kind of
truncated expansion. The idea is to expand enough to
put into display the sources of divergences, typically ``almost local'',
i.e. high frequency, insertions of 2 and 4 point functions in the
$\ph_4^4$ model for instance [R1]. However we must not expand too much in
order to prevent the accumulation of fields at the
same spot and with same frequency.

Previous schemes of phase-space expansions used different
interpolations for the horizontal expansions (in real space) and
the vertical ones (along the frequency axis). The procedure was
inductive and the vertical interpolation nonlinear.
Furthermore a Mayer expansion step has to be performed at every
scale. These features render the previous expansions quite
obscure, and the explicit writing of a formula for the output too difficult.
In this paper we worked out a new type of phase-space cluster expansion,
based on a linear interpolation, which treats horizontal and vertical
expansions on the same footing.
It generalizes the formula of Brydges and Kennedy for horizontal
cluster expansions [BrK, Br2, AR1, A].

This method presents many improvements, since it allows the writing
of explicit formulas, somewhat in the spirit of
Zimmerman's forest formula for perturbative renormalization.
The method is quite general, and robust enough to bear a large
versus small field expansion.
We treat here for simplicity, and as a testing ground, the infrared $\ph_4^4$
model. Note that in this special case such a large field versus
small field expansion is not necessary, unlike the Yang-Mills theory
for instance. We introduce this expansion here for two reasons.
One is that we expand a little more than previous treatments of the
present model and flirt a little closer with the danger represented
by the divergence of the perturbation series.
Indeed the polymers that appear in our expansion
may have gaps in the vertical direction, i.e. they can couple cubes
in the higher frequency slices directly with cubes in the lower ones,
with large gaps in between.
However we still have the factorization of the functional
integrals between polymers that are now more diverse than
in old-fashioned expansions.
One could compare the difference between our vertical expansion
and the previous ones [FMRS] with the difference between
the early horizontal
cluster expansions of [GJS], in constructive field theory, and their
version by
Brydges, Battle and Federbush [Bat, BaF, BrF].
The other reason for introducing a large field versus small field expansion
is that it was never really written in detail when cast
in a multiscale phase-space expansion.
Another improvement is also that our method seems to fit better the intrinsic
combinatoric structure of functional integrals with interactions.
This allows to shunt the intermediate Mayer expansion steps,
which should greatly enhance the clarity of the proofs.
This aspect is however postponed to a next paper [AR2] where the
full renormalized model will be constructed.

In Section 2, we introduce our expansion in full
generality. Specializations to multi-body interactions in lattice
systems, or to the study of
$p$-particle irreducible kernels in field theory as in
[IM], could be done rather naturally in this formalism. However we refer
to [A] for a treatment of these topics. We mention also that a version
of these expansions features in the study of the single slice Anderson model
in [Poi].
Hereupon we specialize our discussion to the example of infrared $\ph_4^4$.
An analog of $p$-particle irreducibility plays a key role here, it
corresponds to parts of the expansion that are free from 2 and 4 point
``almost local'' insertions.

In Section 3 we prove in detail a polymer bound on the
polymers that are free from these insertions. Hence
Theorem 2 is the constructive analog of Weinberg's theorem
on convergent graphs.
However it already involves summing up all orders of perturbation theory and
is the cornerstone of phase-space expansion renormalization group methods.
The proof of a similar statement after the extraction of the local
part of 2 and 4 point functions is the purpose of our next paper [AR2].

We intended the present paper to be particularly pedagogical for a non trivial
result in our discipline; this may account for the length and the level of
formalization.
We painstakingly kept track of numerical constants instead of having
stray $O(1)$'s around, to help the reader check any eventual
misunderstanding. The way combinatorics mix with analysis in such
a proof makes clarity and unambiguousness very difficult to  
attain. We hope that one can, with some endeavor, read these technical
pages and get, in the end, a sound
comprehension of the subject.

\section{An all purpose scheme for cluster expansions}
\markboth{AN ALL PURPOSE SCHEME FOR CLUSTER EXPANSIONS}
{AN ALL PURPOSE SCHEME FOR CLUSTER EXPANSIONS}

\subsection{Interpolation Revisited}

Generally speaking, cluster expansion techniques in constructive
field theory usually stem from a clever application of a Taylor
formula with integral reminder. Writing the full Taylor series
would amount to completely expand the perturbation series which
most often diverges, and therefore should be avoided.
We state in this section, in full generality, the kind of Taylor
formula our expansions are based on.

Let us suppose we have a finite set of indices $L$.
In applications, $L$ will typically be the set of links between
pairs or even larger patches of cubes in the discretization
of space which
is introduced to prevent local accumulation of vertices.
Let $H:{(t_l)}_{l\in L}\mapsto H({(t_l)}_{l\in L})$ be a smooth
function defined on the cube ${[0,1]}^L$ in $\RR^L$.
As an illustration, one should view $H({(t_l)}_{l\in L})$ as the 
partition function of the system where a coupling corresponding
to a link $l$ is weakened by the corresponding parameter $t_l$,
$0\le t_l\le 1$.
We define the particular elements $\bf 0$ and $\bf 1$ of ${[0,1]}^L$
as the vectors with all entries equal to 0 and 1
respectively.
Our expansion which interpolates between $H({\bf 1})$ and
$H({\bf 0})$, inductively generates {\em ordered graphs}.
The idea is that after the explicit production of a sequence of links,
there is a limited possible {\em choice} for the next link to be
derived.
The limitation is the fact that already existing links have connected
the cubes into bigger blocks and it would be redundant and even
dangerous to produce links inside such a block. Indeed we would then
add an arbitrary number of loops and expand too much.

So much for the philosophy, now we denote by $\cG$ the set of finite
ordered {\em sequences} $(l_1,\ldots,l_k)$ made of elements in $L$,
with all possible values of length $k$.
We allow
$k=0$ hence the empty sequence denoted by $\vide$.
We introduce a partial ordering $\le$ on $\cG$ : if $\Gg=(l_1,\ldots,l_k)$
and $\Gg'=({l'}_1,\ldots,{l'}_{k'})$ are two sequences, we have
$\Gg\le \Gg'$ if and only if $k\le k'$ and for any
$a$, $1\le a\le k$, $l_a={l'}_a$ holds,
i.e. if $\Gg$ is an {\em initial segment of} $\Gg'$.
Suppose we have a {\und{choice map}}
$\cC:\cG\rightarrow\cP(L)$, where $\cP(L)$
is the power set
of $L$, i.e. a map such that for any two sequences $\Gg_1$ and $\Gg_2$
satisfying
$\Gg_1\le \Gg_2$, we have $\cC(\Gg_2)\subset \cC(\Gg_1)$.

We define the set of {\und{allowed sequences}} $\cA\cG$,
as the set of all sequences
$\Gg=(l_1,\ldots,l_k)$ such that for any $a$, $ 1\le\ a\le k$,
we have
$l_a\in \cC((l_1,\ldots,l_{a-1}))$.
In particular, we always have $\vide\in\cA\cG$. Besides, any initial
segment of an allowed sequence is allowed.
Finally we suppose that the sequences in $\cA\cG$ are of bounded length.
We can now state the following interpolation lemma.

\begin{lem}
\label{inter}

Under the previous hypothesis we have
\be
H({\bf 1})=
\sum_{{\Gg=(l_1,\ldots,l_k)}\atop{\Gg\in\cA\cG}}
\int_{1\ge h_1\ge\cdots\ge h_k\ge 0}
dh_1\ldots dh_k
{{\partial^k H}\over{\partial t_{l_1}\ldots\partial t_{l_k}}}
\bigl(
T_\Gg({\bf h})
\bigr)
\ \ ,
\label{inter1}
\ee
where $\bf h$ denotes the vector $(h_1,\ldots,h_k)$,
and $T_\Gg({\bf h})$ is the ${(t_l)}_{l\in L}$ vector defined in the
following way:

- if $l\notin \cC(\vide)$ then $t_l=1$,

- if $l\in \cC(\vide)\backslash \cC((l_1))$ then $t_l=h_1$,

- if $l\in \cC((l_1))\backslash \cC((l_1,l_2))$ then $t_l=h_2$,

$\vdots$

- if $l\in \cC((l_1,\ldots,l_{k-1}))\backslash \cC((l_1,\ldots,l_k))$
then $t_l=h_k$,

- if $l\in \cC((l_1,\ldots,l_k))$ then $t_l=0$.

Summation on $k$ includes all possible values.
\end{lem}

\noindent
{\bf Proof:\ \ }
This is done by induction. Using the same notations as before,
let us define the vector ${\til T}_\Gg({\bf h})$ like we did for
$T_\Gg({\bf h})$,
except for the last item, where we set
$t_l=h_k$ if $l\in \cC((l_1,\ldots,l_k))$,
with the convention that $h_0=1$.
Let $n\ge0$ be an integer, we will prove by induction on n the following
formula:
\[
H({\bf 1})=
\sum_{k<n}
\sum_{{\Gg=(l_1,\ldots,l_k)}\atop{\Gg\in\cA\cG}}
\int_{1\ge h_1\ge\cdots\ge h_k\ge 0}
dh_1\ldots dh_k
{{\partial^k H}\over{\partial t_{l_1}\ldots\partial t_{l_k}}}
\bigl(
T_\Gg({\bf h})
\bigr)
\]
\be
+
\sum_{{\Gg=(l_1,\ldots,l_n)}\atop{\Gg\in\cA\cG}}
\int_{1\ge h_1\ge\cdots\ge h_n\ge 0}
dh_1\ldots dh_n
{{\partial^n H}\over{\partial t_{l_1}\ldots\partial t_{l_n}}}
\bigl(
{\til T}_\Gg({\bf h})
\bigr)
\ \ .
\label{inter2}
\ee
Indeed, for $n=0$ it is a tautology.
Besides, the induction step is a consequence of the
fact that, given $\Gg=(l_1,\ldots,l_n)$ in $\cA\cG$ and
${\bf h}=(h_1,\ldots,h_n),1\ge h_1\ge\cdots\ge h_n\ge0$, we can reexpress
\be
{{\partial^n H}\over{\partial t_{l_1}\ldots\partial t_{l_n}}}
\bigl(
{\til T}_\Gg({\bf h})
\bigr)
\ee
as $f(h_{n+1})|_{h_{n+1}=h_n}$.
The function $f(h_{n+1})$ of the new interpolation parameter $h_{n+1}$,
is defined by
\be
f(h_{n+1})=
{{\partial^n H}\over{\partial t_{l_1}\ldots\partial t_{l_n}}}
\bigl(
{(t_l)}_{l\in L}
\bigr)
\ \ ,
\ee
where ${(t_l)}_{l\in L}={\til T}_\Gg({\bf h})$, except for the $t_l$'s with
$l\in \cC((l_1,\ldots,l_n))$ that are set equal to $h_{n+1}$, by definition.
We then write
\be
f(h_{n+1})=f(0)+\int_0^{h_n}dh_{n+1}
{{df}\over{dh_{n+1}}}(h_{n+1})
\ee
\[
=
{{\partial^n H}\over{\partial t_{l_1}\ldots\partial t_{l_n}}}
\bigl(
T_\Gg({\bf h})
\bigr)
\]
\be
+
\int_0^{h_n}
dh_{n+1}
\sum_{l_{n+1}\in \cC((l_1,\ldots,l_n))}
{{\partial^n H}\over{\partial t_{l_1}\ldots\partial t_{l_n}
\partial t_{l_{n+1}}}}
\bigl(
{\til T}_{(\Gg,l_{n+1})}({\bf h},h_{n+1})
\bigr)
\ \ .
\ee
As a result, we get
\[
\int_{1\ge h_1\ge\cdots\ge h_n\ge 0}
dh_1\ldots dh_n
{{\partial^n H}\over{\partial t_{l_1}\ldots\partial t_{l_n}}}
\bigl(
{\til T}_\Gg({\bf h})
\bigr)
\]
\[
=
\int_{1\ge h_1\ge\cdots\ge h_n\ge 0}
dh_1\ldots dh_n
{{\partial^n H}\over{\partial t_{l_1}\ldots\partial t_{l_n}}}
\bigl(T_\Gg({\bf h})
\bigr)
\]
\[
+
\sum_{l_{n+1}|(l_1,\ldots,l_n,l_{n+1})\in\cA\cG}
\int_{1\ge h_1\ge\cdots\ge h_n\ge h_{n+1}\ge 0}
dh_1\ldots dh_n dh_{n+1}
\]
\be
{{\partial^{n+1}H}\over{\partial t_{l_1}\ldots\partial t_{l_n}
\partial t_{l_{n+1}}}}
\bigl(
{\til T}_{(\Gg,l_{n+1})}({\bf h},h_{n+1})
\bigr)
\ \ ,
\ee
thereby proving (\ref{inter2}) at stage $n+1$.
Now, since sequences in $\cA\cG$ have a bounded length, 
(\ref{inter2}) simply reduces
to (\ref{inter1}) for $n$ large enough.
\Endproof

Let us now consider the following specialization and improvement
of the previous lemma. We are given a nonempty set of objects $\cD$,
typically the set of cubes in the cell discretization of real space,
in single
slice models, or of phase-space, in multiscale expansions.

If $p$ is an integer $p\ge 2$, we call {\und{$p$-link}} a map
$l:\cD\rightarrow\NN$ such that $\sum_{\De\in\cD}l(\De)=p$.
This is an unordered combination of $p$ elements amid $\cD$, with possible
repetitions.
In the $\ph^4$ theory, the propagators joining 2 boxes will correspond
to 2-links, while vertices, connecting up to 4 boxes in the phase-space
decomposition, will be associated to 4-links. In this last example
we need to keep record of the multiplicity $l(\De)$ that counts how many
fields does the vertex produce in cube $\De$.
The {\und{support}} of a $p$-link $l$ is the set
${\rm supp}\ l\eqdef\{\De\in\cD|l(\De)\ne0\}$.
The set of $p$-links is denoted by $\cL_p$.

In order to handle situations where several types of links are present,
we suppose $L$ is the disjoint union of copies of sets $\cL_p$.
That is we assume we have a partition  
$\{L_1,\ldots,L_q\}$ of $L$, a sequence of integers $(p_1,\ldots,p_q)$,
$p_\nu\ge 2$, $1\le \nu\le q$, and a map $J:L\rightarrow\cup_{p\ge 2}
\cL_p$, such that for every $\nu$, $J$ restricts on $L_\nu$ to a
bijection with $\cL_{p_\nu}$.
However we will make a slight abuse of notation by forgetting the
distinction between an element of $L$ and the corresponding link
$J(l)$. In case there are several sets of links of type $\cL_p$
for a given $p$,  like in the jungle formulas of 
[AR,KMR] one should be more careful.

The useful version of Lemma \ref{inter} we shall need is that one where
the smooth function $H$ is assumed to be defined, may be not on
the whole ${[0,1]}^L$, but on the following subset $\Om$.
A vector ${(t_l)}_{l\in L}$ is said a {\und{partition vector}}
if there exists a partition $\pi$ of $\cD$ such that $t_l=1$ if
${\rm supp}\ l$ is entirely contained in a block of $\pi$ and $t_l=0$
otherwise.
The set $\Om$ is by definition {\em the convex hull of all the partition
vectors}.
Furthermore, we impose the following restriction on the choice map $\cC$:
it is deduced from a {\und{connectivity map}}
$\Pi:\cG\rightarrow{\rm Part}(\cD)$.
${\rm Part}(\cD)$ is the set of partitions of $\cD$, and ``deduced'' means
that for any $\Gg\in\cG$, we have $\cC(\Gg)={\rm Offdiag}(\Pi(\Gg))$.
$\pi$ being an arbitrary partition of $\cD$, ${\rm Offdiag}(\pi)$
denotes the set
\be
{\rm Offdiag}(\pi)=
\{l\in L|\;\forall B\in\pi,{\rm supp}\ l\not\subset B\}
\ \ .
\ee
In other words, it is the set of links jumping between two or more
blocks of $\pi$.
Note that ${\rm Part}(\cD)$ is equipped with a partial ordering $\le$,
such that two partitions $\pi_1$ and $\pi_2$ satisfy $\pi_1\le\pi_2$,
if and only if $\pi_1$ is obtained from $\pi_2$ by splitting it into
smaller pieces.
Observe that $\pi_1\le\pi_2$ is equivalent to ${\rm Offdiag}(\pi_2)\subset
{\rm Offdiag}(\pi_1)$; we then say that $\pi_1$ is {\em finer} than $\pi_2$.

Inspecting the proof of lemma \ref{inter},
one realizes that its output is still valid, under the present somewhat
weaker hypothesis on $H$.
This is because the involved interpolations never take us outside $\Om$.
Indeed, at each time, we modify the argument ${\bf t}\in\Om$ of $H$ or
its derivatives, according to the following pattern.
We are given $\Gg=(l_1,\ldots,l_n)$ in $\cA\cG$ and we write
${\bf t}={\bf t}_{\rm diag}(\Gg)+h_n {\bf t}_{\rm offdiag}(\Gg)$.
Here ${\bf t}_{\rm diag}(\Gg)$ denotes the vector in ${[0,1]}^L$
equal to ${\bf t}$ except for the entries $t_l$ with $l$
in $\cC(\Gg)={\rm Offdiag(\Pi(\Gg))}$ that are
put to zero. By the same token, ${\bf t}_{\rm Offdiag}(\Gg)$ is
the vector with entries $t_l=\bbbone_{\{l\in\cC(\Gg)\}}$.
We then say ${\bf t}$ is the value at $h_{n+1}=h_n$ of
${\bf t}_{\rm diag}(\Gg)+h_{n+1} {\bf t}_{\rm offdiag}(\Gg)$, and interpolate
in the variable $h_{n+1}$ between $h_n$ and 0. Hence we can find
$\la\in[0,1]$, such that $h_{n+1}=\la h_n$.
Therefore we have
\be
{\bf t}_{\rm diag}(\Gg)+h_{n+1} {\bf t}_{\rm offdiag}(\Gg)
=\la {\bf t}+(1-\la){\bf t}_{\rm diag}(\Gg)\ \ .
\ee
But $\Om$ is convex, and as a consequence ${\bf t}_{\rm diag}(\Gg)
+h_{n+1} {\bf t}_{\rm offdiag}(\Gg)$ will be in $\Om$, as soon as we prove
the similar statement for
${\bf t}_{\rm diag}(\Gg)$.
But this last assertion is true.
Indeed, if ${\bf t}$ is a convex combination of partition
vectors ${\bf t}_1,\ldots,{\bf t}_n$, obtained respectively from
the partitions
$\pi_1,\ldots,\pi_n$, then ${\bf t}_{\rm diag}(\Gg)$ is a
convex combination, with the same weights, of the partition vectors
obtained respectively from $\pi_1\wedge\Pi(\Gg),\ldots,\pi_n\wedge\Pi(\Gg)$.
The symbol $\wedge$ denotes the greatest lower bound in the lattice
${\rm Part}(\cD)$ together with the above mentioned order relation.
In other words, $\pi_1\wedge\pi_2$ is the partition whose blocks are the
non empty sets $B_1\cap B_2$ with $B_1\in\pi_1$ and $B_2\in\pi_2$.
We can restate our result as

\begin{lem}
\label{interuse}
The output of Lemma \ref{inter} is still valid if we only assume
the definiteness and smoothness of $H$ in $\Om$.
\end{lem}

The reason we introduce the partition vectors is the conservation
by the interpolation procedure of the positivity of both the
covariance and the interaction, when we deal with bosonic models.
 To apply our formalism to a concrete situation, one needs to define
the object of interest in term of a function $H$ depending on
coupling
parameters $t_l$ indexed by finite combinations of degrees of
freedom in the system. Then one has to define the connectivity
map $\Gg\mapsto \Pi(\Gg)$, {\em for any sequence} $\Gg$. Lemma \ref{interuse}
will produce
a polymer expansion, whose blocks are those of $\Pi(\Gg)$, $\Gg$ being
now restricted to {\em allowed sequences}.
In single slice cluster expansions with good decay of the
propagator, one needs to consider only 2-links and $\Pi(\Gg)$ is the
partition into ordinary connected components of the graph $\Gg$.
One then recovers, modulo symmetrization in the $t$ parameters,
the Brydges-Kennedy forest formula [BrK, Br2, AR1, A]. 
Another easy application is constructive multi-particle analysis
in the spirit of [IM], albeit more symmetric formulas are
obtained. The idea is to take for $\Pi(\Gg)$
the partition into $p$-particle
irreducible kernels of the graph $\Gg$.
Our method seems also well suited for the treatment of multi-body
interactions in Gibbs random fields on lattices.
We will not develop these aspects here
for which we refer to [A], and tackle the
harder generalization of the
notion of $p$-particle irreducibility
needed for a phase-space analysis of infrared $\ph^4_4$ for instance.

\subsection{Application to infrared $\phi^4_4$: an explicit small field
vs. large field multiscale expansion}

\subsubsection{The model}

We consider a $\phi^4_4$ theory, with a multiscale decomposition of
the bosonic field $\ph$ into $N+1$ slices.
That is we work in a finite cube $\La$ in $\RR^4$, with sides of length
$M^N$, where $M\ge 2$ is a fixed integer.
For each $i$, $0\le i\le N$, we fill $\La$ with cubes whose sides
are for instance semi-open intervals of tne form $[\al,\beta[$ and
of length $M^i$.
The set of such boxes, with cardinal $M^{N-i}$, is denoted by $\cD^{(N)}_i$.
We let $\cD^{(N)}=\cup_{0\le i\le N}\cD^{(N)}_i$, and for any cell
$\De\in\cD^{(N)}$
we define its level $i(\De)$ as the unique $i$, $0\le i\le N$,
with $\De\in\cD^{(N)}_i$.
We also denote its volume, here equal to $M^{4i(\De)}$, by $|\De|$.
If $(x,i)\in\La\times\{0,\ldots,N\}$ we define $\De(x,i)$ as the unique
box of $\cD^{(N)}_i$ containing $x$.
The way we picture the set of boxes $\cD^{(N)}$,
is by stacking the layers $\cD_i$, putting every $\cD^{(N)}_i$
on top of $\cD^{(N)}_{i+1}$,
for any $i$, $0\le i\le N$.
When we say that some box $\De_1$ is \und{above} another cube $\De_2$,
this means that $i(\De_1)<i(\De_2)$ and $\De_1\subset\De_2$.
Two elements of $\cD^{(N)}$ are said \und{vertically neighboring},
if we can order them as $\De_1$, $\De_2$
for which $i(\De_2)=i(\De_1)+1$ and $\De_1\subset\De_2$ hold.
If $\De_1$ is above $\De_2$ and these two cubes are vertically neighboring
we say that $\De_1$ is \und{just above} $\De_2$.
Finally, note that two elements of $\cD^{(N)}$ are either included
one in another or disjoint.
Until Section 3 we will forget about the $N$ dependence so that
$\cD^{(N)}$ is now the set $\cD$ to which we apply the result of Section 2.1.

We let $(\ph_i)_{0\le i\le N}$ be $N+1$ independent scaled Gaussian random
fields on $\La$.
We choose a covariance such that for any $x_1$ and $x_2$ in $\La$
\bea
C(x_1,i_1;x_2,i_2) & = & <\ph_{i_1}(x_1)\ph_{i_2}(x_2)> \\
 & = & \de_{i_1 i_2}\int\frac{d^4 p}{(2\pi)^4}
        \frac{e^{ip(x_1-x_2)}}{p^2}
        \lp e^{-M^{2i_1}p^2}-e^{-M^{2(i_1+1)}p^2}\rp\ \ .\label{defcov}
\eea

This gives for the total field $\ph=\sum_{0\le i\le N}\ph_i$, in the limit
$N\rightarrow\infty$,
the covariance
\be
<\ph(x_1)\ph(x_2)>=\int\frac{d^4 p}{(2\pi)^4}\frac{e^{ip(x_1-x_2)}}{p^2}
e^{-p^2}
\ \ ,
\ee
i.e. the covariance of a Gaussian massless field
with unit UV cut-off.
Note that the fast decrease of the propagator for large momentum,
ensures that our measure is supported on smooth fields, see [E, GJ2].

We write the interaction using a symmetric kernel $K(x_1,i_1;\ldots;
x_4,i_4)$ which is positive in the sense that, for any test fields
$\ph_i(x)$, we have
\be
\sum_{i_1,\ldots,i_4=0}^N
\int_{\La^4}dx_1\ldots dx_4
\,
K(x_1,i_1;\ldots;x_4,i_4)
\phi_{i_1}(x_1)\ldots\phi_{i_4}(x_4)
\ge 0
\ \ .
\label{Pos}
\ee
Here we choose
\be
K(x_1,i_1;\ldots;x_4,i_4)
\eqdef g
\de^4(x_2-x_1)
\de^4(x_3-x_1)
\de^4(x_4-x_1)
\ \ ,
\label{kernel}
\ee
where $g>0$ is the coupling constant. Therefore the positivity condition
(\ref{Pos})
simply reduces to
\be
g\int_\La\lp\sum_{i=0}^N\ph_i(x)\rp^4dx\ge 0
\ \ .
\ee
We illustrate our cluster expansion on the partition function
$Z(\La)$ of the system given by
\be
Z(\La)\eqdef\int d\mu_C\lp(\ph_i)_{0\le i\le N}\rp
\exp\lp-g\int_\La\lp\sum_{i=0}^N\ph_i(x)\rp^4dx\rp
\ \ ,
\ee
or more generally on unnormalized Schwinger functions like
\be
S(x_1,i_1;\ldots;x_n,i_n)\eqdef
\int d\mu_C\lp(\ph_i)_{0\le i\le N}\rp
\ph_{i_1}(x_1)\ldots\ph_{i_n}(x_n)
\exp\lp-g\int_\La\lp\sum_{i=0}^N\ph_i(x)\rp^4dx\rp
\ \ .
\ee
The way we proceed is by finding a suitable function
$H((t_l)_{l\in L})$ such that $H({\bf 1})$ is our quantity of interest.
In this concrete example, we apply the result of Lemma \ref{interuse},
with $\cD$ as a set of objects,
and $L\eqdef\cL_2\cup\cL_4$ as a set of links.
The parameters $(t_l)_{l\in\cL_2}$
serve to decouple the covariance, whose kernel becomes:
\be
C[(t_l)_{l\in\cL_2}](x_1,i_1;x_2,i_2)\eqdef
t_{l[\De(x_1,i_1),\De(x_2,i_2)]}
C (x_1,i_1;x_2,i_2)
\ \ ,
\label{cov}
\ee
where, for any sequence $(\De_1,\ldots,\De_p)$ of cubes, we denote by
$l[\De_1,\ldots,\De_p]$
the $p$-link $l$ defined by $l(\De)\eqdef\#(\{a|\ 1\le a\le p,
\ \De_a=\De\})$, for every $\De\in\cD$.

Likewise, the parameters $(t_l)_{l\in\cL_4}$ are used
to decouple the interaction whose new kernel is:
\be
K[(t_l)_{l\in\cL_4}](x_1,i_1;\ldots;x_4,i_4)\eqdef
t_{l[\De(x_1,i_1),\ldots,\De(x_4,i_4)]}
K(x_1,i_1;\ldots;x_4,i_4)
\ \ .
\ee
Both kernels are symmetric, and their positivity follows from the next
lemma.

\begin{lem}
\label{positivity}
Suppose $L$ contains a copy of $\cL_m$.
Let $\cM(x_1,i_1;\ldots;x_m,i_m)$ be a symmetric kernel, positive in
the sense that for any test fields $\ph_i(x)$ we have
\be 
\sum_{i_1,\ldots,i_m=0}^N
\int_{\La^m}dx_1\ldots dx_m
\cM(x_1,i_1;\ldots;x_m,i_m)
\phi_{i_1}(x_1)\ldots\phi_{i_m}(x_m)
\ge 0
\ \ ;
\ee
then
for any ${\bf t}\in\Om$ the interpolated kernel
\be
\cM[(t_l)_{l\in\cL_m}](x_1,i_1;\ldots;x_m,i_m)\eqdef
t_{l[\De(x_1,i_1),\ldots,\De(x_m,i_m)]}
\cM(x_1,i_1;\ldots;x_m,i_m)
\ee
is also positive.
\end{lem}

\noindent
{\bf Proof:\ \ }
Since the dependence of $\cM[(t_l)_{l\in\cL_m}]$ on ${\bf t}$ is linear
and $\Om$ is the convex hull of partition vectors, it is enough to prove
the assertion for such vectors only.
Therefore we suppose ${\bf t}$ is the partition vector associated to
$\pi\in{\rm Part}(\cD)$.
Then by definition
\[
\sum_{i_1,\ldots,i_m=0}^N
\int_{\La^m}dx_1\ldots dx_m
\ \cM[(t_l)_{l\in\cL_m}](x_1,i_1;\ldots;x_m,i_m)
\phi_{i_1}(x_1)\ldots\phi_{i_m}(x_m)
\]
\[
=\sum_{B\in\pi}
\ \sum_{i_1,\ldots,i_m=0}^N
\int_{\La^m}dx_1\ldots dx_m
\ \bbbone_{\{supp\ l[\De(x_1,i_1),\ldots,\De(x_m,i_m)]\subset B\}}
\]
\be
\cM(x_1,i_1;\ldots;x_m,i_m)
\phi_{i_1}(x_1)\ldots\phi_{i_m}(x_m)
\ee
\be
=\sum_{B\in\pi}
\ \sum_{i_1,\ldots,i_m=0}^N
\int_{\La^m}dx_1\ldots dx_m
\ \cM(x_1,i_1;\ldots;x_m,i_m)
\phi^B_{i_1}(x_1)\ldots\phi^B_{i_m}(x_m)
\ \ ,
\ee
where $\ph^B_i(x)\eqdef\bbbone_{\{\De(x,i)\in B\}}\ph_i(x)$.
The positivity of $\cM$ applied to $\ph^B_i(x)$, and summed over
the blocks $B$ of $\pi$,
just proves the assertion.
\Endproof
As a result, the function $H((t_l)_{l\in L})$ obtained by
replacing, in the partition function or the Schwinger function,
the kernels $C$ and $K$ by their interpolated version, is well defined.
$H$ is also a smooth function. We recall that via an integration
by parts one can prove that a derivation with respect to $t_l$,
$l\in\cL_2$, amounts to introducing a functional differential operator
\be
{1\over 2}\sum_{{(\De_1,\De_2)\in\cD^2}\atop{l[\De_1,\De_2]=l}}
\int_{\De_1}\int_{\De_2}
C(x_1,i(\De_1);x_2,i(\De_2))
\frac{\de}{\de\ph_{i(\De_1)}(x_1)}
\frac{\de}{\de\ph_{i(\De_2)}(x_2)}
\ee
in the functional integral, see [E, GJ2, Br1].

We will be in a position to apply the result of Section 2.1, as soon as we
define
the connectivity map $\Gg\mapsto\Pi(\Gg)$.
However, for technical reasons, we postpone this operation after
introducing a large field versus small field expansion.

\subsubsection{The large field versus small field expansion}

Following [L], for our large field condition we introduce
a $C^\infty$ step function $\chi$

\be
\chi(u)\eqdef
\left\{ \begin{array}{ll}
      1 & {\rm if}\ \  0\le t\le {\frac{1}{2}} \\
      e^2\lp1-e^{-{{1}\over{t-{1\over 2}}}}\rp e^{-{{1}\over{t-1}}} 
      & {\rm if}\ \  {\frac{1}{2}}\le t \le 1 \\
       0 & {\rm if}\ \  t\ge 1
    \end{array}
\right.
\label{smooth}
\ee
that interpolates smoothly between 1 and 0 on the interval
$[0,1]$. 
We choose also a constant $\ep_1>0$.
Given $\De\in \cD_i$, and $j\ge i$ we denote by ${\rm pr}_j(\De)$ the unique
cube in $\cD_j$ that contains $\De$.
Now if $\Ga$ is some subset of $\cD$, and $(\ph_i)$
is a configuration of the fields, we pose
\[
\ch_\Ga((\ph_i))=
\prod_{\De\in\Ga}
(1-\ch)\lp g^{(1+\ep_1)}\int_\De{\lp\sum_{i\in I_\Ga(\De)}\ph_i(x)\rp}^4
dx\rp
\]
\be
\times
\prod_{\De\in\cD\backslash\Ga}
\ch\lp g^{(1+\ep_1)}\int_\De{\lp\sum_{i\in I_\Ga(\De)}\ph_i(x)\rp}^4
dx\rp
\ \ ,
\ee
where $I_\Ga(\De)$ is the largest interval of the form $\{i(\De),
i(\De)+1,\ldots,i\}$ such that for any of its elements $j$, $j>i(\De)$
we have
${\rm pr}_j(\De)\in\Ga$.
$\Ga$ is what we call a {\und{large field region}}.
The cubes of $\Ga$ are {\em large field cubes},
those of $\cD\backslash\Ga$
are {\em small field cubes}.
We now have the identity
\be
1=\sum_{\Ga\subset\cD}
\ch_\Ga((\ph_i))
\label{idcond}
\ee
which is a consequence of the following elementary algebraic lemma.

\begin{lem}
\label{idelem}
Let $r\ge 1$ be an integer and let $R$ be a function that associates
a real number to any pair $(a,\tau_a)$, where $a$ is an integer with
$1\le a\le r$, and $\tau_a$ is a map $\tau_a:\{1,\ldots,a-1\}\rightarrow
\{0,1\}$.

Then
\be
1=\sum_{\tau}\prod_{{1\le a\le r}\atop{\tau(a)=1}}
(1-R)(a,\tau|_{\{1,\ldots,a-1\}})\times
\prod_{{1\le a\le r}\atop{\tau(a)=0}}
R(a,\tau|_{\{1,\ldots,a-1\}})
\ \ ,
\ee
where the sum is over all maps $\tau:\{1,\ldots,r\}
\rightarrow\{0,1\}$.
\end{lem}

\noindent
{\bf Proof:\ \ }
By induction on $r$.
The case $r=1$ is obvious.
If the statement holds for $r\ge1$, then the right hand side
for $r+1$ may be factorized as:
\[
\sum_{\tau|_{\{1,\ldots,r\}}}
\prod_{{1\le a\le r}\atop{\tau(a)=1}}
(1-R)(a,\tau|_{\{1,\ldots,a-1\}})\times
\prod_{{1\le a\le r}\atop{\tau(a)=0}}
R(a,\tau|_{\{1,\ldots,a-1\}})
\]
\be
\times
\lp (1-R)(r+1,\tau|_{\{1,\ldots,r\}})+R(r+1,\tau|_{\{1,\ldots,r\}})\rp
=1
\ \ ,
\ee
by the induction hypothesis at stage $r$ applied to the sum over
the restriction
$\tau|_{\{1,\ldots,r\}}$.
\Endproof

To prove (\ref{idcond}), we introduce an ordering $\De_1,\ldots,\De_r$ for the
elements of $\cD$, such that we enumerate first the cubes of
$\cD_N$, then of $\cD_{N-1}$, and up to $\cD_0$.
We have therefore an identification between subsets $\Ga$
of $\cD$ and their characteristic functions $\tau:\{1,\ldots,r\}\rightarrow
\{0,1\}$
defined by $\tau(a)=1$ if $\De_a\in\Ga$ and 0 otherwise.
We take the function $R$ defined by
\be
R(a,\tau|_{\{1,\ldots,a-1\}})=
\ch\lp g^{(1+\ep_1)}
\int_{\De_a}{\lp\sum_{i\in I_\Ga(\De_a)}\ph_i(x)\rp}^4
dx\rp
\ \ .
\ee
This is a consistent definition.
Indeed by our numbering, $I_\Ga(\De_a)$
is the union of $\{\De_a\}$ with a set that depends only on $\De_a$
and
$\Ga\cap[\cD_{i(\De_a)+1}\cup\cD_{i(\De_a)+2}\cup\ldots
\cup\cD_N]$ whose set of labels is included in $\{1,\ldots,a-1\}$.
Hence $I_\Ga(\De_a)$ only depends on $\tau|_{\{1,\ldots,a-1\}}$.
Finally (\ref{idcond}) follows from Lemma \ref{idelem}.

\subsubsection{The expansion}

Let us study for instance the unnormalized Schwinger function
\[
S(\xi_1,\al_1;\ldots;\xi_n,\al_n)=
\int d\mu_{C[{\bf 1}]}((\ph_i)_{0\le i\le N})
\ \ph_{\al_1}(\xi_1)\ldots
\ph_{\al_n}(\xi_n)
\]
\be
.\exp\lp
-\sum_{j_1,\ldots,j_4=0}^N
\int_{\La^4}dy_1\ldots dy_4
\ K[{\bf 1}](y_1,j_1;\ldots;y_4,j_4)
.\phi_{j_1}(y_1)\ldots\phi_{j_4}(y_4)
\rp
\ \ .
\ee
We begin by inserting the relation (\ref{idcond}) in the exponential so that
\be
S(\xi_1,\al_1;\ldots;\xi_n,\al_n)
=\sum_{\Ga\subset\cD}
H_\Ga({\bf 1})
\ \ ,
\ee
where $H_\Ga((t_l)_{l\in L})$ is the smooth function on $\Om$ defined
by
\[
H_\Ga({\bf t})\eqdef
\int d\mu_{C[{\bf t}]}((\ph_i)_{0\le i\le N})
\ \ph_{\al_1}(\xi_1)\ldots
\ph_{\al_n}(\xi_n)
.\ch_\Ga((\ph_i))
\]
\be
.\exp\lp
-\sum_{j_1,\ldots,j_4=0}^N
\int_{\La^4}dy_1\ldots dy_4
\ K[{\bf t}](y_1,j_1;\ldots;y_4,j_4)
.\phi_{j_1}(y_1)\ldots\phi_{j_4}(y_4)
\rp
\ \ .
\ee
All we need, to interpolate $H_\Ga({\bf 1})$ along the result of Section 2.1,
is to specify the connectivity map $\Gg\mapsto\Pi_\Ga(\Gg)$.
The idea is that blocks of the partition $\Pi_\Ga(\Gg)$ will correspond to
convergent polymers, i.e.
those without internal ``almost local'' 2 and 4 point functions
that are responsible of the divergences in the bare theory, see [R1].
A block of $\Pi_\Ga(\Gg)$ will be a patch of large field regions made
of vertically neighboring cubes, and of isolated small field cubes,
linked together by propagators, and vertices.
Propagators will give a strong connection, whereas vertices will provide
a weak connection. At least five vertices are needed to hook
an almost-local insertion to the background
or low momentum cubes of the polymer
(corresponding
to higher indices $i(\De)$).
Therefore an analog for the vertices of 4-particle irreducibility
appears in this formalism.

To make things more precise, let us first make some combinatorial
definitions.
We introduce a gluing notion between cubes as follows.
Two cubes are said {\und{glued}} if we can order them as $\De_1$,
$\De_2$ with $i(\De_2)=i(\De_1)+1$, $\De_2\in\Ga$,
and $\De_1\subset\De_2$.
Connected components according to gluing are either isolated
small field cubes,
or a \und{dressed large field block} made of vertically neighboring
large field cubes together with the layer of small field
cubes just on top.
Remark that for $\De\in\Ga\cap\cD_0$ there are no such small field
cubes above $\De$; therefore they are not included in the dressed large
field block containing $\De$.
Two elements $\De$ and $\De'$ of $\cD$, such that
a 2 or 4-link $l$ occurs in some fixed ordered graph
$\Gg$, and satisfies $\{\De,\De'\}\subset
{\rm supp}\ l$, are said to be {\und{joined}}.
Suppose $\De$ and $\De'$ are in $\cD$, and there exists a sequence of cubes
$\De_1=\De,\De_2,\ldots,\De_u=\De'$, $u\ge 1$, with
the property that for any $v$,
$1\le v\le u-1$, $\De_v$ and $\De_{v+1}$ are either glued or joined;
then we declare $\De$ and $\De'$ to be
{\em connected} by $\Gg$ and $\Ga$.
If $V\subset\cD$, $V$ is said to be {\em connected} by $\Gg$
and $\Ga$, if
any distinct cubes $\De$ and $\De'$ of $V$ are connected by $\Gg$
and $\Ga$.
The subsequence of $\Gg\in\cG$ where we delete all $p$-link
$l$ with ${\rm supp}\ l\not\subset V$ and
keep the ordering of the remaining links, is denoted by $\Gg_V$.
Finally, $V$ is called a {\und{4-vertex irreducible set}} for
some ordered graph $\Gg$
and a large field region $\Ga$ (we abreviate by 4-VI),
if any graph obtained from
$\Gg_V$ after deleting at most four 4-links still connects $V$.
Now $\Pi_\Ga(\Gg)$ is by definition the set of {\em maximal}
4-VI subsets in $\cD$ for $\Gg$ and $\Ga$.
The blocks of $\Pi_\Ga(\Gg)$ are
called the {\und{4-VI components}}
of $\Br \Gg$ and $\Ga$.
Beware that 4-vertex irreducibility depends on the shape of $\Ga$.
Gluing as well as joining by 2-links provide strong connections,
but joining by 4-links is a weak connection:
at least 5 of them are needed to make up a strong bond. 

\begin{lem}
\label{partition}
$\Pi_\Ga(\Gg)$ thus defined is indeed a partition of $\cD$.
\end{lem}

\noindent
{\bf Proof:\ \ }
Any singleton is 4-VI, therefore $\emptyset\notin
\Pi_\Ga(\Gg)$ and $\Pi_\Ga(\Gg)$ covers $\cD$.
Let $V_1$ and $V_2$ be two elements of $\Pi_\Ga(\Gg)$
with
$V_1\cap V_2\neq\emptyset$.
Let $\Gg'$ be a subsequence obtained from $\Gg_{V_1\cup V_2}$ by deleting
at most four 4-links, then $\Gg'_{V_1}$ is also obtained from $\Gg_{V_1}$
by deleting at most four 4-links.
Since $V_1$ is 4-VI, $\Gg'_{V_1}$ connects $V_1$, and so does $\Gg'$ of which
$\Gg'_{V_1}$ is a subsequence.
Likewise $\Gg'$ connects $V_2$, and since $V_1\cap V_2\neq
\emptyset$, $\Gg'$ finally
connects $V_1\cup V_2$. We have shown that $V_1\cup V_2$ is 4-VI
and by maximality, we get $V_1\cup V_2=V_1=V_2$.
\Endproof

It is obvious that the map $\Gg\mapsto
\cC(\Gg)={\rm Offdiag}(\Pi_\Ga(\Gg))$, is a choice map i.e.
$\Gg_1\le \Gg_2$ implies $\cC(\Gg_2)\subset\cC(\Gg_1)$.
More generally if $\Gg=(l_1,\ldots,l_k)$ and $(l_{j_1},\ldots,l_{j_r})$
is any of its
subsequences, then the partition $\Pi_\Ga((l_{j_1},\ldots,l_{j_r})$
is finer than $\Pi_\Ga(\Gg)$ and thus
\be
{\rm Offdiag}(\Pi_\Ga(\Gg))
\subset
{\rm Offdiag}(\Pi_\Ga((l_{j_1},\ldots,l_{j_r}))
\ \ .
\ee
It is easy to see that we cannot have, in an allowed ordered graph,
a link appearing more than five times. Since the number
of all possible links is finite, this ensures the boundedness of
the length of allowed ordered graphs.

We may now write the full expansion for the Schwinger function,
it is an explicit although titanic formula: 
\[
S(\xi_1,\al_1;\ldots;\xi_n,\al_n)
=\sum_{\Ga\subset\cD}
\sum_{{\Gg=(l_1,\ldots,l_k)}\atop{\Gg\in\cA\cG_\Ga}}
\int_{1\ge h_1\ge\cdots\ge h_k\ge 0}
dh_1\ldots dh_k
\int d\mu_{C[T_{\Gg,\Ga}({\bf h})]}\lp(\ph_i)_{0\le i\le N}\rp
\]
\[
\times
\prod_{{1\le j\le k}\atop{l_j\in\cL_2}}
\lp
\frac{1}{2}\sum_{{(\De^j_1,\De^j_2)\in \cD^2}\atop{l[\De^j_1,\De^j_2]=l_j}}
\int_{\De^j_1}dx^j_1\int_{\De^j_2}dx^j_2
\ C(x_1^j,i(\De^j_1);x_2^j,i(\De^j_2))
\frac{\de}{\de\ph_{i(\De^j_1)}(x_1^j)}\frac{\de}{\de\ph_{i(\De^j_2)}(x_2^j)}
\rp
\]
\[
\times
\ph_{\al_1}(\xi_1)\ldots
\ph_{\al_n}(\xi_n)
\times
\ch_\Ga((\ph_i))
\]
\[
\times
\prod_{{1\le j\le k}\atop{l_j\in\cL_4}}
\lp
-\sum_{{(\De^j_1,\ldots,\De^j_4)\in \cD^4}
\atop{l[\De^j_1,\ldots,\De^j_4]=l_j}}
\int_{\De^j_1}dx^j_1
\ldots
\int_{\De^j_4}dx^j_4
\ K(x^j_1,i(\De^j_1);\ldots;x^j_4,i(\De^j_4))
.\ph_{i(\De^j_1)}(x^j_1)\ldots
\ph_{i(\De^j_4)}(x^j_4)
\rp
\]
\be
\times\exp\lp
-\sum_{i_1,\ldots,i_4=0}^N
\int_{\La^4}
dx_1\ldots dx_4
\ K[T_{\Gg,\Ga}({\bf h})](x_1,i_1;\ldots;x_4,i_4)
.\ph_{i_1}(x_1)\ldots\ph_{i_4}(x_4)
\rp
\ \ .
\label{raw}
\ee
This is the expression we obtain by applying formula (\ref{inter1}) on
$H_\Ga$ for a fixed $\Ga$, then summing over it.
Note that our preferred order in which we compute the derivatives
$\frac{\partial^k}{\partial t_{l_1}\ldots\partial t_{l_k}}$
on $H$ is by applying first the $\partial/\partial t_l$
for $l\in\cL_4$, thus producing explicit vertices,
then doing the derivations for $l\in\cL_2$.
In this way, we produce functional differential operators
that act on the whole integrand consisting in external fields,
explicit vertices, large field conditions and the exponential
of the interpolated interaction.However the raw expression of (\ref{raw})
needs further treatment to display its factorizations
and to rewrite it as a polymer expansion.

If $Y\subset\cD$ is non empty, $Y$ is what we call a {\em polymer}.
Given $(b_1,\ldots,b_m)$, a sequence of not necessarily distinct
cubes of $Y$, and $\ze_1,\ldots,\ze_m$ any points in $\La$ such that
$\ze_q\in b_q$, for every $q$,
$1\le q \le m$, we define the {\em activity}
of $Y$ with insertions $(b_q,\ze_q)_{1\le q\le m}$ as:
\[
\cA(Y;(b_q,\ze_q)_{1\le q\le m})\eqdef
\sum_{\Ga_Y}
\sum_{{{\Gg_Y=(l_1,\ldots,l_k)}\atop{\Gg_Y\in\cA\cG_{\Ga_Y,Y}}}
\atop{Y\ {\rm 4-VI}}}
\int_{1\ge h_1\ge\cdots\ge h_k\ge 0}
dh_1\ldots dh_k
\int d\mu_{C[T_{\Gg_Y,\Ga_Y}({\bf h})]_Y}(\ph_Y)
\]
\[
\times\prod_{{1\le j\le k}\atop{l_j\in\cL_2}}
\lp
\frac{1}{2}\sum_{{(\De^j_1,\De^j_2)\in Y^2}\atop{l[\De^j_1,\De^j_2]=l_j}}
\int_{\De^j_1}dx^j_1\int_{\De^j_2}dx^j_2
\ C(x_1^j,i(\De^j_1);x_2^j,i(\De^j_2))
\frac{\de}{\de\ph^Y_{i(\De^j_1)}(x_1^j)}
\frac{\de}{\de\ph^Y_{i(\De^j_2)}(x_2^j)}
\rp
\]
\[
\times
\ph^Y_{i(b_1)}(\ze_1)\ldots
\ph^Y_{i(b_n)}(\ze_n)
\times
\ch_{\Ga_Y,Y}((\ph_Y))
\]
\[
\times\prod_{{1\le j\le k}\atop{l_j\in\cL_4}}
\lp
-\sum_{{(\De^j_1,\ldots,\De^j_4)\in Y^4}
\atop{l[\De^j_1,\ldots,\De^j_4]=l_j}}
\int_{\De^j_1}dx^j_1
\ldots
\int_{\De^j_4}dx^j_4
\ K(x^j_1,i(\De^j_1);\ldots;x^j_4,i(\De^j_4))
.\ph^Y_{i(\De^j_1)}(x^j_1)\ldots
\ph^Y_{i(\De^j_4)}(x^j_4)
\rp
\]
\[
\times\exp\Biggl(
-\sum_{(\De_1,\ldots,\De_4)\in Y^4}
\int_{\De_1}dx_1
\ldots 
\int_{\De_4}dx_4
\]
\be
K[T_{\Ga_Y,\Gg_Y}({\bf h})](x_1,i(\De_1);\ldots;x_4,i(\De_4))
.\ph^Y_{i(\De_1)}(x_1)\ldots\ph^Y_{i(\De_4)}(x_4)
\Biggr)
\ \ .
\label{activ}
\ee

Here we sum $\Ga_Y$ over all subsets of $Y$ with the property that
if $\De_1\in\cD$, $\De_2\in Y$, $i(\De_2)=i(\De_1)+1$ and $\De_1\subset
\De_2$, then $\De_1\in Y$ also.
This means that every cube just on top of $\Ga_Y$
must be in $Y$.
We denote by $\cA\cG_{\Ga_Y,Y}$ the set of allowed ordered graphs in $Y$,
defined in the same way as we did in $\cD$.
This means that we consider $\cG_Y$,
the set of ordered graphs $\Gg_Y=(l_1,\ldots,l_k)$
such that for any $a$, $1\le a\le k$, $l_a\in L_Y\eqdef
\{l\in L|\ {\rm supp}\ l\subset Y\}$,
then we introduce a connectivity map
$\Pi_{\Ga_Y,Y}:\cG_Y\rightarrow {\rm Part}(Y)$ depending on the large field
region $\Ga_Y$.
The blocks of $\Pi_{\Ga_Y,Y}(\Gg_Y)$ are the 4-VI components of $Y$,
with the gluing and joining notions restricted to cubes of $Y$.
Finally we put $\Gg_Y\in\cA\cG_{\Ga_Y,Y}$ if and only if for any
$a$, $1\le a\le k$, $l_a\in\cC((l_1,\ldots,l_{a-1}))\eqdef
{\rm Offdiag}(\Pi_{\Ga_Y,Y}((l_1,\ldots,l_{a-1})))$.
The sum in (\ref{activ}) is over every
$\Gg_Y\in\cA\cG_{\Ga_Y,Y}$ that makes $Y$
into a single 4-VI component, i.e. such that $\Pi_{\Ga_Y,Y}(\Gg_Y)
=\{Y\}$.
The notation $T_{\Gg_Y,\Ga_Y}({\bf h})$
is for the vector $(t_l)_{l\in L_Y}$ defined
from $(h_1,\ldots,h_k)$ exactly as in Lemma \ref{inter}
but only for $l\in L_Y$.
The functional integral is on a collection $(\phi_Y)$
of fields supported on $Y$, that is 
\be
(\phi_Y)\eqdef
(\phi_i^Y)_{{i\in\{0,\ldots,N\}}\atop{\cD_i\cap Y\ne\emptyset}}
\ \ ,
\ee
where for any $i$,
$\phi_i^Y(x)$ is a random field on $\cup_{\De\in\cD_i\cap Y}\De$.
The measure is Gaussian, with covariance $C[T_{\Gg_Y,\Ga_Y}({\bf h})]_Y$
defined as before in (\ref{cov}) but restricted to entries $(x,i)$ with
$\De(x,i)\in Y$.
Finally we pose:
\[
\ch_{\Ga_Y,Y}((\ph_Y))=
\prod_{\De\in\Ga_Y}
(1-\ch)\lp g^{(1+\ep_1)}\int_\De
{\lp\sum_{i\in I_{\Ga_Y}(\De)}\ph^Y_i(x)\rp}^4
dx\rp
\]
\be
\times
\prod_{\De\in Y\backslash\Ga_Y}
\ch\lp g^{(1+\ep_1)}\int_\De{\lp\sum_{i\in I_{\Ga_Y}(\De)}\ph_i^Y(x)\rp}^4
dx\rp
\ \ .
\label{lfcond}
\ee

Remark that
equality of sequences modulo permutation of the elements is an equivalence
relation in $\cG$.
Any equivalence class is called an {\em unordered graph}.
If $\Gg\in\cG$, the class of $\Gg$ is denoted by 
$\Br \Gg$.
We may also, being a little sloppy with notations,
characterize $\Br \Gg$ as a map
${\Br \Gg}:L\rightarrow\NN$, that simply counts the occurrences of a link
$l\in L$, in an arbitrary representative $\Gg$ of $\Br \Gg$.
One has naturally an ordering ${\Br \Gg}_1\le{\Br \Gg}_2$ if
and only if ${\Br \Gg}_1(l)\le {\Br \Gg}_2(l)$ for any $l\in L$.
One easily sees that for an ordered graph $\Gg$, $\Pi_\Ga(\Gg)$
does not depend on the ordering of links in $\Gg$; however
their multiplicity is relevant. As a result, the map
$\Gg\mapsto\Pi_\Ga(\Gg)$ naturally factorizes through
$\Gg\mapsto{\Br \Gg}\mapsto{\Br \Pi}_\Ga({\Br \Gg})\eqdef
\Pi_\Ga(\Gg)$.
We will hereupon use the same terminology of 4-VI sets and components
for ordered and unordered graphs. 

Given a partition $\pi$ of $\cD$, and a $p$-link $l:\cD\rightarrow\NN$,
for which we must have $\sum_{\De\in\cD}l(\De)=p$,
we define the {\und{reduced $p$-link}} $l^\pi$, on the set $\pi$
instead of $\cD$, as follows: $l^\pi$ is the mapping
$\pi\rightarrow\NN$, such that for all $B\in\pi$,
$l^\pi(B)\eqdef\sum_{\De\in B}l(\De)$.
Intuitively, $l^\pi$ is the link
obtained by contracting any block of $\pi$ to a point.
We also define the {\und{reduced ordered graph}} $\Gg^\pi$ on $\pi$ from an
ordered graph
$\Gg=(l_1,\ldots,l_k)$ of links in $\cD$, by $\Gg^\pi\eqdef(l^\pi_1,\ldots,
l^\pi_k)$. The map $\Gg\mapsto{\Br{\Gg^\pi}}$ factorizes naturally through
a map ${\Br \Gg}\mapsto {\Br \Gg}^\pi\eqdef {\Br{\Gg^\pi}}$ of unordered
graphs.
Now consider a partition $\pi$ of $\cD$, for which there are no pairs
of glued cubes $\De$ and $\De'$ falling in two distinct blocks of
$\pi$. If $G$ is an unordered graph on $\pi$, we can define ${\Br\Pi}^\pi(G)$
the set of 4-VI subsets of $\pi$ with respect to the graph $G$.
Simply we apply the same definition as for $\cD$,
using 2 and 4-links on $\pi$,
but only considering joined elements of $\pi$ to define connections.
The gluing notion for elements of $\pi$, is already embodied in them.
We denote by $0_\pi$ the trivial partition
$\{\{B\}|B\in\pi\}$ of $\pi$.

Given an {\em unordered graph} ${\Br \Gg}$,
we define its symmetry factor $\si({\Br \Gg})\eqdef\prod_{l\in L}({\Br \Gg}
(l))!$, recalling that ${\Br \Gg}(l)$ is the number of occurrences
of $l$ in $\Gg$ or its multiplicity.
Likewise we define for any $l\in L$ the number
\be
\rho(l)\eqdef
\frac{(\sum_{\De\in\cD}l(\De))!}{\prod_{\De\in\cD}(l(\De))!}
\ \ .
\ee

\begin{th}
We can write the following polymer expansion:
\[
S(\xi_1,\al_1;\ldots;\xi_n,\al_n)=
\sum_{\pi\in{\rm Part(\cD)}}
\sum_{{{\Br {\Gg_{\rm ext}}}=
{\Br{(l_1,\ldots,l_k)}}}\atop{{\Br \Pi}^\pi({\Br {\Gg_{\rm ext}}}^\pi)
=0_\pi}}
\frac{1}{\si({\Br {\Gg_{\rm ext}}})}
\prod_{j=1}^k
\lp
\rho(l_j)
\int_{b_1^j}d\ze_1^j
\ldots
\int_{b_4^j}d\ze_1^4
\rp
\]
\be
\prod_{Y\in\pi}
\cA(Y;(b_\nu^j,\ze_\nu^j)_Y,(\De(\xi_q,\al_q),\xi_q)_Y)
\ \ .
\ee

$\Br {\Gg_{\rm ext}}$ is summed over all unordered graphs on $\cD$
made of links in ${\rm Offdiag}(\pi)$ and which satisfy
${\Br \Pi}^\pi({\Br {\Gg_{\rm ext}}}^\pi)=0_\pi$.
This means that the reduced graph of ${\Br {\Gg_{\rm ext}}}$
on $\pi$ has no non
trivial 4-VI components.
Remark that there can be no 2-link in $\Br {\Gg_{\rm ext}},$
since this would create
non trivial 4-VI components in $\pi$. Here
$(l_1,\ldots,l_k)$ is only a choice of ordered representative for
$\Br {\Gg_{\rm ext}}$.
For any $j$, $1\le j\le k$, $(b_1^j,\ldots,b_4^j)$ is a choice of boxes
such that
$l[b_1^j,\ldots,b_4^j]=l_j$. Finally
$(b_\nu^j,\ze_\nu^j)_Y$ is the subfamily of
$(b_\nu^j,\ze_\nu^j)_{{1\le j\le k}\atop{1\le \nu\le 4}}$
corresponding to the entries $\nu$ and $j$ for which $b_\nu^j\in Y$.
By the same token,
$(\De(\xi_q,\al_q),\xi_q)_Y$ is the subfamily of
$(\De(\xi_q,\al_q),\xi_q)_{1\le q\le m}$ corresponding to indices
$q$ verifying $\De(\xi_q,\al_q)\in Y$.
\end{th}

\noindent
{\bf Proof:\ \ }
Starting from (\ref{raw}), the first thing to do is to organize the sum
according to values $\pi$ of $\Pi_\Ga(\Gg)$.
Then if $\pi=\{Y_1,\ldots,Y_r\}$,
we decompose $\Ga$ into $\Ga_{Y_1}\cup\ldots\cup
\Ga_{Y_r}$ where $\Ga_{Y_a}\eqdef\Ga\cap Y_a$.
We see that summing over $\Ga\subset\cD$ is the same as
summing independently on the $\Ga_{Y_a}$'s among subsets of $Y_a$
respectively,
with the restriction that if
$\De_1\in\cD$, $\De_2\in \Ga_{Y_a}$, and $\De_1$ is just above
$\De_2$,
then $\De_1\in Y_a$.
Indeed, $\De_1$ and $\De_2$ would then be glued and
accordingly would have to
fall in the same component
$Y_a$.

The sum over $\Gg=(l_1,\ldots,l_k)$ in $\cA\cG_\Ga$ also splits into
independent sums. Denote by $\Gg_{\rm ext}$ the subsequence
of $\Gg$ that remains after we extract all the disjoint subsequences
$\Gg_{Y_1},\ldots,\Gg_{Y_r}$, that correspond to internal subgraphs
of the
blocks of $\pi$.
To show the independence of the sums over $\Gg_{Y_1},\ldots,\Gg_{Y_r}$
and $\Gg_{\rm ext}$ we need the following lemmas.
The independence of the sums over $\Gg_{Y_1},\ldots,\Gg_{Y_r}$ and
$\Gg_{\rm ext}$ is intuitively trivial, but the formal proof
of such a statement tediously goes along the following
lemmas the trustful reader may safely skip.

\begin{lem}
\label{intrinsic}
Let $\Gg=(l_1,\ldots,l_k)$ be in $\cG$,
not necessarily allowed, and $V$ be a non empty union of blocks from
$\Pi_\Ga(\Gg)$.
Denote by $\Ga_V$ the intersection of $\Ga$ with $V$.
$\Pi_{\Ga_V,V}(\Gg_V)$ is by definition a partition of $V$, namely that of
4-VI components of $V$,
made from the large field region $\Ga_V$ and the ordered graph $\Gg_V$
obtained from $\Gg$ by keeping the internal links of $V$.

We then have the following equalities:
\be
\Pi_{\Ga_V,V}(\Gg_V)=
\Pi_\Ga(\Gg)\cap\cP(V)=
\Pi_\Ga(\Gg)|_V
\ \ ,
\ee
where $\cP(V)$ denotes the power set of $V$, and $\pi|_V$ denotes the
partition of $V$ made by the non empty sets of the form $A\cap V$,
$A\in\pi$.
\end{lem}

\noindent
{\bf Proof:\ \ }
Since $V$ is a non empty union of blocks $\Pi_\Ga(\Gg)$, it is clear that
$\Pi_\Ga(\Gg)\cap\cP(V)=\Pi_\Ga(\Gg)|_V$;
therefore $\Pi_\Ga(\Gg)\cap\cP(V)$ is a partition of $V$.
Now let $W$ be a non empty subset of $V$;
we claim that $W$ is 4-VI for $\Gg_V$ and $\Ga_V$
if and only if $W$ is 4-VI for $\Gg$ and $\Ga$.

Suppose $W$ is 4-VI in $V$ for the graph $\Gg_V$ and the large
field region $\Ga_V$, and let $\Gg'$ by any subsequence obtained from
$\Gg_W$ by deleting at most four 4-links.
Since $W\subset V$ we have $\Gg_W=(\Gg_V)_W$, and therefore $\Gg'$ is
obtained from $(\Gg_V)_W$ by deleting at most four 4-links.
Since $W$ is 4-VI for $\Gg_V$ and $\Ga_V$, we can connect any
pair of cubes in $W$, by a chain of elements in $W$ that are successively
glued by $\Ga_V$ or joined by $\Gg'$.
But $\Ga_V\subset\Ga$, so that the cubes of the chain are
successively glued by
$\Ga$ or joined by $\Gg'$.
We have thus proven that $W$ is 4-VI for $\Gg$ and $\Ga$.

Suppose $W$ is 4-VI for $\Gg$ and $\Ga$, and let $\Gg'$ be any subsequence
obtained from $(\Gg_V)_W$ by deleting at most four 4-links.
Since $(\Gg_V)_W=\Gg_W$ and $W$ is 4-VI for $\Gg$ and $\Ga$,
we conclude that any pair of cubes in $W$ is connected by
a chain of cubes in $W$ that are successively glued by $\Ga$ or joined
by $\Gg'$. But if two elements of $W$ are glued by $\Ga$,
we can for instance denote them by $\De_1$, $\De_2$ with
$\De_2\in\Ga$ and $\De_1$ just above $\De_2$.
Since $\De_2\in W\subset V$, we obtain by definition
of $\Ga_V$, that $\De_2\in \Ga_V$. Therefore $\De_1$ and $\De_2$ are
glued by $\Ga_V$. It follows from this remark that $W$ will be 4-VI
for $\Gg_V$ and $\Ga_V$.

Now let $W\in\Pi_\Ga(\Gg)\cap\cP(V)$, $W$ is 4-VI for $\Gg$ and $\Ga$,
and therefore also for $\Gg_V$ and $\Ga_V$ since $W\subset V$.
Let $W'$ be a subset of $V$, containing $W$, that is
4-VI for $\Gg_V$ and $\Ga_V$. By the other implication of the
above proven equivalence we obtain that $W'$ is 4-VI for
$\Gg$ and $\Ga$. By maximality of $W$ for this last
property due to $W\in\Pi_\Ga(\Gg)$, we get $W'=W$. As a result,
$W$ is maximal with respect to 4-vertex irreducibility for $\Gg_V$ and
$\Ga_V$, and thus $W\in\Pi_{\Ga_V,V}(\Gg_V)$.

We have proven $\Pi_\Ga(\Gg)\cap\cP(V)\subset\Pi_{\Ga_V,V}(\Gg_V)$,
but both are partitions of $V$. This mere property entails
$\Pi_\Ga(\Gg)\cap\cP(V)=\Pi_{\Ga_V,V}(\Gg_V)$.
\Endproof

\begin{lem}
\label{pardec}
Let $\Gg=(l_1,\ldots,l_k)$ be a sequence in $\cG$, $\Ga$ be a large field
region, and let $\Pi_\Ga=\{Y_1,\ldots,Y_r\}$.
Then we have the following equivalence:
\be
\Gg\in\cA\cG_\Ga\;\Leftrightarrow\;
\forall a,\ 1\le a\le r,\;
\Gg_{Y_a}\in\cA\cG_{\Ga_{Y_a},Y_a}
\ \ .
\ee
\end{lem}

\noindent
{\bf Proof:\ \ }
Suppose $\Gg\in\cA\cG_\Ga$ and for some $a$,
$1\le a\le r$, that $\Gg_{Y_a}$ is the subsequence $(l_{j_1},\ldots,
l_{j_u})$ of $\Gg$.
For every $v$, $1\le v\le u$, since $\Pi_\Ga((l_{j_1},l_{j_2},\ldots,
l_{j_{(u-1)}}))\le\Pi_\Ga(\Gg)$, we infer that $Y_a$ is a union of blocks in
$\Pi_\Ga((l_{j_1},l_{j_2},\ldots,l_{j_{(u-1)}}))$.
As a result, Lemma \ref{intrinsic} applies and gives
\be
\Pi_{\Ga_{Y_a},Y_a}((l_{j_1},l_{j_2}.\ldots,l_{j_{(u-1)}}))=
\Pi_\Ga((l_{j_1},l_{j_2},\ldots,l_{j_{(u-1)}}))|_{Y_a}
\ \ .
\label{partint}
\ee
But $\Gg\in\cA\cG_\Ga$ entails
\be
l_{j_v}\in{\rm Offdiag}(
\Pi_\Ga((l_1,l_2,\ldots,l_{(j_v)-1})))
\subset
{\rm Offdiag}(\Pi_\Ga((l_{j_1},l_{j_2},\ldots,l_{j_{(u-1)}})))
\ \ .
\ee
Now (\ref{partint}) and $supp\ l_{j_v}\subset Y_a$, imply
\be
l_{j_v}\in
{\rm Offdiag}(\Pi_{\Ga_{Y_a},Y_a}((l_{j_1},l_{j_2}.\ldots,l_{j_{(u-1)}})))
\ \ .
\ee
By definition, we have just checked that $\Gg_{Y_a}\in\cA\cG_{\Ga_{Y_a},Y_a}$.

Conversely, suppose that for any $a$, $1\le a\le r$,
$\Gg_{Y_a}\in\cA\cG_{\Ga_{Y_a},Y_a}$, and
consider j, any integer such that $1\le j\le k$.
If $l_j$ is in the subsequence $\Gg_{\rm ext}$ of $\Gg$ then
\be
l_j\in{\rm Offdiag}(\Pi_\Ga(\Gg))
\subset
{\rm Offdiag}(\Pi_\Ga((l_1,l_2,\ldots,l_{j-1})))
\ \ .
\ee
Else, $l_j$ must belong to some subsequence $\Gg_{Y_a}=
(l_{\mu_1},\ldots,l_{\mu_u})$ of $\Gg$.
Suppose $j=\mu_v$ for some $v$, $1\le v\le u$.
Since $\Pi_\Ga((l_1,l_2,\ldots,l_{j-1}))\le\Pi_\Ga(\Gg)$,
$Y_a$ is a union of blocks of $\Pi_\Ga((l_1,l_2,\ldots,l_{j-1}))$.
Therefore Lemma \ref{intrinsic} applies and shows that
\bea
\Pi_\Ga((l_1,l_2,\ldots,l_{j-1}))|_{Y_a} & = &
\Pi_{\Ga_{Y_a},Y_a}((l_1,l_2,\ldots,l_{j-1})_{Y_a}) \\
 & = & \Pi_{\Ga_{Y_a},Y_a}((l_{\mu_1},l_{\mu_2},\ldots,l_{\mu_{(v-1)}}))
\ \ .\label{indep}
\eea
Since $\Gg_{Y_a}\in\cA\cG_{\Ga_{Y_a},Y_a}$, we have
$l_j=l_{\mu_v}\in
{\rm Offdiag}
(\Pi_{\Ga_{Y_a},Y_a}((l_{\mu_1},l_{\mu_2},\ldots,l_{\mu_{(v-1)}})))$,
and from (\ref{indep}) we conclude that
$l_j\in{\rm Offdiag}(\Pi_\Ga((l_1,l_2,\ldots,l_{j-1})))$.
We have thus checked that $\Gg\in\cA\cG_\Ga$.
\Endproof

\begin{lem}
If $\Gg$ is a sequence in $\cG$, $\pi=\Pi_\Ga(\Gg)$ and $\Gg_{\rm ext}$
is the subsequence of $\Gg$ obtained by keeping the links that belong
to ${\rm Offdiag}(\pi)$, then
\be
\Pi^\pi(\Gg_{\rm ext}^\pi)=0_\pi
\ \ .
\ee
\end{lem}

\noindent
{\bf Proof:\ \ }
Let $\cV$ be a 4-VI subset of $\pi$ for the reduced graph $\Gg_{\rm ext}^\pi$,
and define $V=\cup_{B\in\cV}B$.
We claim that $V$ is a 4-VI subset of $\cD$ for $\Gg$ and $\Ga$.
To prove this, we consider $\Gg'$, any subsequence of
$\Gg_V$ obtained by deleting at most four 4-links.
For any $B\in\cV$, ${\Gg'}_B$ is obtained by deleting at most four
4-links from $(\Gg_V)_B=\Gg_B$.
But $B$ is 4-VI for $\Gg$ and $\Ga$, therefore ${\Gg'}_B$ and hence
$\Gg'$ connect $B$.

Remark that $(\Gg')^\pi$ is obtained from $(\Gg_V)^\pi$ by deleting
at most four 4-links, over $\pi$ this time. Remark also that
$(\Gg_V)^\pi=(\Gg^\pi)_\cV$; and since $\cV$ is 4-VI for
$\Gg_{\rm ext}^\pi$ and consequently for $\Gg^\pi$, we deduce
that $(\Gg')^\pi$ connects $\cV$.

If $\De\in V$, we denote by $B(\De)$ the unique block $B\in\cV$
such that $\De\in B$. Now let $\De$ and $\De'$ be two elements in $V$.
Since $(\Gg')^\pi$ connects $\cV$, there exists a chain
$B_1,\ldots,B_u$, $u\ge 1$, in $\cV$ with $B_1=B(\De)$,
$B_u=B(\De')$, and such that for any $v$, $1\le v\le u-1$,
there is a link $l_j$ in $\Gg'$ such that
$B_v$ and $B_{v+1}$ belong to the support of $l_j^\pi$.
For any $v$, $1\le v\le u-1$, pick a cube $\De_v^\leftarrow$
in $B_v$ and a cube $\De_{v+1}^\rightarrow$ in $B_{v+1}$,
such that $\De_v^\leftarrow$ and $\De_{v+1}^\rightarrow$
belong to the support of the above mentioned link $l_j$.
For any $v$, $\De_v^\leftarrow$ and $\De_{v+1}^\rightarrow$
are joined by $\Gg'$.
Since $\Gg'$ connects any $B\in\cV$, we have that
$\De$ and $\De_1^\leftarrow$ are connected by $\Gg'$, also that
for any $v$, $2\le v\le u-1$, $\De_v^\rightarrow$ and
$\De_v^\leftarrow$ are connected by $\Gg'$, and finally
that $\De_u^\rightarrow$ and $\De'$ are connected by $\Gg'$.
Therefore $\Gg'$ connects $\De$ and $\De'$.
This shows that $V$ is 4-VI for $\Gg$ and $\Ga$,
and thus cannot be made by more than one block $B\in\pi$,
so that $\cV$ has to be a singleton.
Finally $\Pi^\pi(\Gg_{\rm ext}^\pi)=0_\pi$, the trivial partition.
\Endproof

We need also the converse statement

\begin{lem}
Let $\pi$ be some partition of $\cD$. Let for any $Y\in\pi$,
$\Ga_Y$ be a subset of $Y$ satisfying the condition that any cube, just above
a cube of $\Ga_Y$, is in $Y$.
Let also $\Gg_Y$ be a sequence of links with support in $Y$, such that
$\Pi_{\Ga_Y,Y}(\Gg_Y)=\{Y\}$. Let $\Gg_{\rm ext}$ be a sequence made of links
in ${\rm Offdiag}(\pi)$, such that $\Pi^\pi(\Gg_{\rm ext}^\pi)=0_\pi$.
Finally, let $\Ga=\cup_{Y\in\pi}\Ga_Y$.
Then for any sequence $\Gg$ obtained by arbitrary intertwining of the
$\Gg_Y$, for $Y\in\pi$, together with $\Gg_{\rm ext}$ we have
\be
\Pi_\Ga(\Gg)=\pi
\ \ .
\ee
\end{lem}

\noindent
{\bf Proof:\ \ }
Any $Y\in\pi$ is 4-VI for $\Gg_Y$ and $\Ga_Y$, hence it is also 4-VI for
$\Gg$ and $\Ga$.
From the argument in the proof of lemma (\ref{partition}),
one sees that any subset
$V$ of $\cD$ that
is 4-VI for $\Gg$ and $\Ga$, and is maximal for this property, must be a
union of $Y$'s belonging to a subset $\cV$ of $\pi$.
We claim that $\cV$ is 4-VI for $\Gg_{\rm ext}^\pi$.
First note that any subsequence of $(\Gg_{\rm ext}^\pi)_\cV$
obtained by deleting
at most four 4-links, is of the form
$(\Gg')^\pi$ where $\Gg'$ is a subsequence of $(\Gg_{\rm ext})_V$ obtained
also by deleting at most four 4-links.
Now $\Gg_V$ is obtained by intertwining of $(\Gg_{\rm ext})_V$
and the $\Gg_Y$, for $Y\in\cV$.
Therefore if $\Gg''$ denoted the subsequence obtained from $\Gg_V$
by deleting the same 4-links as in the extraction of $\Gg'$ from
$(\Gg_{\rm ext})_V$, then $\Gg''$ is obtained from
$\Gg_V$ by deleting at most four 4-links
and is an intertwining of $\Gg'$ and the $\Gg'_Y$, $Y\in\cV$.
Now since $V$ is 4-VI for $\Gg$ and $\Ga$, $\Gg''$ must connect $V$.
If $Y$ and $Y'$ are two elements of $\cV$,
pick a cube $\De$ in $Y$ and a cube $\De'$ in $Y'$.
There exists a chain $\De_1,\ldots,\De_u$, $u\ge 1$, of cubes such
that $\De_1=\De$, $\De_u=\De'$ and for any $v$, $1\le v\le u-1$, $\De_v$ and
$\De_{v+1}$ are either glued by $\Ga$ or joined by $\Gg''$.
By assumption on the $\Ga_Y$, there can be no gluing between cubes belonging
to different $Y$'s.
As a result, there can be no gluing by $\Ga$ between a cube
in $V$ and a cube in $\cD\backslash V$.
This, together with the fact that $\Gg''$ is made of links whose supports
ly in $V$, compels the cubes $\De_v$, $1\le v\le u$, to remain in $V$.
Let for every $v$, $1\le v\le u$, $Y(\De_v)$ denote the
unique $Y\in\cV$ containing $\De_v$.
Now for any $v$, $1\le v\le u-1$, either $\De_v$ and $\De_{v+1}$
are glued by $\Ga$ and therefore have to belong to the same $Y$,
i.e. $Y(\De_v)=Y(\De_{v+1})$, or $\De_v$ and $\De_{v+1}$ are joined
by some link $l_j$ of $\Gg''$.
If $l_j$ appears in some $\Gg_Y$, $Y\in\cV$, then again we have
$Y(\De_v)=Y(\De_{v+1})$.
Else, $l_j$ must be in $(\Gg_{\rm ext})_V$ and therefore $(l_j)^\pi$ appears
in $(\Gg')^\pi$ and joins $Y(\De_v)$ with $Y(\De_{v+1})$.
The chain $Y(\De_1)=Y$, $Y(\De_2),\ldots,Y(\De_u)=Y'$ shows that
$Y$ and $Y'$ are connected by $(\Gg')^\pi$.
To conclude, we have that $\cV$ is 4-VI for $(\Gg_{\rm ext})^\pi$,
and since $\Pi^\pi((\Gg_{\rm ext})^\pi)=0_\pi$,
$\cV$ must be a singleton.
Thus $V\in\pi$ and this completes our proof that $\Pi_\Ga(\Gg)=\pi$.
\Endproof

This lemma closes the series of verifications that ensure that
given a partition $\pi$ of $\cD$, summing over $\Gg\in\cA\cG_\Ga$, such that
$\Pi_\Ga(\Gg)=\pi$,
is the same as summing independently over $\Gg_Y\in\cA\cG_{\Ga_Y,Y}$
for any $Y\in\pi$,
with the requirement that $\Pi_{\Ga_Y,Y}(\Gg_Y)=\{Y\}$,
together with summing over $\Gg_{\rm ext}$, made of links in $Offdiag(\pi)$
and such that $\Pi^\pi(\Gg_{\rm ext}^\pi)=0_\pi$.
Finally, one has to sum over all possible intertwinings of $\Gg_{\rm ext}$
with the $\Gg_Y$, $Y\in\pi$, to make up $\Gg$.

Remark that this last sum over intertwinings allows the factorization
of the integrations in the $h$ parameters.
Indeed in (\ref{raw}) each parameter $h_j$
is better labeled by the corresponding
link $l_j$ of $\Gg$. In that way, the ordering
of links inside $\Gg$ impose the corresponding ordering
$h_1\ge\ldots\ge h_k$ over the parameters.
Summing over the intertwinings just recomposes a domain of integration
on the $h$ parameters
where the ordering constraints, between parameters attached to
links from different subsequences $\Gg_Y$ or $\Gg_{\rm ext}$,
have been relaxed.
Finally remark that the condition on $\Gg_{\rm ext}$ does not depend
on its ordering.
Therefore the sum over $\Gg_{\rm ext}$ boils down to a sum
over $\Br{\Gg_{\rm ext}}$, the corresponding unordered graph,
then a sum over the choice of
ordered representative $\Gg_{\rm ext}=(l_1,\ldots,
l_{k_{\rm ext}})$.
This last sum is performed by summing over the permutations of $k_{\rm ext}$
elements, and dividing by the symmetry factor $\si({\Br{\Gg_{\rm ext}}})$
due to repetition of identical links.
The sum over permutations relaxes the ordering
constraints over the $h$ parameters corresponding to links of $\Gg_{\rm ext}$,
that are now integrated independently form 0 to 1.
Besides, it results from the rule in Lemma (\ref{inter})
for defining $T_\Gg({\bf h})$,
that nothing in the integrand of (\ref{raw}) depend on the $h$ parameters of
$\Gg_{\rm ext}$.
Indeed, if $\Gg=(l_1,\ldots,l_k)$ and for some $j$, $1\le j\le k$,
$l_j$ is in $\Gg_{\rm ext}$,
then $\Pi_\Ga((l_1,\ldots,l_{j-1}))=
\Pi_\Ga((l_1,\ldots,l_j))$.
This results from the fact that those two partitions are 
finer than $\Pi_\Ga(\Gg)$, so that a subset $V$ of $\cD$ that is
4-VI for $(l_1,\ldots,l_{j-1})$ or $(l_1,\ldots,l_j)$, has to be included
in a block of $\pi$; but then ${\rm supp}\ l_j\not\subset V$
and hence $(l_1,\ldots,l_{j-1})_V=(l_1,\ldots,l_j)_V$.
Finally the $h$ parameters of $\Gg_{\rm ext}$ are
integrated out, yielding a factor 1.

Now the nice feature of (\ref{raw}), that allows the factorization stated
in Theorem 1, is that, by definition,
for any $l\in{\rm Offdiag}(\Pi_\Ga(\Gg))$,
$(T_\Gg({\bf h}))_l=0$.
Therefore in the covariance as well as in the interaction, there
can be no coupling between fields supported by different blocks
of $\pi=\Pi_\Ga(\Gg)$.
Hence, the fields $(\ph_i)_{0\le i\le N}$
on $\La$ can be integrated separately as independent
sets $(\ph_Y)$, $Y\in\pi$, as explained in the definition of the activities.

It is also easily seen, by application of Lemma (\ref{intrinsic}),
that the part of the interaction and the covariance
concerning a block $Y$ is interpolated only with the parameters $h_j$,
such that $l_j\in L_Y$, forming the vector ${\bf h}_Y$. In fact, the couplings
are weakened by the components of the vector
$T_{\Gg_Y,\Ga_Y}({\bf h}_Y)$.

A few minor remarks remain in order to complete the proof of the theorem.
First, note the factorization
\be
\ch_\Ga((\ph_i)_{0\le i\le N})=
\prod_{Y\in\pi}
\ch_{\Ga_Y,Y}((\ph_Y))
\ \ ,
\ee
of the large field conditions, since the stack
of large field cubes under some cube $\De$, whose scales range through
the set $I_\Ga(\De)$, is glued together with $\De$. Thus they have to belong
to the component $Y$ of $\De$, where gluing is done by $\Ga_Y$ only.

Second, for any link $l_j$ in $\Gg_{\rm ext}$ that must be a
4-link, we have chosen a sequence $(b_1^j,\ldots,l_4^j)$
of boxes such that $l[b_1^j,\ldots,b_4^j]=l_j$,
instead of summing like in (\ref{raw}) over all such equences.
Therefore we had to compensate by the combinatoric factor $\rh(l_j)$.
Third, the insertions $\ph_{\al_q}(\xi_q)$ are gathered together
according to the block $Y$ they belong to.
For such a block, we have to consider also the insertions
$\ph_{i(b_\nu^j)}(\ze_\nu^j)$, with $b_\nu^j\in Y$, that come from
the links of $\Gg_{\rm ext}$.
The proof of the theorem is now complete.
\Endproof

To conclude with the algebraic considerations, we state a lemma
to control the lenght of an allowed graph by the size of its supporting
polymer. In analogy with trees we get a bound of linear type.

\begin{lem}
\label{fivefor}
If $Y$ is some polymer, $\Ga_Y$ is a large field region in $Y$,
and $\Gg_Y=(l_1,\ldots,l_k)$ is an allowed graph
in $Y$, with respect to $\Ga_Y$;
then we have the following bound
\be
k\le 6\#(Y)-6
\ \ .
\ee
\end{lem}

\noindent{\bf Proof:\ \ }
We consider $\Gg_{Y,1}$ the subsequence of $\Gg_Y$ made by the 2-links,
and $\Gg_{Y,2}$ made by the 4-links. From the proof of lemma (\ref{pardec}),
it is clear that subsequences of allowed graphs are allowed.
Being allowed, for a graph of 2-links only, is easily seen from
the definition, to imply that it is a forest, i.e. a graph of ordinary
links with no loops. Therefore if $k_1$ denotes the length of
$\Gg_{Y,1}$, we must have $k_1\le \#(Y)-1$.
It remains to show that if $k_2$ denotes the lenght of $\Gg_{Y,2}$,
we have $k_2\le 5\#(Y)-5$.
Remark that we can define the notion of 5-VI subsets with respect to
an ordered graph of 2 and 4-links and a large field region in some polymer,
exactly as we did for 4-VI subsets. Namely, we ask that it remains
connected when we remove up to five 4-links from the graph.
The lemma will be established as soon as we prove the two following
statements. First, an allowed graph $\Gg_Y=(l_1,\ldots,l_k)$,
made of 4-links only,
is such that the only 5-VI
subsets $V$ of $Y$ are contained in the blocks of $\pi_Y$,
the partition of connected components in $Y$ when gluing bonds
are taken into account only.
Second, any graph with the last property and such that
each of its links is in ${\rm Offdiag}(\pi_Y)$,
has a lenght no greater than
$5\#(Y)-5$.

To prove the first statement, consider $(\Gg_Y)_V=(l_{a_1},
\ldots,l_{a_r})$ the subsequence
of $\Gg_Y$, made by the links whose support is entierly contained
in $V$. Suppose that $r\ge 1$. Since $(\Gg_Y)_V$ connects $V$ even after
removing up to five 4-links, we conclude that the initial segment
$(l_{a_1},l_{a_2},\ldots,l_{a_{(r-1)}})$
connects $V$ even after removing up to
four 4-links. As a result $V$ is 4-VI with respect to
$(l_{a_1},l_{a_2},\ldots,l_{a_{(r-1)}})$,
and thus to the larger graph $(l_1,l_2,\ldots,l_{(a_r)-1})$.
But by the allowedness of
$\Gg_Y$, $l_{a_r}$ must belong to ${\rm Offdiag}(
\Pi_{\Ga_Y}((l_1,l_2,\ldots,l_{(a_r)-1})))$.
However $l_{a_r}$ has support in $V$ which is included in a component of
$\Pi_{\Ga_Y}((l_1,l_2,\ldots,l_{(a_r)-1}))$. This leads to a contradiction.
Therefore we have $r=0$, i.e. the only links that are internal to $V$
are the gluing links. Since $V$ is 5-VI, it is connected, and thus
contained in a block of $\pi_Y$, as wanted. Note that a link $l_a$ of $\Gg_Y$
is in ${\rm Offdiag}(\Pi_{\Ga_Y}((l_1,l_2,\ldots,l_{a-1})))$ and thus in
${\rm Offdiag}(\pi_Y)$, since $\pi_Y$ is finer than
$\Pi_{\Ga_Y}((l_1,l_2,\ldots,l_{a-1}))$.

We now prove the second statement by induction on $\#(Y)\ge 1$.
If $\#(Y)=1$, then $\pi_Y$ is reduced to $\{Y\}$, and since
the links of $\Gg_Y$ must be in ${\rm Offdiag}(\pi_Y)$, the only
possibility is $\Gg_Y=\emptyset$. The bound is satisfied in this case.
If $\#(Y)\ge 1$ but $\pi_Y=\{Y\}$, the same conclusion applies.
Now suppose $\#(Y)\ge 1$ and $\pi_Y\neq\{Y\}$. Since $Y$ is not contained in
a block of $\pi_Y$, $Y$ is not 5-VI.
As a result we can remove no more than five 4-links in the graph
$\Gg_Y$ and thus disconnect $Y$ into nonempty disjoint
components $Y_1,\ldots,Y_q$, with $q\ge 2$.
Consider for any $r$, $1\le r\le q$, the subsequence $\Gg_{Y_r}$
of $\Gg_Y$
made of the links that are internal to $Y_r$. It is easy to see
that $\Gg_{Y_r}$ satisfies the hypothesis of the statement we are
proving, with respect to the polymer $Y_r$ and its
partition $\pi_{Y_r}$ into connected components for the gluing
bonds. By the induction hypothesis, the lenght $k_r$ of
$\Gg_{Y_r}$, is bounded by $5\#(Y_r)-5$.
This implies for the lenght $k$ of $\Gg_Y$
\be
k\le 5+\sum_{r=1}^q k_r
\le 5+\sum_{r=1}^q(5\#(Y_r)-5)
\le 5\lp\sum_{r=1}^q\#(Y_r)\rp\ -5
=5\#(Y)-5
\ \ ,
\ee
since $q\ge 2$.
This completes the inductive proof of the second statement,
and the lemma follows.
\Endproof

\section{The bound on convergent polymers}
\markboth{THE BOUND ON CONVERGENT POLYMERS}
{THE BOUND ON CONVERGENT POLYMERS}
\subsection{The main theorem}

This section will be entierly devoted to the proof of the following theorem.
It is a polymer bound in the spirit of [Br1 ,R1], that is uniform in the
volume and the infrared cut-offs.
We consider activities for polymers that are due to 4-VI graphs
which do require the renormalization procedure.
The theorem is a constructive analog of Weinberg's theorem
on convergent graphs, with the difference that we have to sum
all orders of perturbation theory.
This result we view as a cornerstone of the constructive approach via
phase-space expansions.

\begin{th}
There is a constant $C>0$ such that, for any $\et>0$
and $K>0$, there exists a $g_0>0$, such that for any $g$ satisfying
$0<g\le g_0$, and for any $N\ge 0$ we have:
for any $\De_{\rm org}\in\cD^{(N)}$, and for any finite
family $(\De_s^{\rm ext},\ze_s^{\rm ext})_{s\in\cS}$
such that for every $s\in\cS$, $\De_s^{\rm ext}\in\cD^{(N)}$
and $\ze_s^{\rm ext}\in\De_s^{\rm ext}$, the following
bound:
\be
\sum_{{{Y|Y\subset\cD^{(N)}}\atop{\De_{\rm org}\in Y}}
\atop{\{\De_s^{\rm ext}|s\in\cS\}\subset Y}}
|\cA(Y;(\De_s^{\rm ext},\ze_s^{\rm ext})_{s\in\cS})|.K^{\#(Y)}
\le\et.C^{\#(\cS)}.\prod_{\De\in\cD^{(N)}}(E_\De!)^{1/2}
.\prod_{s\in\cS}M^{-i(\De_s^{\rm ext})}
\ \ .
\label{polybound}
\ee
Here $E_\De$ denotes the number $\#(\{s\in\cS|\De_s^{\rm ext}=\De\})$
that counts the external sources in $\De$.
\end{th}

Remark that the case $\cS=\emptyset$, corresponding to vacuum polymers,
is included.
In the case where $\cS\neq\emptyset$ we usually do not have a fixed
origin $\De_{\rm org}$ to break translational invariance. To apply the
theorem we just have then to choose one of the $\De_s^{\rm ext}$'s to be
$\De_{\rm org}$.

Prior to any other consideration, one can rewrite the interaction $\cI$
inside the exponential in the formula (\ref{activ})
for the activity of a polymer, in
a form where the positivity is more patent.
Let us denote the contribution of a 4-link $l$ in $Y$ by
\be
\Theta_l\eqdef
\sum_{{(\De_1,\ldots,\De_4)\in Y^4}
\atop{l[\De_1,\ldots,\De_4]=l}}
\int_{\De_1}dx_1
\ldots
\int_{\De_4}dx_4
\ K(x_1,i(\De_1);\ldots;x_4,i(\De_4))
.\ph_{i(\De_1)}(x_1)\ldots
\ph_{i(\De_4)}(x_4)
\ \ .
\label{deftheta}
\ee
Since we will be working always in some definite polymer $Y$,
without having to distinguish it from other blocks of some partition
like in Section 2, we will
drop the $Y$ subscript from $\Ga_Y$, $\Gg_Y$, $T_{\Gg_Y,\Ga_Y}$,
$\cA\cG_{\Ga_Y,Y}$, $\Pi_{\Ga_Y,Y}$ and the collection
of fields $\ph_Y$.
We define for any $a$, $1\le a\le k$,
\be
\Xi_a\eqdef
\sum_{{l\not\in{\rm Offdiag}(\Pi_\Ga((l_1,\ldots,l_a)))}
\atop{l\in{\rm Offdiag}(\Pi_\Ga((l_1,\ldots,l_{a-1})))}}
\Theta_l
\ee
and
\be
\Xi_0=\sum_{l\not\in{\rm Offdiag}(\Pi_\Ga(\emptyset))}
\Theta_l
\ \ ,
\ee
and we introduce the convention $h_0=1$ and $h_{k+1}=0$.
Then by definition of $T_{\Gg,\Ga}$ we have
\be
\cI=g\sum_{a=0}^k
h_a.\Xi_a
\ \ .
\ee
We now perform an Abel transformation
\bea
\cI & = & g\sum_{a=0}^k\lp\sum_{a'=a}^k(h_{a'}-h_{a'+1})\rp\Xi_a \\
    & = & g\sum_{a'=0}^k(h_{a'}-h_{a'+1})\lp\sum_{a=0}^{a'}\Xi_a\rp\ \ .
\eea
But since ${\rm Offdiag}\Pi_\Ga((l_1,\ldots,l_{a-1}))\supset
{\rm Offdiag}\Pi_\Ga((l_1,\ldots,l_a))$
for every $a$, $1\le a \le k$, we have:
\bea
\sum_{a=0}^{a'}\Xi_a & = &
\sum_{l\not\in{\rm Offdiag}(\Pi_\Ga((l_1,\ldots,l_{a'})))}
\Theta_l\\
 & = & \sum_{B\in\Pi_\Ga((l_1,\ldots,l_{a'}))}
\ \sum_{l|{\rm supp}\ l\subset B}\Theta_l
\eea
For any $a'$, $0\le a'\le k$, $B\in\Pi_\Ga((l_1,\ldots,l_{a'}))$
and $x\in\La$ we introduce the field
\be
\Phi^B(x)\eqdef
\sum_{i=0}^N
\bbbone_{\{\De(x,i)\in B\}}
\ph_i(x)
\ \ .
\ee
This notation and equation (\ref{deftheta}) allow us to write
\be
\sum_{l|{\rm supp}\ l\subset B}\Theta_l=
\int_\La
{\lp\ph^B(x)\rp}^4dx
\ \ .
\label{rewint}
\ee
Finally the interaction becomes:
\be
\cI=g\sum_{a=0}^k(h_a-h_{a+1})
\sum_{B\in\Pi_\Ga((l_1,\ldots,l_a))}\int_\La{\lp\ph^B(x)\rp}^4dx
\ee
Since $1\ge h_1\ge\ldots\ge h_k\ge0$, the positivity of $\cI$,
already proven in general in lemma (),
appears here very explicitly.

We see from (\ref{polybound}) and the expression
(\ref{activ}) for the activity that we
roughly have the following successive sums to control
\be
\sum_{Y}
\sum_{\Ga}
\sum_{\Gg}
\sum_{\GP}\;
{\mbox{``individual contribution''}}
\ \ ,
\ee
where the last one is over the functional derivation procedures $\GP$.
The way we bound the sums we perform first, determine what is left
to bound the remaining sums.
Therefore we start by explaining the last sum namely on $Y$, because
it rules the upstream bounding of the individual contributions.

\subsection{The sum over the location of polymers}

The sum over the position of the polymer $Y$, in the lattice
$\cD^{(N)}$ with a number $N+1$ of slices, with the restriction
to polymers containing a fixed box $\De_{\rm org}$, is done thanks to
the following proposition that provides \und{uniform} bounds in $N$,
that is in the infrared cut off,and also in the volume cut-off
$\La$.
We formulate this proposition in an arbitrary dimension $d$,
and as a general closed result that can be applied directly
in other situations where phase space cluster expansions
are needed.
The requirements of the lemma seem to be among the very minimal we can
demand.
Indeed we need a just summable power-like decay of the horizontal links,
as well as a factor $M^{-(d+\ep)}$ per connected component, in a very weak
sense, of high frequency slices, instead of the stronger
hypothesis of a factor $M^{-(d+\ep)}$ per cube.
We do not ask $M$ to be large either.
If $Y$ is a polymer, i.e. a finite non empty subset of $\cD$,
we pose
\be
i_{\rm max}(Y)=\max\{i|Y\cup\cD_i^{(N)}\neq\emptyset\}
\ \ .
\ee
We suppose $F$ is an ordinary graph on $Y$.
In the whole section a link means an unordered pair $\{\De_1,\De_2\}$
made of two distinct cubes in $\cD^{(N)}$.
We do not use here the notion of $p$-link of Section 2.
We suppose also that $F$ has no loops.
$F$ is then a union of non overlapping trees, i.e. a forest.
We ask that $F$ is made only of horizontal bonds, i.e. that for any
$l=\{\De_1,\De_2\}$ in $F$ we have $i(\De_1)=i(\De_2)\eqdef i(l)$.
In that case we pose also
\be
{\rm dist}_2(l)\eqdef{\rm dist}_2(\De_1,\De_2)\eqdef\inf
\{d_2(x_1,x_2)|x_1\in\De_1,x_2\in\De_2\}
\ \ .
\ee
The subscript 2 refers to the usual Euclidean distance.

Given $i$, $0\le i\le N$, we let $\cD_{\le i}^{(N)}$ denote
$\cup_{0\le j\le i}\cD_j^{(N)}$, and we define a projection map
${\rm pr}_i:\cD_{\le i}^{(N)}\rightarrow\cD_i^{(N)}$
that associates to any box $\De$, with $i(\De)\le i$,
the unique box
${\rm pr}_i(\De)$ of $\cD_i^{(N)}$ containing $\De$.
We denote by $Y^{(i)}$ the subset ${\rm pr}_i(Y\cap\cD_{\le i}^{(N)})$
of $\cD_i^{(N)}$.
We consider the graph $F^{(i)}$ on $Y^{(i)}$ made by all links
$\{\De_1,\De_2\}$ for which there exists $\De'_1$, $\De'_2$ in
$Y\cap\cD_{\le i}^{(N)}$ with ${\rm pr}_i(\De'_1)=\De_1$,
${\rm pr}_i(\De'_2)=\De_2$ and $\{\De'_1,\De'_2\}\in F$.
Remark that $F^{(i)}$ need not be a forest.
Let $\cR_i(Y,F)$ be the set of connected components of $Y^{(i)}$
with respect to the graph $F^{(i)}$.
We say that $F$ is {\em satisfying} to $Y$ if
$\#(\cR_{i_{\rm max}(Y)}(Y,F))=1$.
A polymer $Y$ for which there exists a satisfying $F$, is called
{\em admissible}.

Next we pose
\be
\cR(Y,F)\eqdef
\cup_{i<i_{\rm max}(Y)}\cR_i(Y,F)
\ \ .
\ee
For some fixed $\ep_2>0$, we introduce
\be
\cT_{\ep_2}(Y,F)=
M^{-(d+\ep_2)\#(\cR(Y,F))}
.\prod_{l\in F}
(1+M^{-i(l)}{\rm dist}_2(l))^{-(d+\ep_2)}
\ \ ,
\ee
and
\be
\cT_{\ep_2}(Y)=\max_{F\ {\rm satisfying\ to}\ Y}\cT_{\ep_2}(Y,F)
\ \ ,
\ee
if $Y$ is admissible.
We now have the following result.

\begin{prop}
\label{PropY}
There exists a constant $K_1(d,M,\ep_2)>0$ such that,
for any integer $Q\ge 1$ and for any fixed $\De_{\rm org}$
in $\cD^{(N)}$, we have
\be
\sum_{{{Y\subset\cD^{(N)}}\atop{Y\ {\rm admissible}}}
\atop{\De_{\rm org}\in Y,\ \#(Y)=Q}}
\cT_{\ep_2}(Y)
\le
K_1(d,M,\ep_2)^Q
\ \ .
\ee
\end{prop}

\noindent
{\bf Proof:\ \ }
We have to organize the bounding term $\cT_{\ep_2}(Y,F)$ into a
tree decay allowing to sum over $Y$.
Having defined ${\Br Y}\eqdef\cup_{i\le i_{\rm max}(Y)}Y^{(i)}$;
the first step is to collect a factor less than one per cube of
${\Br Y}$.

A cube $\De$ in ${\Br Y}$ is called a \und{working} cube if it
falls in one of the following categories:

a) - $\De$ is a cube of $Y$

b) - $\De$ is a \und{fork} of ${\Br Y}$ that is a cube in $\cD_i^{(N)}\cap
{\Br Y}$ for which there exists two distinct cube $\De_1$,
$\De_2$ in $\cD_{i-1}^{(N)}\cap{\Br Y}$ and contained in $\De$.

c) - $\De$ is in some $Y^{(i)}$ and there exists $\De'\in Y^{(i)}$
such that ${\rm dist}_2(\De,\De')>0$ and
$\{\De,\De'\}\in F^{(i)}$.

The set of working cubes is denoted by ${\Br Y}^W$.
Let us choose $\de$ an integer $\de\ge 1$, such that $M^\de\ge 3^d$.
We say that a cube $\De$ of ${\Br Y}$ is an
\und{active} cube, if there exists
a working cube $\De'$ contained in $\De$ with $i(\De)-i(\De')\le\de$.
The set of active cubes is denoted by ${\Br Y}^A$.
Since such a $\De'$ can be associated to at most $\de$ bigger cubes
$\De$, we have the trivial bound
\be
\#({\Br Y}^A)\le\de.\#({\Br Y}^W)
\ \ .
\label{trivbound}
\ee
A cube in ${\Br Y}^I\eqdef{\Br Y}\backslash{\Br Y}^A$ is called
an \und{idle} cube.
Small factors will come from the working cubes and the components in
$\cR(Y,F)$, so the first thing to show is a bound on the number
of idle cubes, that is linear in $\#({\Br Y}^W)$ and
$\#(\cR(Y,F))$.

Let $i\le i_{\rm max}(Y)$; the links of $F^{(i)}$
whose extremities are idle cubes, define
connected components among the set $Y^{(i)}\cap{\Br Y}^I$.
Those are themselves embedded in larger connected components
of the full graph $F^{(i)}$, namely in some element of $\cR_i(Y,F)$.

Let $X$ be such a component of idle cubes; we have the following
(far from optimal) geometric bound.

\begin{lem}
$\#(X)\le 3^d$
\end{lem}

\noindent
{\bf Proof:\ \ }
Suppose there exist $\De$ and $\De'$ in $X$ with
${\rm dist}_2(\De,\De')>0$.
Since $X$ is connected there is a sequence $\De_1,\ldots,\De_\ga$,
$\ga\ge 1$ of distinct cubes in $X$ such that $\De_1=\De$, $\De_\ga=\De'$
and $\{\De_\beta,\De_{\beta+1}\}\in F^{(i)}$ for any $\beta$,
$1\le \beta\le\ga-1$.
Since the cubes $\De_\beta$, $1\le \beta\le\ga$, are in $X$, they are idle
and therefore for every $\beta$, $1\le \beta\le\ga-1$, we have
${\rm dist}_2(\De_\beta,\De_{\beta+1})=0$.
Since ${\rm dist}_2(\De_1,\De_\ga)>0$ we can consider $\al$ the smallest
integer, $2\le \al\le\ga$, such that ${\rm dist}_2(\De_1,\De_\al)>0$.
We will use this to derive a contradiction.

First, notice that if $\De$ is an idle cube then $i(\De)>\de$.
Indeed, $\De\in{\Br Y}$ implies the existence of some $\De^1\in Y$,
with $\De^1\subset\De$. Since by a), $\De^1\in{\Br Y}^W$ but
on the other hand
$\De\in{\Br Y}^I$ we must have $i(\De)>i(\De^1)+\de\ge\de$.

By descending induction on $j$, $i(\De)-\de\le j\le i(\De)$,
we show that there exists a unique cube $\hat{\De}^j$ in $Y^{(j)}$ and
contained in $\De$.
For $j=i(\De)$, $\hat{\De}^j$ has to be $\De$ itself.
Assume the statement is proven for $j$, $i(\De)-\de< j\le i(\De)$.
Since $i(\De)-j<\de$ and $\De\in {\Br Y}^I$, $\hat{\De}^j$ cannot
be in $Y$, hence there exists a cube $b$ of $Y$ with $b\subset
\hat{\De}^j$ and $i(b)<j$.
Now define $\hat{\De}^{j-1}$ as ${\rm pr}_{j-1}(b)$.
We have to show that there is no other cube $b'$ in $Y^{(j-1)}$
and contained in $\De$.
If there was such a $b'$, ${\rm pr}_j(b')$ would be in $Y^{(j)}$
and contained in $\De$, therefore by the induction hypothesis ${\rm pr}_j
(b')=\hat{\De}^j$.
Now since $b'$ and $\hat{\De}^{j-1}$ are distinct cubes
of $Y^{(j-1)}$ contained in $\hat{\De}^j\in Y^{(j)}$, by b)
$\hat{\De}^j$ has to be a fork.
But $i(\De)-i(\hat{\De}^j)<\de$ and $\De\in{\Br Y}^I$,
therefore we have a contradiction.
This induction argument allow us to define for any idle cube $\De$, the
unique cube $\hat\De$ of $Y^{i(\De)-\de}$ contained in $\De$.
Simply put $\hat{\De}=\hat{\De}^{i(\De)-\de}$ with the notations of the
previous proof.

We return now to the sequence $\De_1,\ldots,\De_\al$, and remark that
for any $\beta$, $1\le \beta\le \al-1$,
$\{\De_\beta,\De_{\beta+1}\}\in F^{(i)}$, and therefore
there exists $\De'_\beta$ and $\De'_{\beta+1}$ in $Y$ with
$\De'_\beta\subset\De_\beta$,
$\De'_{\beta+1}\subset\De_{\beta+1}$ and $\{\De'_\beta,\De'_{\beta+1}\}
\subset F$.
By the idleness of $\De_\beta$ and $\De_{\beta+1}$,
we must have
\be
i(\De'_\beta)=i(\De'_{\beta+1})<i-\de=i(\hat{\De}_\beta)
=i(\hat{\De}_{\beta+1})
\ \ .
\ee
Note that ${\rm pr}_{i-\de}(\De'_\beta)$ satisfies the defining conditions
of $\hat{\De}_\beta$ and by the unicity, they
must be equal, and thus $\De'_\beta\subset\hat{\De}_\beta$.
Likewise we have $\De'_{\beta+1}\subset\hat{\De}_{\beta+1}$
and, as a by-product, we obtain $\{\hat{\De}_\beta,\hat{\De}_{\beta+1}\}
\in F^{(i-\de)}$.
Since $\De_\beta$, $\De_{\beta+1}$ are idle, $\hat{\De}_\beta$ and
$\hat{\De}_{\beta+1}$ cannot be working cubes and by c) we finally get
${\rm dist}_2(\hat{\De}_\beta,\hat{\De}_{\beta+1})=0$.
This last statement is equivalent is equivalent to the same with the
distance ${\rm dist}_\infty(x,y)\eqdef\sup_{1\le \mu\le d}|x^\mu-y^\mu|$.

Up to now what we have is that for any $\beta$, $1\le\beta\le \al-1$,
${\rm dist}_\infty(\hat{\De}_\beta,\hat{\De}_{\beta+1})=0$ and
${\rm dist}_\infty(\De_1,\De_\beta)=0$, besides ${\rm dist}_\infty
(\De_1,\De_\al)>0$.
Since $\De_1$ and $\De_\al$ are in the grid $\cD_i^{(N)}$ of mesh
$M^i$, we infer that ${\rm dist}_\infty(\De_1,\De_\al)\ge M^i$.
If ${\rm diam}_\infty(A)$ denotes the diameter for the distance
${\rm dist}_\infty$ of a bounded subset of
$\RR^d$, we have the elementary inequality
\be
{\rm dist}_\infty(A,C)\le
{\rm dist}_\infty(A,B)+
{\rm dist}_\infty(B,C)+
{\rm diam}_\infty(B)
\ \ .
\ee
By iteration we derive
\[
{\rm dist}_\infty(\hat{\De}_1,\hat{\De}_\al)\le
{\rm dist}_\infty(\hat{\De}_1,\hat{\De}_2)
+{\rm dist}_\infty(\hat{\De}_2,\hat{\De}_3)
+\ldots+
{\rm dist}_\infty(\hat{\De}_{\al-1},\hat{\De}_\al)
\]
\be
+{\rm diam}_\infty(\hat{\De}_2)+\ldots+
{\rm diam}_\infty(\hat{\De}_{\al-1})
\ \ ,
\ee
and thus
\be
M^i\le{\rm dist}_\infty(\De_1,\De_\al)
\le
{\rm dist}_\infty(\hat{\De}_1,\hat{\De}_\al)
\le
(\al-2)M^{(i-\de)}
\ \ ,
\ee
implying
\be
M^\de\le\al-2
\ \ .
\label{inegdel}
\ee
However, since the cubes $\De_2,\ldots,\De_{\al-1}$ are distinct and
verify ${\rm dist}_2(\De_1,\De_\beta)=0$ for any $\beta$,
$2\le\beta\le\al-1$, by the geometric constraint that a cube has only
$3^d-1$ neighbors in the lattice $\cD_i^{(N)}$, we conclude
that $\al-2\le 3^d-1$. This together with (\ref{inegdel})
and the assumption $M^\de\ge
3^d$ made at the beginning, leads to a contradiction.

As a conclusion for every $\De$ and $\De'$ in $X$, a connected
component of idle cubes swimming in some larger component in
$\cR_i(Y,F)$, we have ${\rm dist}_2(\De,\De')=0$. Therefore $\#(X)
\le 3^d$. Note that with further work one can show that
$\#(X)\le 2^d$, besides by refining the previous argument,
the assumption $M^\de\ge 3^d$ can be weakened to $M^\de\ge 2^d+1$,
however we are not looking here for optimal bounds.
\Endproof

Now if $X$ is a connected component of idle cubes in $Y^{(i)}$.
Either $X$ is isolated that is $X\in\cR_i(Y,F)$,
or $X$ is embedded in a strictly bigger element $V$
of $\cR_i(Y,F)$.
In the second case there must exist an active cube $\De^a$ in
$V$ linked by $F^{(i)}$ to some cube $\De^b$ of $X$.
Note that ${\rm dist}_2(\De^a,\De^b)=0$, for if it were not so,
$\De^a$ would fall in the category c) of working cubes,
which is forbidden by its assumed idleness.
Since distinct components of idle cubes are disjoint and a 
$\De^a$ can have at most $3^d-1$ neighbors, we see that at most
$3^d-1$ components $X$ can be associated in that way to a single active
$\De^a$.

From these facts and (\ref{trivbound})
we infer that the total number of idle components $X$
is bounded by
\be
\sum_{i\le i_{\rm max}(Y)}
\#(\cR_i(Y,F))
+(3^d-1).\#({\Br Y}^A)
\le
[\#(\cR(Y,F))+1]+\de.(3^d-1).\#({\Br Y}^W)
\ \ .
\ee
From the lemma we deduce
\be
\#({\Br Y}^I)
\le3^d(3^d-1)\de\#({\Br Y}^W)
+3^d(\#(\cR(Y,F))+1)
\ \ ,
\ee
and since $\#({\Br Y}^W)\ge\#(Y)\ge 1$, and $\de\ge 1$, we obtain
\bea
\#({\Br Y}) & \le & \de.\#({\Br Y}^W)+\#({\Br Y}^I)\\
            & \le & (9^d+1).\de.\#({\Br Y}^W)+3^d\#(\cR(Y,F))\ \ .
                \label{Ybar}
\eea

Now define
\be
\cU_{\ep_2}(Y,F)\eqdef M^{-\frac{\ep_2}{2}\#(\cR(Y,F))}.
\prod_{l\in F}
{\lp
1+M^{-i(l)}{\rm dist}_2(l)
\rp}^{-\frac{\ep_2}{2}\#(\cR(Y,F))}
\ \ ,
\ee
so that
\be
\cT_{\ep_2}(Y,F)=
\cU_{\ep_2}(Y,F).
\cT_{\frac{\ep_2}{2}}(Y,F)
\ \ .
\label{decsumY}
\ee
One can easily see that for each $l=\{\De_1,\De_2\}\in F$, the
set
\be
\{i|N\ge i\ge i(l),\ {\rm dist}_2({\rm pr}_i(\De_1),{\rm pr}_i(\De_2))
=0\}
\ee
is non empty.
Let us denote its minimum by $i^\ast(l)$.
If $\De\in{\Br Y}$ is a cube containing either $\De_1$ or $\De_2$, such that
$i(l)\le i(\De)<i^\ast(l)$, since
${\rm dist}_2({\rm pr}_{i(\De)}(\De_1),{\rm pr}_{i(\De)}(\De_2))>0$
and $\De\in\{{\rm pr}_{i(\De)}(\De_1),{\rm pr}_{i(\De)}(\De_2)\}\in
F^{(i(\De))}$, we obtain that $\De$ is a working
cube of type c).
We then say that $\De$ is {\em produced} by the link $l$.
Note that many links may produce the same cube, however any type c)
working cube is produced by at least one link $l$.
We denote by ${\Br Y}^{W,b}$ and ${\Br Y}^{W,c}$ the sets of working
cubes of type b) and c) respectively,
and by $c(l)$ the number of such cubes produced by the link $l$.

Now if $l$ produces some working cubes, we must have $i^\ast(l)>i(l)$ and
therefore
\be
{\rm dist}_2(l)\ge{\rm dist}_2
({\rm pr}_{i^\ast(l)-1}(\De_1),{\rm pr}_{i^\ast(l)-1}(\De_2))
\ge M^{i^\ast(l)-1}
\ \ ,
\ee
and thus
\be
{\lp
1+M^{-i(l)}{\rm dist}_2(l)
\rp}^{-\frac{\ep_2}{2}}\le
M^{-\frac{\ep_2}{2}(i^\ast(l)-i(l)-1)}\le M^{-\frac{\ep_2}{4}c(l)}.
M^{\frac{\ep_2}{2}}
\ \ .
\ee
As a consequence, we have
\be
\cU_{\ep_2}(Y,F)\le M^{-\frac{\ep_2}{2}\#(\cR(Y,F))}.
M^{\frac{\ep_2}{2}\#(Y)}.M^{-\frac{\ep_2}{4}\#({\Br Y}^{W,c})}
\ \ ,
\label{calU}
\ee
thereby extracting a factor $M^{-\frac{\ep_2}{4}}$, less than 1, per
cube of ${\Br Y}^{W,c}$.
Note that we have used the fact that $F$ is a forest on $Y$, so that
$\#(F)\le\#(Y)-1$.

Now consider ${\Br Y}$, we can map it into the power set of 
$Y$ as follows: to each $\De\in{\Br Y}$, we associate
$S(\De)\eqdef\{\De'\in Y|\De'\subset\De\}$.
The range of $S$ is a set of non empty subsets of $Y$
that are either included
in one another or disjoint; this is what we call a \und{forest of subsets}
in $Y$.

\begin{lem}
\label{subsfor}
If $\cE$ is a forest of subsets in some finite non empty set $E$, then
we have the bound
$\#(\cE)\le2\#(E)-1$.
\end{lem}

\noindent
{\bf Proof:\ \ }
By induction on $\#(\cE)$. If $E$ is a singleton $\{e\}$, $\cE$ must
be empty or equal to $\{\{e\}\}$ so that the inequality holds.

If $\#(E)>1$, consider the forest of subsets $\cE'=\cE\backslash
\{E\}$. If $ \cE'=\emptyset$ we are done, otherwise let $E_1,\ldots,
E_k$, be the maximal elements of $\cE'$, and
$\cE_j=\cE'\cap\cP(E_j)$, $1\le j\le k$.
Since $\#(E_j)<\#(E)$, by the induction hypothesis we have
\be
\#(\cE)\le
\sum_{j=1}^k\#(\cE_j)\;+1
\le\sum_{j=1}^k(2\#(E_j)-1)\;+1
\label{forsub}
\ee
If $k\ge 2$, by the disjointness of the $E_j$, (\ref{forsub}) leads to
\be
\#(\cE)\le2\#(E)-k+1\le
2\#(E)-1
\ee
If $k=1$, then since $\#(E_1) < \#(E)$, (\ref{forsub}) becomes
\be
\#(\cE)\le 2\#(E_1)-1+\le 2(\#(E)-1)<2\#(E)-1
\ \ .
\ee
\Endproof

Remark that $S:{\Br Y}^{W,b}\rightarrow\cP(Y)$ is injective.
In fact if $\De_1$ and $\De_2$ are elements of ${\Br Y}^{W,b}$,
with $S(\De_1)=S(\De_2)$,
then there exists a $\De\in Y$ with $\De\subset\De_1\cap\De_2$.
This implies for instance $\De_1\subset\De_2$. Suppose that
the inclusion is strict.
Then since $\De_2$ is a fork, there exists $\De_3\in{\Br Y}\cap
\cD_{i(\De_2)-1}^{(N)}$ included in $\De_2$ and distinct from
$\De_4={\rm pr}_{i(\De_2)-1}(\De_1)\in{\Br Y}\cap\cD_{i(\De_2)-1}^{(N)}$.
Therefore $\De_3\cap\De_4=\emptyset$, and the non empty $S(\De_3)$
verifies $S(\De_3)\subset S(\De_2)$ but
$S(\De_3)\cap S(\De_1)\subset S(\De_3)\cap S(\De_4)=\emptyset$.
This contradicts $S(\De_1)=S(\De_2)$.

The injectivity allows us to claim that
\be
\#({\Br Y}^{W,b})=\#(S({\Br Y}^{W,b})).
\ee
Now $S({\Br Y}^{W,b})$ is a forest of subsets of $Y$. But, it may be viewed
also as a forest of subsets of the set $\cM$ of minimal elements of
$S({\Br Y}^{W,b})$ itself.
Remark that for any fork, $\De$ $\#(S(\De))\ge 2$,
and thus every one of those minimal subsets contain at least two
elements. They are also disjoint, and this forces $\#(\cM)\le
\frac{1}{2}\#(Y)$.
By Lemma (\ref{subsfor}), we obtain
\be
\#(S({\Br Y}^{W,b}))\le 2\#(\cM)-1
\ \ ,
\ee
and finally
\be
\#({\Br Y}^{W,b})\le\#(Y)-1
\ \ .
\label{forks}
\ee
Consequently
\be
\#({\Br Y}^W)\le2\#(Y)+\#({\Br Y}^{W,c})
\ \ ,
\ee
and by (\ref{Ybar}),
\be
\#({\Br Y})\le 2.\de.(9^d+1)\#(Y)+\de.(9^d+1)\#({\Br Y}^{W,c})
+3^d\#(\cR(Y,F))
\ \ .
\ee
This together with (\ref{calU}) proves
\be
\cU_{\ep_2}(Y,F)\le
M^{-\ep_3\#({\Br Y})}.C_1^{\#(Y)}
\ \ ,
\label{calU2}
\ee
where $\ep_3\eqdef\frac{\ep_2}{4(9^d+1)}$ and $C_1\eqdef
M^{\ep_2}$.
\medskip

The second step in the proof of the proposition is to organize
$\cT_{\frac{\ep_2}{2}}(Y,F) .M^{-\ep_3\#({\Br Y})} .C_1^{\#(Y)}$
into a tree decay.
Let us introduce $G$ the ordinary graph on $\Br Y$, whose links are
those of $F$, plus all pairs $\{\De_1,\De_2\}$ where $\De_1$ and $\De_2$
belong to $\Br Y$, and one of the two,
is just on top of the other.
For any $i\le i_{\rm max}(Y)$ and for any $A\in\cR_i(Y,F)$, denote
\be
A^\uparrow\eqdef\{\De\in{\Br Y}|
\exists\De'\in A,\ \De\subset\De'\}
\ \ .
\ee
Pose also
\be
\cR^\uparrow_i(Y,F)\eqdef\{A^\uparrow|A\in\cR_i(Y,F)\}
\ \ .
\ee
Since for $A\in\cR(Y,F)$ we have $A\subset A^\uparrow$, we obtain
that the map $A\mapsto A^\uparrow$ is injective.
Beside it is surjective by definition, therefore
$\cR^\uparrow(Y,F)\eqdef\cup_{i<i_{\rm max}(Y)}\cR^\uparrow_i(Y,F)$
a has cardinal equal to $\#(\cR(Y,F))$.
For any $i\le i_{\rm max}(Y)$, we define
\be
G_i\eqdef\{\{\De_1,\De_2\}\in G|\ i(\De_1)\le i,\ i(\De_2)\le i\}
\ \ ,
\ee
{\em the truncation of} $G$ {\em above frequency} $i$.
Note that every
$\De$ in ${\Br Y}\cap\cD_{\le i}^{(N)}$ is connected to ${\rm pr}_i(\De)$
by $G_i$ since for every $j$, $i(\De)\le j < i$,
$\{{\rm pr}_j(\De),{\rm pr}_{j+1}(\De)\}\in G_i$.
From this property it follows easily that any $A^\uparrow\in
\cR^\uparrow_i(Y,F)$ is connected by $G_i$. Besides if $\De_1$ and
$\De_2$ are connected by $G_i$,
then either ${\rm pr}_i(\De_1)={\rm pr}_i(\De_2)$ or
$\{{\rm pr}_i(\De_1),{\rm pr}_i(\De_2)\}\in F^{(i)}$ so that $\De_1$
and $\De_2$ fall in the same $A^\uparrow$ of $\cR^\uparrow_i(Y,F)$.
Since the $A^\uparrow$ cover ${\Br Y}\cap\cD_{\le i}^{(N)}$,
we conclude that $\cR^\uparrow_i(Y,F)$ is the set of connected components
of ${\Br Y}\cap\cD_{\le i}^{(N)}$ with respect to the graph $G_i$.

Considering the fact that $G_{i_{\rm max}(Y)}$ connects the
unique element $\Br Y$ of ${\cR^\uparrow}_{i_{\rm max}(Y)}(Y,F)$, we can
construct a tree $\Gt_0\subset G_{i_{\rm max}(Y)}$ connecting $\Br Y$,
with the property that for every $i\le i_{\rm max}$ and every
$A^\uparrow\in\cR^\uparrow_i(Y,F)$, the set of links of $\Gt_0$
that are internal to $A^\uparrow$ form a tree that connects $A^\uparrow$.
The procedure is inductive on $i$.
For $i=i_{\rm min}(Y)\eqdef\min\{i|Y\cap\cD_i^{(N)}\neq\emptyset\}=
\min\{i|{\Br Y}\cap\cD_i^{(N)}\neq\emptyset\}$,
we choose out of $G_i$ a tree that connects $A^\uparrow$, for every
$A^\uparrow\in\cR^\uparrow_i(Y,F)$. The union of such trees is denoted
by $\Gt^i_0$.
Now suppose $i_{\rm min}(Y)\le i<i_{\rm max}(Y)$ and we have built
$\Gt^i_0\subset G_i$, a forest on $\cD_{\le i}^{(N)}$ whose connected
components are the elements of $\cR_i^\uparrow(Y,F)$.
If $A^\uparrow\in\cR_{i+1}^\uparrow(Y,F)$, we just add to the subforest of
$\Gt_0^i$ made by the links that are internal to $A^\uparrow$,
any choice of links from $G_{i+1}$ to get a connecting tree on $A^\uparrow$.
$\Gt_0^i$ plus the newly added links form the set we denote by $\Gt_0^{i+1}$.
Finally we take $\Gt_0=\Gt_0^{i_{\rm max}(Y)}$,
and it satisfies the required property.

The rule for summation depends on the {\em root} $\De_{\rm org}\in{\Br Y}$.
For any $\De\in{\Br Y}$ we define its {\em level} ${\rm lv}(\De)$
as the minimal number of links in $\Gt_0$ to connect it to the root.
To any link
$l=\{\De_1,\De_2\}\in\Gt_0$
such that ${\rm lv}(\De_1)={\rm lv}(\De_2)-1$ we define 
$f(l)\eqdef\De_1$ or the {\em father}, and $s(l)\eqdef\De_2$ the {\em son}.
A link $l$ is now oriented from the son to the father.
Hence we say that $l$ goes {\em downward} if $i(f(l))=i(s(l))+1$, and it
goes {\em upward} if $i(f(l))=i(s(l))-1$.
The only other possibility is that $l$ be a horizontal link i.e.
$l\in F$.
For such a link we can extract from $\cT_{\ep_2/2}(Y,F)$ a factor
$(1+M^{-i(l)}d(l))^{-(d+\ep_2/2)}$
that is enough to sum $s(l)$ knowing $f(l)$.
Note that by construction of $G$ for any $\De\in{\Br Y}$
there is at most one
link $\{\De,\De'\}\in G\backslash F$ with $i(\De')>i(\De)$.
Therefore $\#(G\backslash F)\le \#({\Br Y})$, and as a result
for any upward link we can extract from $M^{-\ep_3\#({\Br Y})}$ a factor
$M^{-\ep_3}$.
We now have to show that from $\cT_{\frac{\ep_2}{2}}(Y,F)$ we can extract
also a factor $M^{-(d+\ep_2/2)}$ per downward link.

Denote by $\Gt_0^{\rm down}$ the set of downward links of $\Gt_0$.
To such a link $l$ we associate the component
$B(l)\in\cR_{i(s(l))}^\uparrow(Y,F)$ that contains $s(l)$.
Note that $i(s(l))<i_{\rm max}(Y)$; therefore $l\mapsto B(l)$ defines
a map from $\Gt_0^{\rm down}$ to $\cR^\uparrow(Y,F)$.
Now we claim that the map $l\mapsto B(l)$ is injective.
This is a consequence of the following lemma about trees.

\begin{lem}
\label{treestruc}
Suppose $\Gt$ is a tree connecting some finite set $E$, whose root
we denote by $e_{\rm root}$.
Given two elements $e_1$ and $e_2$, a minimal subset of $\Gt$, that
contains enough links to connect $e_1$ and $e_2$, is called a path
between $e_1$ and $e_2$.
By definition of a tree such a path exists and is unique, and is denoted by
$\Gp(e_1,e_2)$.
We define the level of some element $e\in E$ as ${\rm lv}(e)
\eqdef\#(\Gp(e,e_{\rm root}))$.
Any link $l\in\Gt$ is of the form $\{e_1,e_2\}$ with
${\rm lv}(e_2)={\rm lv}(e_1)+1$, and again we pose $s(l)\eqdef e_2$
and $f(l)\eqdef e_1$. We now have the following result.
If $F$ is a subset of $E$ such that the forest
\be
\Gt_F\eqdef\{l\in\Gt|l\subset F\}
\ee
is a tree connecting $F$, then there can be at most one link $l\in\Gt$
such that $s(l)\in F$ and $f(l)\not\in F$.
\end{lem}

\noindent
{\bf Proof:\ \ }
We need first to make the following remarks.
Given some $e\in E$, any vertex along $\Gp(e,e_{\rm root})$
has a level at most equal to ${\rm lv}(e)$.
Besides there can be at most one link $l\in\Gt$, with $s(l)=e$.
Indeed if there were two such links $l_1$ and $l_2$, then $f(l_1)\neq
f(l_2)$, and from $\Gp(f(l_1),e_{\rm root})\cup
\Gp(f(l_1),e_{\rm root})$ we could
extract a path between $f(l_1)$ and $f(l_2)$.
Such a path would have all its vertices of levels at most
${\rm lv}(f(l_1))={\rm lv}(f(l_2))={\rm lv}(e)-1$.
But $\{l_1,l_2\}$ would be a path between $f(l_1)$ and $f(l_2)$
having a vertex, namely $e$, of level ${\rm lv}(e)$.
The existence of these two distinct paths between $f(l_1)$ and $f(l_2)$,
in the tree $\Gt$,
produces a contradiction.

Now let $l_1$ and $l_2$ be two distinct links satisfying the hypothesis of the
lemma with respect to some subset $F$.
From the previous remarks we infer that
$s(l_1)\neq s(l_2)$. Now there exists a sequence $e_1,\ldots,
e_\al$, $\al\ge 2$, of distinct elements of $F$,
such that $e_1=s(l_1)$, $e_\al=s(l_2)$ and
\be
\Gp(s(l_1),s(l_2))=\{\{e_1,e_2\},\{e_2,e_3\},
\ldots,\{e_{\al-1},e_\al\}\}\subset \Gt_F
\ \ .
\ee
Let $\beta$, $1\le \beta\le \al$ be such that ${\rm lv}(e_\beta)$ is
maximal. Then $\beta$ cannot satisfy $1<\beta<\al$,
for we would have two distinct links $\{e_{\beta-1},e_\beta\}$
and $\{e_\beta,e_{\beta+1}\}$ with $s(\{e_{\beta-1},e_\beta\})=
s(\{e_\beta,e_{\beta+1}\})=e_\beta$ which is excluded by the previous
remarks.

Let for instance $\beta=1$, then $\{e_1,e_2\}\subset\Gt_F$
and $l_1\not\subset\Gt_F$ are two distinct links
with $s(\{e_1,e_2\})=s(l_1)=e_1$, this is also impossible.
The case $\beta=\al$ is ruled out in the same manner.
\Endproof

Now let $l_1$ and $l_2$ be in $\Gt_0^{\rm down}$, with $B(l_1)=
B(l_2)=B\in\cR_i^\uparrow(Y,F)$.
We must have $i(s(l_1))=i(s(l_2))=i$; and therefore since
$l_1$ and $l_2$ go downward, $i(f(l_1))=i(f(l_2))=i+1$,
Thus $f(l_1)$ and $f(l_2)$ are not elements of $B$.
But, by our construction of $\Gt_0$, the subset
$(\Gt_0)_B$ of all bonds of $\Gt_0$ that ly in $B$, is a tree connecting
$B$. The lemma now allows us to conclude that $l_1=l_2$.
The injectivity of $l\mapsto B(l)$ implies that
\be
\#(\Gt_0^{\rm down})\le
\#(\cR^\uparrow(Y,F))=\#(\cR(Y,F))
\ \ .
\ee
Let $\Gt_0^{\rm up}$ denote the set of upward links of $\Gt_0$,
and $\Gt_0^{\rm hor}$ the set of horizontal links of $\Gt_0$.
Now $\#(\Gt_0^{\rm up})\le \#({\Br Y})$, (\ref{calU}),
and (ref{decsumY}) show that
\be
\cT_{\ep_2}(Y,F)\le
C_1^{\#(Y)}
.M^{-\ep_3\#(\Gt_0^{\rm up})}
.M^{-(d+\frac{\ep_2}{2})\#(\Gt_0^{\rm down})}
.\prod_{l\in\Gt_0^{\rm hor}}
{\lp 1+M^{-i(l)}{\rm dist}_2(l)\rp}^{-(d+\frac{\ep_2}{2})}
\ \ .
\label{treedecay}
\ee
The right hand side is already a summable tree decay for $\Gt_0$,
however the number of its vertices that are not elements of $Y$
may be too big.
In order to control that number we have to prune the tree $\Gt_0$. a little.
The result will be a tree $\hat\Gt$ whose support $\hat Y$
verifies $Y\subset \hat{Y}\subset\Br{Y}$, and has a
cardinal bounded linearly by $\#(Y)$.

Two distinct links in $\Gt_0^{\rm up}\cup\Gt_0^{\rm down}$
are said {\em consecutive} if they are of the form
$l_1=\{\De_1,\De_2\}$ and $l_2=\{\De_2,\De_3\}$
with $\De_2\not\in Y$,
$\De_1$ just above $\De_2$, $\De_2$ just above $\De_3$,
and such that no other link of $\Gt_0$ contains $\De_2$.
$\De_2$ is then called the {\em intermediate} cube of $l_1$
and $l_2$.
Consecutiveness defines a graph on $\Gt_0^{\rm up}\cup\Gt_0^{\rm down}$
whose set of connected components
we denote by $\cX$. The links that belong to some component $\Gx\in\cX$
must all be going upward or all going downward.
Otherwise, there would be two consecutive links $l_1$ and $l_2$ in
$\Gx$ going in different directions.
Keeping the previous notations,
the case $l_1\in\Gt_0^{\rm up}$, $l_2\in\Gt_0^{\rm down}$ is ruled out since
$\De_2=s(l_1)=s(l_2)$, and $l_1\neq l_2$ contradicts the remark made
at the beginning of the proof of Lemma (\ref{treestruc}).
The remaining case where $l_1\in\Gt_0^{\rm down}$
and $l_1\in\Gt_0^{\rm up}$ is also forbidden. Indeed, since the path
in $\Gt_0$ going from
$\De_2$ to the root $\De_{\rm org}$ goes through vertices with
levels no greater than ${\rm lv}(\De_2)$, and since ${\rm lv}
(\De_1)={\rm lv}(\De_3)={\rm lv}(\De_2)+1$ and there is no link $l\in
\Gt_0$ apart from $l_1$ and $l_2$ that contain $\De_2$, we conclude
that $\De_2=\De_{\rm org}\in Y$.
But this is also excluded in the definition of consecutiveness.

We define the subset $Y_1$ of $Y$ obtained by removing
all the intermediate cubes of consecutive links.
We define the following graph $\Gt_1$ on $Y_1$
\[
\Gt_1\eqdef
(\Gt_0^{\rm hor}\cap\cP(Y_1))
\cup\{\{\De_1,\De_2\}\in\cD(Y_1)|\De_1\neq\De_2,
\]
\be
\exists\Gx\in\cX,\exists
l_1\in\Gx,\exists l_2\in\Gx,\De_1\in l_1,\De_2\in l_2\}
\ \ .
\ee
$\Gt_1$ is simply the graph obtained by replacing maximal linear
chains of consecutive vertical links between nearest neighbors by
a single link joining their extremities.
It is not difficult to see that $\Gt_1$ is a tree connecting $Y_1$.
With the choice of root $\De_{\rm org}$, we can define
as before for any vertex $\De$ of $Y_1$ its level ${\rm lv}(\De)$ and for any
link $l\in \Gt_1$ the cubes $s(l)$ and $f(l)$ with respect to the
tree structure of $\Gt_1$. We can also
make the distinction between links going upward or backward, namely weather
$i(s(l))>i(f(l))$ or $i(s(l))<i(f(l))$ respectively.

Now define the following function on $\cD^{(N)}\times\cD^{(N)}$
\be
F(\De_1,\De_2)\eqdef
\left\{ \begin{array}{ll}
      0 & {\rm if}\ \  \De_1=\De_2 \\
      {\lp 1+M^{-i(\De_1)}{\rm dist}_2(\De_1,\De_2)\rp}^{-(d+\ep_2/2)}
        & {\rm if}\ \  i(\De_1)=i(\De_2),\ \De_1\neq\De_2 \\
      M^{-\ep_3[i(\De_2)-i(\De_1)]}
        & {\rm if}\ \  i(\De_1)<i(\De_2) \\
      M^{-(d+\ep_2/2)[i(\De_1)-i(\De_2)]}
        & {\rm if}\ \  i(\De_1)>i(\De_2)
    \end{array}
\right.
\ \ .
\ee

From (\ref{treedecay}) we readily deduce
\be
\cT_{\ep_2}(Y,F)\le
C_1^{\#(Y)}
.\prod_{l\in\Gt_1}
F(f(l),s(l))
\ \ .
\ee
Indeed, the horizontal links of $\Gt_0$ and $\Gt_1$are the
same so that we still have for them the factors
\be
{\lp 1+M^{-i(l)}{\rm dist}_2(l)\rp}^{-(d+\ep_2/2)}=F(f(l),s(l))
\ \ .
\ee
If $l$ is a vertical link of $\Gt_1$ going upward
obtained from the component $\Gx\in\cX$, then since all links in
$\Gx\subset\Gt_0$ must go upward, we collect from (\ref{treedecay}) the factor
\be
M^{\ep_3\#(\Gx)}=M^{-\ep_3[i(s(l))-i(f(l))]}=F(f(l),s(l))
\ \ .
\ee
Likewise if $l$ is a downward link of $\Gt_1$ obtained from some
$\Gx$, we collect from () the factor
\be
M^{-(d+\ep_2/2)\#(\Gx)}
=M^{-(d+\ep_2/2)[i(f(l))-i(s(l))]}
=F(f(l),s(l))
\ee
The last step of pruning is to cut off the leafs
of $\Gt_1$ that are in $Y_1\backslash Y$.
A leaf of $\Gt_1$ is a cube $\De\in Y_1$ such that there is a unique
$l\in\Gt_1$ with $\De\in l$.
Now we denote by $\hat Y$ the subset of 
$Y_1$ obtained by removing the leafs of $\Gt_1$ that are not in $Y$.
We finally put $\hat\Gt\eqdef\Gt_1\cap\cP({\hat Y})$. It
is obvious that $\hat \Gt$ is a tree connecting $\hat Y$, and that
$\hat Y$ contains $Y$. Besides (\ref{treedecay}) trivially entails
\be
\cT_{\ep_2}(Y,F)\le
C_1^{\#(Y)}
.\prod_{l_\in\hat{\Gt}}
F(f(l),s(l))
\ \ .
\ee
Now the point is that we have the following bound.

\begin{lem}
\label{prune}
\be
\#(\hat{Y})\le 2\#(Y)-1
\ \ .
\ee
\end{lem}

\noindent{\bf Proof:\ \ }
If $\De\in \hat{Y}\backslash Y$, then $\De$ cannot be a leaf of $\Gt_1$,
therefore there exist two distinct links of $\Gt_1$ arriving at $\De$.
These are vertical links, since horizontal
links are between cubes of $Y$ only.
They have to be produced by two distinct
and thus disjoint components $\Gx_1$ and $\Gx_2$ in $\cX$.
If $l_1\in\Gx_1$, $l_2\in\Gx_2$ and $\De\in l_1\cap l_2$,
then $l_1\neq l_2$.
We have showed that at least two vertical links in $\Gt_0$ arrive at
$\De$.
Now we claim that $\De$ has to be a {\em fork} in $\Br Y$.
Indeed, vertical links in $\Gt_0$ are between vertically neighboring cubes.
Therefore if $l_1=\{\De_1,\De\}$ and $l_2=\{\De_2,\De\}$, then
having both $\De_1$ and $\De_2$ below $\De$ is impossible by $\De_1\neq\De_2$.
Besides, if we suppose that $\De$ is not a fork, the case where $\De_1$
and $\De_2$ are above $\De$ is also ruled out.
The only remaining possibility is that, for instance,
$\De_1$ be above and $\De_2$ below $\De$.
But then, since $\De\not\in Y$ and $\De$ is supposed not to be a fork,
$l_1$ and $l_2$ must be consecutive.
This contradicts $l_1\in\ch_1$, $l_2\in\ch_2$ and $\ch_1\neq\ch_2$.

Finally, we have already proven in (\ref{forks})
that the number of forks is bounded
by $\#(Y)-1$, and the lemma follows.
\Endproof

Now it is easy to see that there is a constant $K_2(d,M,\ep)$ such that
\be
\sup_{\De_1\in\cD^{(N)}}
\sum_{\De_2\in\cD^{(N)}}
F(\De_1,\De_2)
\le
K_2(d,M,\ep)
\ \ ,
\ee
uniformly in $N$.

We now have at our disposal all the needed ingredients to sum
over $Y$.
If $Y$ is an admissible set in $\cD^{(N)}$ containing $\De_{\rm org}$, and
such that $\#(Y)=Q$, we choose a forest $F(Y)$ that is satisfying to $Y$
and such that $\cT_{\ep_2}(Y)=\cT_{\ep_2}(Y,F(Y))$.
Then starting from $Y$ and $F(Y)$, we construct, as described previously,
a set $\hat{Y}$ containing $Y$
and a tree $\hat{\Gt}$ connecting $\hat{Y}$.
Then the sum over $Y$ is bounded by
\be
\sum_Y
\cT_{\ep_2}(Y)
\le
C_1^Q\times
\sum_{{Q\le\hat{Q}}\atop{\le 2Q-1}}
\sum_{\hat{Y}|\#(\hat{Y})=\hat{Q}}
\sum_{\hat{\Gt}}
\sum_{Y\subset\hat{Y}}
\prod_{l\in\hat{\Gt}}
F(f(l),s(l))
\ \ .
\ee
The sum over $\hat{Q}$ is over the cardinal
of $\hat{Y}$ that, according to Lemma (\ref{prune}),
has to range through the interval $\{Q,\ldots,2Q-1\}$.
The sum over $Y$, as a subset of $\hat{Y}$, costs a factor $2^{\#(\hat{Y})}
<4^Q$.
The sum over $\hat{\Gt}$ and $\hat{Y}$ is done in the usual
way [GJ,B,R] thanks to the tree decay $\prod_{l\in\hat{\Gt}}
F(f(l),s(l))$.

For each $\hat{Y}$ there exist an injection $\tau:\{1,2,\ldots,\hat{Q}\}
\rightarrow \cD^{(N)}$ whose range is $\hat{Y}$.
Summing over $\hat{Y}$ is the same as summing over $\tau$ and
dividing by $\hat{Q}!$. Now $\tau$ allows
to transport the tree structure $\hat{\Gt}$ of the varying set $\hat{Y}$
onto the fixed set
$\{1,2,\ldots,\hat{Q}\}$, yielding a tree $\til{\Gt}$, and a root $\til{r}$.
We now have
\be 
\sum_{\hat{Y}|\#(\hat{Y})=\hat{Q}}
\sum_{\hat{\Gt}}
\prod_{l\in\hat{\Gt}}
F(f(l),s(l))
=
\frac{1}{\hat{Q}!}
\sum_{\til{\Gt}}
\sum_{\til{r}}
\sum_{\tau|\tau(\til{r})=\De_{\rm org}}
\prod_{\til{l}\in\til{\Gt}}
F(\tau(f(\til{l})),\tau(s(\til{l})))
\ \ .
\ee
The sum over $\tau$ such that $\tau(\til{r})=\De_{\rm org}$, is
done by summing over $\tau(j)$ where $j\in\{1,\ldots,\hat{Q}\}$
is a leaf of $\til{\Gt}$, then for the ancestors all the way up to
the root $\til{r}$.
We then obtain a factor $K_1(d,M,\ep_2)^{\hat{Q}-1}$;
the sum over $\til{r}$ costs a factor $\hat{Q}$.
The sum over $\til{\Gt}$ is bounded, with the help of Cayley's theorem, 
by $\hat{Q}^{\hat{Q}-2}$.
Finally
\be
\sum_{Y}
\cT_{\ep_2}(Y)\le
(4C_1)^Q\times
\sum_{Q\le\hat{Q}\le 2Q-1}
\frac{1}{\hat{Q}!}
\hat{Q}^{\hat{Q}-1}
K_2(d,M,\ep_2)^{\hat{Q}-1}
\ \ .
\ee
Note that for any integer $n\ge 1$, $n^n\le e^n n!$, therefore
\be
\sum_Y
\cT_{\ep_2}(Y)\le
K_1(d,M,\ep_2)^Q
\ \ ,
\ee
with $K_1(d,M,\ep_2)=4C_1.e^2.K_2(d,M,\ep_2)^2$.
\Endproof

\subsection{A toolkit for the bounds}

We gather in this section a few lemmas that will be useful in the
bounding process.
The first two are standard see [GJ2 ,R1], but for completeness
we prove them in detail.

\begin{lem}
\label{propbd}
For any $r\ge 0$, there exists a constant $K_3(r)\ge 1$ such that,
for any $x_1$ and $x_2$ in $\La$ and $i_1$ and $i_2$ in
$\{0,1,\ldots.N\}$
and for any multi-indices $\al_1$ and $\al_2$ of length not greater
than 2, we have
\be
|\partial_{x_1}^{\al_1}\partial_{x_2}^{\al_2}
C(x_1,i_1;x_2,i_2)|\le
\de_{i_1,i_2}K_3(r)
\lp 1+M^{-i_1} d_2(x_1,x_2)\rp^{-r}
M^{-i_1(2+|\al_1|+|\al_2|)}
\ \ .
\label{bdprop}
\ee
As customary, a multi-index $\al$ is a quadruplet $(\al^1,\ldots,\al^4)$
of nonnegative integers, whose length is denoted by $|\al|
\eqdef \al^1+\ldots+\al^4$. The notation $\partial_x^\al$ is for
$\partial^{|\al|}/(\partial x^1)^{\al^1}\ldots
(\partial x^4)^{\al^4}$, where upper indices label the coordinates of $x$.
\end{lem}

\noindent{\bf Proof:\ \ }
We have by (\ref{defcov})
\be
\partial_{x_1}^{\al_1}\partial_{x_2}^{\al_2}
C(x_1,i_1;x_2,i_2)
=
\de_{i_1,i_2}
\int\frac{d^4 p}{(2\pi)^4}
\frac{e^{ip(x_1-x_2)}}{p^2}
i^{|\al_1|-|\al_2|}
p^{\al_1+\al_2}
\lp e^{-M^{2i_1}p^2}-e^{-M^{2(i_1+1)}p^2}\rp
\ \ ,
\ee
where given a multi-index $\al=(\al^1,\ldots,\al^4)$, $p^\al$
denotes the product $\prod_{\mu-1}^4
(p^\mu)^{\al^\mu}$, where the $p^\mu$'s are the components of
the vector $p$. We make the change of variables $p=M^{-i_1}\til{p}$
so that
\be
|\partial_{x_1}^{\al_1}\partial_{x_2}^{\al_2}
C(x_1,i_1;x_2,i_2)|\le
\de_{i_1,i_2}
M^{-i_1(2+|\al_1|+|\al_2|)}
\left|
\int\frac{d^4 \til{p}}{(2\pi)^4}
\frac{e^{i\til{p}X}}{\til{p}^2}
\til{p}^{\al_1+\al_2}
\lp e^{-\til{p}^2}-e^{-M^2\til{p}^2}\rp
\right|
\ \ .
\ee
where $X\eqdef M^{-i_1}(x_1-x_2)$.
For any values of $\al_1$ and $\al_2$ ranging over a finite set,
the integral in the right hand side 
is the Fourier transform of a smooth function in $\til{p}$, therefore
it decreases faster than any positive power of $\frac{1}{(1+|X|_2)}$.
In particular there exists a constant $K_3(r,\al_1,\al_2)\ge 1$,
such that
\be
\left|
\int\frac{d^4 \til{p}}{(2\pi)^4}
\frac{e^{i\til{p}X}}{\til{p}^2}
\til{p}^{\al_1+\al_2}
\lp e^{-\til{p}^2}-e^{-M^2\til{p}^2}\rp
\right|
\le 
K_3(r,\al_1,\al_2). (1+|X|_2)^{-r}
\ \ .
\ee
Now take $K_3(r)=\max\{K_3(r,\al_1,\al_2)|\ |\al_1|\le 2,
|\al_2|\le 2\}$
and the lemma follows.
\Endproof

Note that, with our choice of cut-off, we have even the exponential
decrease of the propagator, but we prefer to use only a power law to 
obtain the bounds on the cluster expansion.

\begin{lem}
\label{Gauss}
{\bf The Gaussian bound - The principle of local factorials}

There exists a constant $K_4\ge 1$ such that for any $N\ge 0$, and any
polymer $Y$,
large field region $\Ga$, allowed graph $\Gg$,
and vector of interpolating parameters $h=(h_1,\ldots,h_k)$,
we have the following bound
\be
\left|
\int
d\mu_{C[T_{\Gg,\Ga}({\bf h})]}(\ph) 
\prod_{j\in J}
\partial_{x_j}^{\al_j}\ph_{i_j}(x_j)\right|
\le
\prod_{j\in J}\lp K_4M^{-i_j(1+|\al_j|)}\rp
\times\prod_{\De\in Y}(n(\De)!)^{\frac{1}{2}}
\ \ .
\label{Gaussbd}
\ee
Here $(x_j)_{j\in J}$ is any family of points in $\La$,
$(i_j)_{j\in J}$ is any corresponding family of scale indices in
$\{0,1,\ldots,N\}$, and $(\al_j)_{j\in J}$ is any family of multi-indices
such that for every $j$, $|\al_j|\le 2$.
We require that for each $j\in J$,
$x_j$ be interior to the cube $\De(x_j,i_j)$.
Finally we have introduced the notation
$n(\De)\eqdef\#(\{j\in J|\De(x_j,i_j)=\De\})$.
\end{lem}

\noindent{\bf Proof:\ \ }
Since the measure is Gaussian, we can integrate thanks to Wick's
theorem. We have a sum over all the contractions $c$,
i.e. over the involutions of $J$ without fixed points,
between the fields.
Remark also that a derivative $\partial_{x_j}^{\al_j}$
acts now on the extremity of the propagator attached to the contracted field
$\partial_{x_j}^{\al_j}\ph_{i_j}(x_j)$.
Besides the propagators are those of the interpolated covariance
$C[T_{\Gg,\Ga}({\bf h})]$ defined in (\ref{cov}).
Since we have assumed the points $x_j$ to be interior to the corresponding
cubes $\De(x_j,i_j)$, the derivatives $\partial_{x_j}^{\al_j}$
do not act on the $h$ dependence that is in general
discontinuous across the boundary of a cube $\De$.
Since the multiplying $h$ parameter is between 0 and 1,
we bound the derivated propagator of the interpolated
covariance by the corresponding derivated propagator of the
original covariance $C$.
Then Lemma \ref{propbd} applies and gives with $r=5$
\[
\left|
\int
d\mu_{C[T_{\Gg,\Ga}({\bf h})]}(\ph) 
\prod_{j\in J}
\partial_{x_j}^{\al_j}\ph_{i_j}(x_j)\right|
\le
\prod_{j\in J}
\lp
\sqrt{K_3(5)}.M^{-i_j(1+|\al_j|)}
\rp
\]
\be
\times
\sum_c
\prod_{{\{j,j'\}\subset J}\atop{j'=c(j)}}
\de_{i_j.i_{j'}}
\lp
1+M^{-i_j}d_2(x_j,x_{j'})\rp^{-5}
\ \ .
\label{locfa}
\ee
The sum over $c$ is done inductively following a classical argument.
Remark that there exists a constant $K_5\ge 1$ such that for any
$\De\in\cD_0^{(N)}$,
\be
\sum_{\De'\in\cD_0^{(N)}}
\lp 1+d_2(\De,\De')\rp^{-5}
\le K_5
\ \ .
\label{defK5}
\ee
Since the boxes of scale $i$ have side $M^i$ we have also that
for any $\De\in Y$
\be
\sum_{\De'\in Y}
\de_{i(\De),i(\De')}\lp 1+M^{-i(\De)}d_2(\De,\De')\rp^{-5}
\le K_5
\ \ .
\ee
We denote $\De_j\eqdef \De(x_j,i_j)$ for every $j\in J$.
Suppose we have ordered $J$ as $\{j_1,\ldots,j_s\}$
such that $n(\De_{j_1})\ge n(\De_{j_2})\ge\ldots\ge
n(\De_{j_s})$. To sum over $c(j_1)$ , we sum first over $\De_{c(j_1)}$,
then over $c(j_1)$ knowing $\De_{c(j_1)}$.
The sum over $\De_{c(j_1)}$ is done thanks to the following factor
in (\ref{locfa})
\be
\de_{i_{j_1},i_{c(j_1)}}
\lp 1+M^{-i_{j_1}}d_2(x_{j_1},x_{c(j_1)})\rp^{-5}
\le
\de_{i(\De_{j_1}),i(\De_{c(j_1)})}
\lp 1+M^{-i(\De_{j_1})}d_2(\De_{j_1},\De_{c(j_1)})\rp^{-5}
\ \ ,
\ee
and costs a factor $K_5$.
The sum over $c(j_1)$ knowing $\De_{c(j_1)}$ costs
a factor
\be
n(\De_{c(j_1)})\le\sqrt{n(\De_{j_1})n(\De_{c(j_1)})}
\ \ ,
\ee
by our choice of the ordering of $J$.
We now pick the element $j$ with the smallest label in
$J\backslash\{j_1,c(j_1)\}$ and sum over its image by $c$ in the
same manner as before thus obtaining a factor
\be
K_5\times\sqrt{n(\De_j)n(\De_{c(j)})}
\ee
and continue the process.
Since a square root $\sqrt{n(\De_j)}$ will appear
exactly once by definition of the contraction
scheme $c$, in the end we have that
\be
\sum_c\prod_{{\{j,j'\}\subset J}\atop{j'=c(j)}}
\de_{i_j,i_{j'}}
\lp 1+M^{-i_j}d_2(x_j,x_{j'})\rp^{-5}
\le
\prod_{j\in J}\sqrt{(K_5n(\De_j)}
=K_5^{\frac{1}{2}\#(J)}\prod_{{\De\in Y}\atop{n(\De)\neq 0}}
\sqrt{n(\De)^{n(\De)}}
\ee
\be
\le \lp\sqrt{K_5 e}\rp^{\#(J)}\prod_{\De\in Y}\sqrt{n(\De)!}
\ \ ,
\ee
since for any integer $n$, $n\ge 1$, we have $n^n\le e^n n!$.
The lemma follows with $K_4=\sqrt(e.K_5.K_3(5))$.
\Endproof

\begin{lem}
\label{displ}
{\bf The displacement of local factorials}

Suppose $\cO$ and $\cE$ are finite sets,
$S_1$ and $S_2$ are two maps from $\cO$ to $\cE$, and
$G$ is a function from $\cE\times\cE$ to $[0,+\infty[$ such that
\be
\sup_{\De_1\in\cE}
\sum_{\De_2\in\cE}
G(\De_1,\De_2)\le 1
\ \ .
\ee
We introduce for every $\De\in\cE$, the notation
\be
n_1(\De)\eqdef\#(\{a\in \cO|S_1(a)=\De\})
\ \ ,
\ee
\be
n_2(\De)\eqdef\#(\{a\in \cO|S_2(a)=\De\})
\ \ .
\ee
If for every $a\in \cO$, $G(S_1(a),S_2(a))>0$, then
\be
\prod_{\De\in\cE}
{n_1(\De)!}
\le e^{\#(\cO)}\times\prod_{\De\in\cE}{n_2(\De)!}
\times
\prod_{a\in\cO}
G(S_1(a),S_2(a))^{-1}
\ \ .
\ee
\end{lem}

\noindent{\bf Proof:\ \ }
Remark that the left hand side is the number of permutations
$\si:\cO\rightarrow\cO$ such that $S_1=S_1\circ\si$.
Given such a $\si$, we define the map $\psi_\si:\cO\rightarrow\cE$,
by $\psi_\si\eqdef S_2\circ\si$.
Now
\be
\sum_{\si} 1=
\sum_\psi
\#(\{\si|\psi_\si=\psi\})
\ \ ,
\ee
where the sum over $\psi$ ranges through all the maps $\psi:\cO
\rightarrow\cE$ for which there exists a $\si$ such that $\psi=\psi_\si$.
For a given $\psi$, looking for $\si$ such that $\psi_\si=\psi$
implies looking for $\si(a)$ such that $S_2(\si(a))=\psi(a)$,
for every $a\in\cO$.
There are at most $n_2(\psi(a))$ possible choices, therefore
\be
\#(\{\si|\psi_\si=\psi\})\le\prod_{a\in\cO}
n_2(\psi(a))
\ \ .
\ee
However we supposed that there exists at least one $\si_0$
such that $\psi=\psi_{\si_0}$
and thus
\be
\prod_{a\in\cO}n_2(\psi(a))=\prod_{a\in\cO}n_2(S_2(\si_0(a)))
=\prod_{a\in\cO}n_2(S_2(a))
\ \ ,
\ee
since $\si_0:\cO\rightarrow\cO$ is a permutation.
Finally since
\be
\prod_{a\in\cO}n_2(S_2(a))=\prod_{{\De\in\cE}\atop{n_2(\De)\neq 0}}
n_2(\De)^{n_2(\De)}
\le
\prod_{{\De\in\cE}\atop{n_2(\De)\neq 0}}
\lp
e^{n_2(\De)}.n_2(\De)!\rp
\ \ ,
\ee
we have
\be
\#(\{\si|\psi_\si=\psi\})\le
e^{\#(\cO)}\times\prod_{\De\in\cE}n_2(\De)!
\ \ .
\label{dis1}
\ee
Let $\psi$ be of the form $\psi_\si$, for some permutation $\si$
such that $S_1\circ\si=S_1$, we then say that $\psi$ is {\em admissible}.
This entails
\bea
\prod_{a\in\cO}G(S_1(a),\psi(a)) & = &
\prod_{a\in\cO}G(S_1(a),S_2(\si(a)))\\
 & = & \prod_{a\in\cO}G(S_1(\si(a)),S_2(\si(a)))\\
 & = & \prod_{a\in\cO}G(S_1(a),S_2(a))\ \ ,\label{rearreng}
\eea
since $\si$ is a permutation of $\cO$.
As a consequence, if for every $a\in\cO$, $G(S_1(a),S_2(a))>0$,
then we have also $G(S_1(a),\psi(a))>0$.
Therefore we can write
\be
\sum_{\psi\ {\rm admissible}} 1=
\sum_{\psi\ {\rm admissible}}\ \prod_{a\in\cO}G(S_1(a),\psi(a))\times
\prod_{a\in\cO}G(S_1(a),\psi(a))^{-1}
\ee
\be
\le
\lp
\sum_{\psi\ {\rm admissible}}\ \prod_{a\in\cO}G(S_1(a),\psi(a))
\rp
. \sup_{\psi\ {\rm admissible}}
\lp\prod_{a\in\cO}G(S_1(a),\psi(a))^{-1}\rp
\ \ .
\ee
But by the argument used to prove (\ref{rearreng}) we conclude,
since there exists some admissible $\psi$ (take $\psi_\si$ with $\si$ the
identity), that
\be
\sup_{\psi\ {\rm admissible}}
\lp\prod_{a\in\cO}G(S_1(a),\psi(a))^{-1}\rp=
\prod_{a\in\cO}G(S_1(a),S_2(a))^{-1}
\ \ .
\label{dis2}
\ee
Remark also that by assumption on $G$,
\be
\sum_{\psi\ {\rm admissible}}\ \prod_{a\in\cO}G(S_1(a),\psi(a))
\le
\prod_{a\in\cO}
\lp\sum_{\De\in\cE}G(S_1(a),\De)\rp\le 1
\ \ .
\label{dis3}
\ee
Now putting (\ref{dis1}), (\ref{dis2}) and (\ref{dis3})
together just proves the lemma.
\Endproof

We now state a lemma that allows to sum small cubes in big ones,
with a less than summable vertical decay, typically like $M^{-(4+\ep)
|i-j|}$. This miracle is possible since the small cubes are
restricted to an already known polymer $Y$.
The price to pay is a product of local factorials for the big cubes
of reference, and a small factor per cube of $Y$.
The idea is to distribute, for each cube of $Y$, a piece of the associated
small factor among the cubes below.
This motivates the somewhat poetic name of the lemma.

\begin{lem}
\label{rainlem}
{\bf The rain of small factors}

Let $K_6\ge 1$ and $\ep_4>0$ be two chosen constants.
Suppose ${\Br{\De_1}},\ldots,{\Br{\De_p}}$
are \und{distinct} boxes in $\cD^{(N)}$, and $n_1,\ldots,n_p$ are
positive integers.
For a given polymer $Y$ in $\cD^{(N)}$, define
\be
\cY_Y({\Br{\De_1}},\ldots,{\Br{\De_p}};n_1,\ldots,n_p)
\eqdef
\prod_{k=1}^p
\lp
\sum_{\De_k\in Y,\De_k\subset{\Br{\De_k}}}
M^{-\ep_4(i({\Br{\De_k}})-i(\De_k))}
\rp^{n_k}
\ \ .
\label{rainres}
\ee
Then we have the following bound
\be
\cY_Y({\Br{\De_1}},\ldots,{\Br{\De_p}};n_1,\ldots,n_p)
\le K_6^{\#(Y)}
\lp
(1-M^{-\frac{\ep_4}{2}})^2\log K_6
\rp^{-(n_1+\ldots n_p)}
.\prod_{k=1}^p n_k!
\ \ .
\ee
\end{lem}

\noindent{\bf Proof:\ \ }
We pose $q=M^{-\frac{\ep_4}{2}}$. Since $(1-q)\sum_{i=0}^{+\infty}
q^i=1$
we can write
\be
1= K_6^{\#(Y)}.
\prod_{\De\in Y}
\lp
K_6^{-(1-q)\sum_{i=0}^{+\infty}q^i}
\rp
\ee
\be
\le
K_6^{\#(Y)}.
\prod_{\De\in Y}
\lp
\prod_{{\Br{\De}}\in\cD^{(N)},\De\subset{\Br{\De}}}
K_6^{-(1-q)q^{[i({\Br{\De}})-i(\De)]}}
\rp
\ee
\be
\le
K_6^{\#(Y)}.
\prod_{{\Br{\De}}\in\cD^{(N)}}
\lp
\prod_{\De\in Y,\De\subset{\Br{\De}}}
K_6^{-(1-q)q^{[i({\Br{\De}})-i(\De)]}}
\rp
\ee
\be
\le
K_6^{\#(Y)}
\prod_{k=1}^p
\lp
\prod_{\De\in Y,\De\subset{\Br{\De_k}}}
K_6^{-(1-q)q^{[i({\Br{\De_k}})-i(\De)]}}
\rp
\ \ .
\label{rainpour}
\ee
Therefore we define for any ${\Br{\De}}\in\cD^{(N)}$ and $n\ge 1$,
\be
\om({\Br{\De}},n)\eqdef
\prod_{\De\in Y,\De\subset{\Br{\De}}}
\lp
K_6^{-(1-q)q^{[i({\Br{\De}})-i(\De)]}}
\rp
\times
\lp
\sum_{\De\in Y,\De\subset{\Br{\De}}}
M^{-\ep_4[i({\Br{\De}})-i(\De)]}
\rp^n
\ee
for which we can write, recalling that $q=M^{-\frac{\ep_4}{2}}$,
\be
\om({\Br{\De}},n)=
\sum_{i_1,\ldots,,i_n=0}^{i({\Br{\De}})}
\prod_{j=1}^n
q^{[i({\Br{\De}})-i_j]}
\cA_{i_1,\ldots,i_n}
\ \ .
\label{defomeg}
\ee
Here $\cA_{i_1,\ldots,i_n}$ denotes
\be
\cA_{i_1,\ldots,i_n}
\eqdef
\sum_{\De_1,\ldots,\De_n}
\prod_{j=1}^n
q^{[i({\Br{\De}})-i_j]}
.
\prod_{\De\in Y,\De\subset{\Br{\De}}}
\lp
K_6^{-(1-q)q^{[i({\Br{\De}})-i(\De)]}}
\rp
\ \ ,
\ee
where, for any $j$, $1\le j \le n$,
the sum over $\De_j$, $1\le j\le n$, ranges through
all the cubes $\De_j\in Y\cap\cD_{i_j}^{(N)}$ contained in $\Br{\De}$.
Suppose the set $\{i_1,\ldots,i_n\}$ can be written as
$\{a_1,\ldots,,a_m\}$
with $m\ge 1$ and $a_1,\ldots,a_m$
all distinct.
Now let $\nu_r$ $1\le r\le m$ denote the number of $j$'s,
$1\le j\le n$ such that $a_r=i_j$.
Then for every $r$, $1\le r\le m$, we have $\nu_r\ge 1$,
beside $\sum_{1\le r\le m} \nu_r=n$.
If for any $i$, $0\le i\le N$,
$\rh_i$ denotes $\#(\{\De\in Y| i(\De)=i,\De\subset{\Br{\De}}\})$, then
\be
\cA_{i_1,\ldots,i_n}
=
\rh_{a_1}^{\nu_1}\ldots
\rh_{a_m}^{\nu_m}
.\prod_{\De\in Y,\De\subset{\Br{\De}}}
\lp
K_6^{-(1-q)q^{[i({\Br{\De}})-i(\De)]}}
\rp
.\prod_{r=1}^m
q^{[i({\Br{\De}})-a_r]\nu_r}
\ee
\be
\le
\rh_{a_1}^{\nu_1}\ldots
\rh_{a_m}^{\nu_m}
.\prod_{r=1}^m
\lp
K_6^{-(1-q)q^{[i({\Br{\De}})-a_r]}}
\rp^{\rh_{a_r}}
.\prod_{r=1}^m
q^{[i({\Br{\De}})-a_r]\nu_r}
\ \ .
\label{rainA}
\ee
Indeed, we have kept only the powers of $K_6^{-1}$ that come from cubes
$\De\in Y$, $\De\in{\Br{\De}}$ that
have a scale among $a_1,\ldots,a_m$.
(\ref{rainA}) now reads
\be
\cA_{i_1,\ldots,i_n}
\le
\prod_{r=1}^m
\lp
\lp 
q^{[i({\Br{\De}})-a_r]}\rh_{a_r}
\rp^{\nu_r}
\exp\lp
-(1-q)(\log K_6)q^{[i({\Br{\De}})-a_r]}\rh_{a_r}\rp
\rp
\ee
\be
\le
\lp
(1-q)(\log K_6)
\rp^{-n}
\times\prod_{r=1}^m\lp y_r^{\nu_r}\exp(-y_r)\rp
\ \ ,
\ee
where
$y_r\eqdef(1-q)(\log K_6)
q^{[i({\Br{\De}})-a_r]}\rh_{a_r}$, and we have used
$\sum_{1\le r\le m}\nu_r=n$.
Now the bound $y^\nu e^{-y}\le\nu!$, for any $y\ge 0$,
shows that
\be
\cA_{i_1,\ldots,i_n}
\le
\lp
(1-q)(\log K_6)
\rp^{-n}.\prod_{r=1}^m\nu_r!
\ \ .
\ee
By property of the multinomial coefficients, since
$\sum_{1\le r\le m}\nu_r=n$, we have
\be
\frac{n!}{\nu_1!\ldots\nu_m!}\ge 1
\ \ .
\label{multinom}
\ee
and therefore
\be
\cA_{i_1,\ldots,i_n}
\le
\lp
(1-q)(\log K_6)
\rp^{-n}.n!
\ \ .
\ee
This together with (\ref{defomeg}), implies
\be
\om({\Br{\De}},n)\le
\lp
(1-q)(\log K_6)
\rp^{-n}.n!.
\sum_{i_1,\ldots,i_n=0}^{i({\Br{\De}})}
\prod_{j=1}^n
q^{[i({\Br{\De}})-i_j]}
\ee
\be
\le
\lp
(1-q)(\log K_6)
\rp^{-n}.n!.\prod_{j=1}^n\lp\sum_{i=0}^{+\infty}q^i\rp
\ee
\be
\lp
(1-q)^2(\log K_6)
\rp^{-n}.n!
\ \ .
\label{ombound}
\ee
Finally by (\ref{rainpour}) and (\ref{defomeg}), we have
\be
\cY_Y({\Br{\De_1}},\ldots,{\Br{\De_p}};n_1,\ldots,n_p)\le
K_6^{\#(Y)}.
\prod_{k=1}^p
\om({\Br{\De_k}},n_k)
\ \ ,
\ee
which, together with the bound (\ref{ombound}) proves the lemma.
\Endproof

A basic tool that seems unavoidable for just
renormalizable models is what we call \und{domination} [FMRS,R].
It is the use of the positivity of the interaction to bound low momentum
fields. An essential ingredient is to compare
a field $\ph(x)$ to its average on a volume containing $x$.
The average can be bounded by the interaction, or using a small field
condition in $\ch_\Ga$, whereas the difference term called {\em fluctuation}
is expressed with double gradients of the field that are well behaved
regarding bounds.
This motivates the following lemma,
with a flavor of Sobolev inequalities.

\begin{lem}
\label{avrSobo}
There exists a constant $K_7\ge 1$ such that for any $N\ge 0$,
for any pair of cubes $\De_1$, $\De_2$ in $\cD^{(N)}$
such that $\De_1\subset\De_2$, and for any coordinate indices
$\mu$ and $\nu$, there exist smooth functions
\be
f_{\De_1,\De_2}^{\mu,\nu}:
\De_2\times(\De_1\times\De_1\times[0,1]\times[0,1])
\rightarrow\RR
\ee
and a smooth function $\cX_{\De_1,\De_2}$ with the same
domain and range $\De_2$, such that the following requirements are satisfied.
First we demand that
\be
\sup_{x\in\De_2}\sum_{\mu,\nu=1}^4
\int_{\De_1\times\De_1\times[0,1]\times[0,1]}d{\bf w}
|f_{\De_1,\De_2}^{\mu,\nu}(x,{\bf w})|\le 16 M^{2i(\De_2)}
\ \ ,
\label{L1bd}
\ee
and second that for any smooth function $\ph$ on $\De_2$ and for any
point $x$ of $\De_2$, we have the bound:
\[
\left|
\ph(x)-
\int_{\De_1\times\De_1\times[0,1]\times[0,1]}d{\bf w}
f_{\De_1,\De_2}^{\mu,\nu}(x,{\bf w})
\partial_\mu\partial_\nu\ph
(\cX_{\De_1,\De_2}(x,{\bf w}))
\right|
\]
\be
\le
K_7.\lp\frac{1}{|\De_1|}\int_{\De_1}dy\ \ph(y)^4\rp^{\frac{1}{4}}
M^{i(\De_2)-i(\De_1)}
\ \ .
\label{avrbd}
\ee
\end{lem}

\noindent{\bf Proof:\ \ }
Consider a fixed smooth function $\de_0$ on $\RR^4$ that
takes nonnegative values, has support inside $]0,1[^4$, and verifies
\be
\int_{\RR^4}dy\ \de_0(y)=1
\ee
Now if the cube $\De_1$ is of the form $\ze_1+[0,M^{i(\De_1)}[^4$
with $\ze_1\in\RR^4$, we define a smooth function $\de_{\De_1}$
as follows. For any $y\in\RR^4$ we pose
\be
\de_{\De_1}(y)\eqdef
M^{-4i(\De_1)}
\de_0\lp(y-\ze_1)M^{-i(\De_1)}\rp
\ \ .
\ee
Then $\de_{\De_1}$ takes also nonnegative values, has support
in the interior of $\De_1$ and verifies
\be
\int_{\De_1}dy \de_{\De_1}(y)=1
\ \ .
\label{normal}
\ee
Now let $x\in\De_2$, then use twice Taylor's formula and once
(\ref{normal}) to write
\be
\ph(x)=\frac{1}{|\De_1|}\int_{\De_1}du \ph(x)
\ee
\be
=\frac{1}{|\De_1|}\int_{\De_1}du
\lp
\ph(u)+\int_0^1ds(x-u)^\nu\partial_\nu\ph((1-s)u+sx)\rp
\ee
\be
=\frac{1}{|\De_1|}\int_{\De_1}du\ \ph(u)
+\frac{1}{|\De_1|}\int_{\De_1}du
\int_{\De_1}dv \int_0^1ds\ \de_{\De_1}(v)
(x-u)^\nu\partial_\nu\ph((1-s)u+sx)
\ee
\[
=\frac{1}{|\De_1|}\int_{\De_1}du\ \ph(u)
+\frac{1}{|\De_1|}\int_{\De_1}du
\int_{\De_1}dv \int_0^1ds\ \de_{\De_1}(v)
(x-u)^\nu
\]
\be
\lp
\partial_\nu\ph(v)+\int_0^1dt\ ((1-s)u+sx-v)^\mu
\partial_\mu\partial_\nu\ph((1-t)v+t(1-s)u+tsx)
\rp
\ee
\[
=\frac{1}{|\De_1|}\int_{\De_1}du\ \ph(u)
+
\int_{\De_1\times\De_1\times[0,1]\times[0,1]}d{\bf w}
f_{\De_1,\De_2}^{\mu,\nu}(x,{\bf w})
\partial_\mu\partial_\nu\ph
(\cX_{\De_1,\De_2}(x,{\bf w}))
\]
\be
+
\frac{1}{|\De_1|}\int_{\De_1}du
\int_{\De_1}dv\int_0^1ds\ \de_{\De_1}(v)(x-u)^\nu
\ \ .
\label{avr}
\ee
Here we have introduced
\be
f_{\De_1,\De_2}^{\mu,\nu}(x,u,v,s,t)\eqdef
\frac{1}{|\De_1|}\de_{\De_1}(v)(x-u)^\nu
((1-s)u+sx-v)^\mu
\ee
and
\be
\cX_{\De_1,\De_2}(x,u,v,s,t)\eqdef
(1-t)v+t(1-s)u+tsx
\ \ .
\ee
Now remark that by Holder's inequality
\be
\left|
\frac{1}{|\De_1|}\int_{\De_1}du\ \ph(u)
\right|\le
\lp\frac{1}{|\De_1|}\int_{\De_1}du\ \ph(u)^4\rp^{\frac{1}{4}}
\ee
The third expression in equation (\ref{avr}) we denote by $R$ and transform
by an integration by parts on $v$. Actually, we have
\be
R=\frac{1}{|\De_1|}\int_{\De_1}du
\int_{\De_1}dv\int_0^1ds
\ (-\partial_\nu\de_{\De_1}(v))(x-u)^\nu\ph(v)
\ \ .
\ee
There is no boundary term, since the support of $\de_{\De_1}$ is
in the interior of $\De_1$.
Since $(x-u)^\nu$ is bounded by the size $M^{i(\De_2)}$
of the cube $\De_2$, we have
\be
|R|\le M^{i(\De_2)}\int_{\De_1}dv\ |\partial_\nu\de_{\De_1}(v)||\ph(v)|
\ \ .
\ee
By applying the Cauchy-Schwarz then the Holder inequalities, we obtain
\be
|R|\le M^{i(\De_2)}
\lp \int_{\De_1}dv\ (\partial_\nu\de_{\De_1}(v))^2\rp^{\frac{1}{2}}
\lp \int_{\De_1}dv\ \ph(v)^2\rp^{\frac{1}{2}}
\ee
\be
\le M^{i(\De_2)+2i(\De_1)}
\lp \int_{\De_1}dv\ (\partial_\nu\de_{\De_1}(v))^2\rp^{\frac{1}{2}}
\lp \frac{1}{|\De_1|}\int_{\De_1}dv\ \ph(v)^4\rp^{\frac{1}{4}}
\ \ .
\ee
Note that, by the change of variable $v=\ze_1+M^{i(\De_1)}v_0$,
we readily derive
\be
\int_{\De_1}dv\ (\partial_\nu\de_{\De_1}(v))^2
=M^{4i(\De_1)}
\int_{[0,1[^4}dv_0\ M^{-10i(\De_1)}(\partial_\nu\de_0(v_0))^2
\ \ ,
\ee
and therefore
\be
|R|\le M^{i(\De_2)-i(\De_1)}
\lp \frac{1}{|\De_1|}\int_{\De_1}dv\ \ph(v)^4\rp^{\frac{1}{4}}
\lp \int_{[0,1[^4}dv_0\ (\partial_\nu\de_0(v_0))^2 \rp^{\frac{1}{2}}
\ \ .
\ee
As a result if we choose as our constant
\be
K_7\eqdef
1+\max_{1\le \nu\le 4}
\lp \int_{[0,1[^4}dv_0\ (\partial_\nu\de_0(v_0))^2 \rp^{\frac{1}{2}}
\ \ ,
\ee
the bound (\ref{avrbd}) follows.
Finally, for any coordinate indices $\mu$ and $\nu$ and any $x\in\De_2$,
we have
\be
\int_{\De_1}du\int_{\De_1}dv
\int_0^1ds\int_0^1dt
\left|
f_{\De_1,\De_2}^{\mu,\nu}(x,u,v,s,t)
\right|
\le
M^{2i(\De_2)}\int_{\De_1}dv\ \de_{\De_1}(v)=M^{2i(\De_2)}
\ \ ,
\ee
so the proof is complete.
\Endproof

\subsection{Giving every cube its small factor}

Usually, a small factor per cube comes from the coupling constant appearing
in explicitly derived vertices whose number is roughly proportional to the
cardinal of the polymer. However, this proportionality no longer holds for
large field versus small field expansions, since a large 
field region behaves like a single cube as far as the combinatorics of the 
expansion are concerned. To obtain a small factor per cube now involves the 
extraction of a probabilistic estimate on the large field regions that decays
exponentially in its volume. We summarize this large deviation argument by 
saying that ``large field regions are typically rare and small''. 
This is embodied more precisely in the following technical proposition:

\begin{prop}
\label{propsmf}
Let $\ep_5$ be some chosen positive constant, then
there exists a function $U:]0,1[\to ]0,1[$ such that
$\lim_{g\to 0} U(g) =0$
and such that
for any $N\ge 0$, any polymer $Y$ in $\cD ^{(N)}$, any large field region
$\Ga$ in $Y$, any allowed graph $\Gg$
making $Y$ 4-vertex irreducible, and for any
value of the $h$ parameters, the following inequality holds for $0<g<1$:
\[
\int d\mu_{C_{[T_{{\Gg},\Ga}({\bf h})]}} (\ph) \prod_{\De \in \Ga} {\bbbone}_{
\bigl\{
\int_{\De}\ph^{4}_{I_{\Ga}(\De)}\ \ge \ \half g^{-(1+\ep_{1})}\bigr\}}
.\prod_{\De \in Y \backslash \Ga} {\bbbone}_{\bigl\{
\int_{\De}\ph^{4}_{I_{\Ga}(\De)}\ \le \ g^{-(1+\ep_{1})}\bigr\}} 
\]
\be
\prod_{\De\ {\rm isolated}} g^{\ep_5} \ .\ exp (-\cI)  \le U(g) ^{\# (Y)}
\ \ .
\label{smfact}
\ee
We used here the notations of Section 2, and have written $\ph_{I_{\Ga}(\De)}$
instead of $\sum_{i\in I_{\Ga}(\De)}\ph_{i}$ for simplicity.
$\cI$ denotes again the interpolated interaction as reexpressed in
(\ref{rewint}).
Finally a cube $\De$ in $Y$ is called \und{isolated} if
$I_\Ga(\De)=\{i(\De)\}$.
This means that the large field block containing $\De$ is reduced to $\De$,
hence no other cube is glued to $\De$ by $\Ga$. Note that sharp characteristic
functions are used in this bound, and denoted by a $\bbbone$, instead
of the previously introduced smooth functions $\ch_{\Ga}$. Indeed this bound 
will be used after the smooth functions have been bounded by the sharp ones.
\end{prop}

\prf The task of looking for small factors per cubes of $Y$ can be reduced 
to the extraction of such factors for a special category of cubes. This is 
why we define first the notion of $summital$ cubes of $Y$. A cube
$\De$ of $Y$
is said $summital$ if either $i(\De)\ge 1$ and there exists a cube 
$\tilde\De \in \cD^{(N)}_{i(\De)-1}$ that contains no cube of $Y$,
or $i(\De)=0$
in which case we choose $\tilde\De =\De$. 

We denote by $Y_{S}$ the set of summital cubes of $Y$. We have:
\begin{lem}
\label{roof}
The following inequalities relate the cardinals of $Y$ and $Y_{S}$:
\be
\# (Y_{S}) \le \# Y \le {M^{4}\over M^{4}-1} \# (Y_{S}) - {1 \over M^{4}-1}
\ \ .
\ee
\end{lem}

\medskip

\prf The first inequality is tautological, therefore we concentrate on 
the second. If $\De$ is any cube of $\cD^{(N)}$, we define 
$n_{Y,\De}=\#(\{\De'
\in Y | \De' \subset \De \} ) $ and $n_{Y_{S},\De}=\#(\{\De'
\in Y_{S} | \De' \subset \De \} ) $. We prove by induction on $i(\De)$,
that if $n_{Y,\De}>0$ then $n_{Y_{S},\De}>0$ and
\be
n_{Y,\De}  \le {M^{4}\over M^{4}-1} n_{Y_{S},\De} - {1 \over M^{4}-1}
\ \ .
\label{summit}
\ee

This will prove the lemma since $Y$ is non empty and is contained, 
together with $Y_{S}$, in the unique cube of the last slice $\cD^{(N)}_{N}$.

For $i(\De)=0$: if $n_{Y,\De}>0$, then $\De\in Y$ is the only possibility,
and by definition $\De$ has to be a summital cube of $Y$. Therefore 
$n_{Y,\De}=n_{Y_{S},\De}=1$ and we have equality in (\ref{summit}).

Suppose the statement is proven for any $\De$ with $i(\De)=i$, for some fixed
$i\ge 0$. Let now $\De$ be a cube in the following layer, i.e. with 
$i(\De)=i+1$, satisfying $n_{Y,\De}>0$. Consider the set $J$ of cubes $\De'$
such that $\De' \subset \De$, $i(\De')=i$ and $n_{Y,\De'}>0$.

By the induction hypothesis, for any $\De'\in J$, $n_{Y_{S},\De'}>0$
and 
\be
n_{Y,\De'}  \le {M^{4}\over M^{4}-1} n_{Y_{S},\De'} - {1 \over M^{4}-1}
\ \ .
\label{summit2}
\ee
A few cases are to be distinguished:

\noindent{\bf -Case 1:} $\De \not\in Y$. We have by (\ref{summit2})
\be
n_{Y,\De} =\sum_{\De'\in J}n_{Y,\De'}
 \le \sum_{\De' \in J}
\biggl(  {M^{4}\over M^{4}-1} n_{Y_{S},\De'} - {1 \over M^{4}-1}\biggr)
=  {M^{4}\over M^{4}-1} n_{Y_{S},\De} - {\# (J) \over M^{4}-1} \ \ .
\label{summit3}
\ee
Since $n_{Y,\De}>0$, $J$ must be non-empty
and (\ref{summit3}) with $\# (J) \ge 1$ implies (\ref{summit}).

\noindent{\bf -Case 2:} $\De \in Y-Y_{S}$. Since $\De$ is not summital, 
all the cubes $\De'$ just above $\De$ contain a cube of $Y$, and therefore
satisfy $n_{Y,\De'}>0$. Therefore $\# (J)=M^{4}$, and we can write
\be
n_{Y,\De}= 1 + \sum_{\De'\in J}n_{Y,\De'} \le 1 + \sum_{\De'\in J}
\biggl(  {M^{4}\over M^{4}-1} n_{Y_{S},\De'} - {1 \over M^{4}-1}\biggr)
\ee
\be
= 1 +  {M^{4}\over M^{4}-1} n_{Y_{S},\De} - {M^{4} \over M^{4}-1}
=  {M^{4}\over M^{4}-1} n_{Y_{S},\De} - {1 \over M^{4}-1}\ \ ,
\ee
as wanted.

\noindent{\bf -Case 3:} $\De \in Y_{S}$. Here $J$ can possibly be empty,
however we have again:
\be
n_{Y,\De}\le  1 + \sum_{\De'\in J}
\biggl(  {M^{4}\over M^{4}-1} n_{Y_{S},\De'} - {1 \over M^{4}-1}\biggr)
= 1 +  {M^{4}\over M^{4}-1}( n_{Y_{S},\De}-1) - {\# (J) \over M^{4}-1}
\ee
\be
\le  {M^{4}\over M^{4}-1} n_{Y_{S},\De} +1 - {M^{4} \over M^{4}-1}
=  {M^{4}\over M^{4}-1} n_{Y_{S},\De} -{1 \over M^{4}-1} \ \ ,
\ee
This completes the proof of the lemma. \Endproof

Thanks to the lemma, our goal is now limited to extract a small factor per
summital non isolated cube of $Y$.
Let $\De$ be such a cube, we consider 
$P(\De)=
\{\De' \in Y | \De \subset\De'\}$ and the partitions
${\Gp}$ of $P(\De)$ into connected
components with respect to gluing by $\Ga$. If $X$
is such a component, let $i(X)$ denote $\{i(\De') | \De' \in X\}$. Finally
when $X$ ranges through ${\Gp}$, the collection of the $i(X)$
can be written $\{I_{0}(\De), ...,I_{\mu(\De)}(\De)\} $, $\mu(\De)\ge 0$,
where for any $\nu$, $0\le \nu\le \mu(\De)$, $I_{\nu}(\De)$ is connected,
namely of the form $\{i_{\nu}(\De),i_{\nu}(\De)+1,...,j_{\nu}(\De)\}$,
with $j_{\nu}(\De)\ge i_{\nu}(\De)$. Besides the ordering has been chosen
so that for any $\nu$, $j_{\nu}(\De)< i_{\nu +1}(\De)$. Note that 
$i_{0}(\De) =i(\De)$ and since $\De$ is not isolated, $j_{0}(\De)>i_{0}(\De)$.
We denote $\ph _{I_{\nu}(\De)}= \sum_{i=i_{\nu}(\De)}^{j_{\nu}(\De)}\ph_{i}$.

Now let $D$ be some positive constant integer, to be specified later.
For any summital non isolated cube $\De$ of $Y$, we introduce
in the left hand side of (\ref{smfact}) the identity
\be
1=\prod_{{i_0(\De)\le\beta\le i_0(\De)+D}\atop{{\rm pr}_\beta(\De)\in Y}}
\lp
\bbbone_{\{\int_{\tilde \De} \ph_{\beta}^{4} \le K_{8}g^{-(1+\ep_{1})}\}}
+ 
\bbbone_{\{\int_{\tilde \De} \ph_{\beta}^{4} > K_{8}g^{-(1+\ep_{1})} \}}
\rp
\ \ ,
\label{fstdec}
\ee
where $\tilde \De$ is the above
chosen cube in $\cD^{(N)}$ such that $\tilde\De 
\subset\De$ and $\tilde \De$ contains no cube of $Y$.
Here $K_{8}$ is some positive constant that will be fixed later.

Next we expand the product of such factors.
The cubes $\De$, such that at least one term of the form
$\bbbone_{\{\int_{\tilde \De} \ph_{\beta}^{4} > K_{8}g^{-(1+\ep_{1})} \}}$
is selected, are called of type I. Let $\beta(\De)$ be a choice
of index $\beta$, $i_0(\De)\le\beta\le i_0(\De)+D$,
${\rm pr}_\beta(\De)\in Y$,
satisfying the last
property. We can bound the characteristic functions of either form,
chosen for the other indices $\beta$, by one. Remark that one would
have to sum over all possible choices in this first expansion,
i.e. to pay a factor $2^{D+1}$ per non isolated summital cube.

For the cubes such that for any $\beta$ the selected term is
$\bbbone_{\{\int_{\tilde \De} \ph_{\beta}^{4} \le K_{8}g^{-(1+\ep_{1})} \}}$,
which we call of type II, we further introduce
the factor
\be
1=\bbbone_{\{\int_{\tilde \De} \Ph_{I_0(\De)}^{4} 
\le K_{9}g^{-(1+\ep_{1})}\}}+
\bbbone_{\{\int_{\tilde \De} \Ph_{I_0(\De)}^{4} 
>K_{9}g^{-(1+\ep_{1})}\}}
\ \ .
\label{decK9}
\ee

Here again $K_{9}$ is a constant to be specified later.
Now we expand the product of the expressions in the right hand side of
(\ref{decK9}) when $\De$ ranges over type II cubes.
If the first term of (\ref{decK9})
is chosen we call $\De$ of type II.1, otherwise
$\De$ is called of type II.2. For this last category of cubes
two cases are possible.

In case $j_0(\De)\le i_0(\De)+D$, we say that the cube is of type II.2.1,
else we say it is of type II.2.2.
Finally for every II.2.2 cube $\De$, we introduce 
a last chopping of the functional integral, using 
this time a majoration:
\[
1 \le \bbbone_{\bigl\{\max\limits_
{1\le \nu\le\mu(\De)} \lp a_{\nu}(\De)^{16/3}
\int_{\tilde \De} \ph_{I_{\nu}(\De)}^{4}\rp \le K_{10}g^{-(1+\ep_{1})}
\bigr\}}
\]
\be
+ \sum_{
1 \le \nu\le\mu(\De)}\bbbone_{\bigl\{ a_{\nu}(\De)^{16/3}\int_{\tilde \De}
 \ph_{I_{\nu}(\De)}^{4} > K_{10}g^{-(1+\ep_{1})} \bigr\}} 
\ \ ,
\label{decK10}
\ee
where $a_{\nu}(\De)= M^{\frac{3}{40}(i_{\nu}(\De)-i(\tilde\De))}$,
and $K_{10}$ is a constants to be fixed later.

We now expand the product of the right hand side of (\ref{decK10}),
where $\De$
ranges over all cubes of type II.2.2. If the first term of (\ref{decK10}) is
chosen, $\De$ is called of type II.2.2a. If a term
$\bbbone_{\bigl\{ a_{\nu}(\De)^{16/3}\int_{\tilde \De}
 \ph_{I_{\nu}(\De)}^{4} > K_{10}g^{-(1+\ep_{1})} \bigr\}}$
has been chosen, we say that $\De$ is of type II.2.2b.$\nu$.

We see that we have a tree of possible choices, on which we have to sum
to get a majorant of the whole functional integral in (\ref{smfact}).
Note however that the distinction between type II.2.1 and
type II.2.2 cubes is not to be summed over; one has to take the supremum
over the output of both cases.
Now let us concentrate on a particular term in this majorant expression.
Such a term is of the form:

\[
{\GC} = \int  d\mu_{C_{[T_{{\Gg},\Ga}({\bf h})]}} (\ph)
\ \prod_{\De \in \Ga} {\bbbone}_{
\bigl\{\int_{\De}\ph^{4}_{I_{\Ga}(\De)}\ \ge \ \half g^{-(1+\ep_{1})}\bigr\}}
\times\prod_{\De \in Y \backslash \Ga} {\bbbone}_{\bigl\{
\int_{\De}\ph^{4}_{I_{\Ga}(\De)}\ \le \ g^{-(1+\ep_{1})}\bigr\}} 
\]
\[
\times \exp (-\cI)\times\prod_{\De\ {\rm isolated}} g^{\ep_5}
\times\prod_{\De \ {\rm type \ I}}
\bbbone_{\{\int_{\tilde \De} \ph_{\beta(\De)}^{4} > K_{8}g^{-(1+\ep_{1})} \}}
\]
\[
\times\prod_{\De \ {\rm type \ II.1}} \lp
\lp\prod_{{i_0(\De)\le\beta\le i_0(\De)+D}\atop{{\rm pr}_\beta(\De)\in Y}}
\bbbone_{\{\int_{\tilde \De} \ph_{\beta}{4} 
\le K_{8}g^{-(1+\ep_{1})} \}}\rp
.\bbbone_{\{\int_{\tilde \De} \ph_{I_{0}(\De)}^{4}
\le K_{9}g^{-(1+\ep_{1})} \}}
\rp
\]
\[
\times\prod_{\De \ {\rm type \ II.2.1}} \lp
\lp\prod_{{i_0(\De)\le\beta\le i_0(\De)+D}\atop{{\rm pr}_\beta(\De)\in Y}}
\bbbone_{\{\int_{\tilde \De} \ph_{\beta}{4} 
\le K_{8}g^{-(1+\ep_{1})} \}}\rp
.\bbbone_{\{\int_{\tilde \De} \ph_{I_{0}(\De)}^{4}
> K_{9}g^{-(1+\ep_{1})} \}}
\rp
\]
\[
\times\prod_{\De \ {\rm type \ II.2.2a}}\Biggl(
\lp\prod_{{i_0(\De)\le\beta\le i_0(\De)+D}\atop{{\rm pr}_\beta(\De)\in Y}}
\bbbone_{\{\int_{\tilde \De} \ph_{\beta}{4} 
\le K_{8}g^{-(1+\ep_{1})} \}}\rp
.\bbbone_{\{\int_{\tilde \De} \ph_{I_{0}(\De)}^{4}
> K_{9}g^{-(1+\ep_{1})} \}}
\]
\[
.\bbbone_{\bigl\{\max\limits_{1
\le \nu\le\mu(\De)} \lp a_{\nu}(\De)^{16/3}
\int_{\tilde \De} \ph_{I_{\nu}(\De)}^{4}\rp \le K_{10}g^{-(1+\ep_{1})}
\bigr\}} 
\Biggr)
\]
\[
\times\prod_{\De \ {\rm type \ II.2.2b.}\nu}\Biggl(
\lp\prod_{{i_0(\De)\le\beta\le i_0(\De)+D}\atop{{\rm pr}_\beta(\De)\in Y}}
\bbbone_{\{\int_{\tilde \De} \ph_{\beta}{4} 
\le K_{8}g^{-(1+\ep_{1})} \}}\rp
.\bbbone_{\{\int_{\tilde \De} \ph_{I_{0}(\De)}^{4}
> K_{9}g^{-(1+\ep_{1})} \}}
\]
\be
.\bbbone_{\bigl\{ a_{\nu}(\De)^{16/3}\int_{\tilde \De}
 \ph_{I_{\nu}(\De)}^{4} > K_{10}g^{-(1+\ep_{1})} \bigr\}} \Biggr)
\ \ .
\label{contsmf}
\ee

For each $\De$ of type I, we use the majoration
$1 \le K_{8}^{-1}g^{(1+\ep_{1})} \int_{\tilde\De}\ph_{\beta(\De)}^{4}$
which is valid in the domain of integration defined by the characteristic 
functions.
For  $\De$ of type II.1, let $\hat\De$ be the box in
$\cD_{i(\De)+1}$ containing $\De$ (i.e.
the box just below).
Since we supposed that the cube $\De$ is non isolated,
we have $\hat{\De}\in\Ga$.
Therefore by definition of $\ch_\Ga$, we have the constraint 
the constraint
\be
\half g^{-(1+\ep_{1})} \le \int_{\hat\De}
\ph_{{\hat I}(\De)}^{4}
\ \ ,
\ee
where $\ph_{{\hat I}(\De)}\eqdef
\sum_{i_0(\De)<i\le j_0(\De)}\ph_{i}$.
But for any $x\in \hat \De$, we can use Lemma \ref{avrSobo} to write
\be
\ph_{{\hat I}(\De)}(x)  =
\ph_{{\hat I}(\De)}(x) - {\rm Fluct}(\hat\De,
\ph_{{\hat I}(\De)}) (x) + {\rm Fluct}(\hat\De,
\ph_{{\hat I}(\De)}) (x) 
\ \ ,
\ee
where 
\be
{\rm Fluct}(\hat\De,\ph_{{\hat I}(\De)}) (x) \eqdef 
\int_{\tilde\De^{2}\times [0,1]^{2}}\ d {\bf w} 
f_{\tilde\De, \hat\De}^{\mu,\nu} (x, {\bf w} ) 
\partial_{\mu}\partial_{\nu}\ph_{\hat{I}(\De)} \bigl(\cX_{\tilde\De, \hat\De} 
(x, {\bf w} ) \bigr)
\ee
satisfies, since $i(\hat\De)-i(\tilde\De)\le 2$,
\be
\big| \ph(x) - {\rm Fluct}(\hat\De,
\ph_{{\hat I}(\De)}) (x) \big|
\le K_{7} \biggl({1\over |\tilde \De |} \int_{\tilde\De} dy 
\ \ph_{{\hat I}(\De)}
(y)^{4} \biggr)^{1/4} M^{2}
\ \ .
\ee

As a result we obtain from the elementary inequality 
$(a+b)^{4}\le 8(a^{4}+b^{4})$,

\be
\half g^{-(1+\ep_{1})} \le \int_{\hat\De} dx\ 8 \biggl[ \biggl(  
\ph_{{\hat I}(\De)}(x) 
- {\rm Fluct}(\hat\De,
\ph_{{\hat I}(\De)}) (x)
\biggr)^{4} + \biggl( {\rm Fluct}(\hat\De,
\ph_{{\hat I}(\De)}) (x) 
\biggr)^{4}\biggr]
\ee
\be
\le 8 \int_{\hat\De} dx  \biggl({\rm Fluct}(\hat\De,
\ph_{{\hat I}(\De)}) (x)
\biggr)^{4} + 8{|\hat\De |\over |\tilde \De |} K_{7}^{4} M^{8}
\int_{\tilde\De} dy
\ \ph_{{\hat I}(\De)} (y)^{4}
\ee
\be
\le  8 \int_{\tilde\De} dx  \biggl({\rm Fluct}(\hat\De,
\ph_{{\hat I}(\De)}) (x)
\biggr)^{4} + 8 M^{16} K_{7}^{4}\int_{\tilde\De} dy 
\Bigl(\ph_{I_{0}(\De)} (y)  - \ph_{i(\De)}(y) \Bigr)^{4}
\ee
\be
\le  8 \int_{\tilde\De} dx  
\biggl({\rm Fluct}(\hat\De,
\ph_{{\hat I}(\De)}) (x)
\biggr)^{4} + 64 M^{16} K_{7}^{4}\biggl(\int_{\tilde\De} 
\ph_{I_{0}(\De)}^{4}   + \int_{\tilde\De}
\ph_{i(\De)}^{4} \biggr) \ \ .
\ee
But for type II.1 cubes, we have also the constraints
$\int_{\tilde \De} \ph_{i(\De)}^{4} \le K_{8} g^{-(1+\ep_{1})}$
and $\int_{\tilde \De} \ph_{I_{0}(\De)}^{4} \le K_{9} g^{-(1+\ep_{1})}$,
so as a result 
\be
\int_{\tilde \De}dx  \biggl({\rm Fluct}(\hat\De,
\ph_{{\hat I}(\De)}) (x)
\biggr)^{4} \ge K_{11} g^{-(1+\ep_{1})}
\ \ ,
\ee
with $K_{11}=(\frac{1}{8}[(1/2)-64 M^{16}K_{7}^{4}(K_{8}+K_{9})]$.
We suppose we have chosen $K_{8}$ and $K_{9}$ such that $K_{11}>0$. Complete
fixing of the constants is postponed to the end of the proof.
In the integral (\ref{contsmf}) we can now 
readily introduce the majoration 
\be
1 \le K_{11}^{-1} g^{1+\ep_{1} } \int_{\tilde \De} dx 
\biggl({\rm Fluct}(\hat\De,
\ph_{{\hat I}(\De)}) (x)\biggr)^{4}
\ee
that is valid in the domain of integration.

Consider now the case where $\De$ is of type II.2.1.
As we did for type II.1 cubes and with the same notations
we can write
\be
\half g^{-(1+\ep_{1}) }
\le
\int_{\hat\De} dx  \biggl({\rm Fluct}(\hat\De,
\ph_{{\hat I}(\De)}) (x)
\biggr)^{4} + 8{|\hat\De |\over |\tilde \De |} K_{7}^{4} M^{8}
\biggl(\int_{\tilde\De} dy
\ph_{{\hat I}(\De)} (y)^{4}
\biggr)
\ \ .
\ee
But by H\"older's inequality, for any $y\in\til\De$ we have
\be 
\lp
\ph_{{\hat I}(\De)} (y)\rp^4
\le
\lp
\sum_{i_0(\De)<i\le j_0(\De)}1
\rp^3
\lp
\sum_{i_0(\De)<i\le j_0(\De)}\ph_i(y)^4
\rp
\ee
and thus
\be
\int_{\tilde\De} dy
\ \ph_{{\hat I}(\De)} (y)^{4}
\le
D^3
\sum_{i_0(\De)<i\le j_0(\De)}
\int_{\tilde\De} dy
\ \ph_i(y)^4
\le
D^4 K_8 g^{-(1+\ep_1)}
\ \ .
\ee
Denote now
\be
K_{12}\eqdef
\frac{1}{8}
\lp\frac{1}{2}-8M^{16}K_7^4 D^4K_8\rp
\ \ .
\ee
If we assume we have fixed the constants such that $K_{12}>0$,
then we can introduce in the functional integral
the majoration
\be
1 \le K_{12}^{-1} g^{1+\ep_{1} } \int_{\tilde \De} dx 
\biggl({\rm Fluct}(\hat\De,
\ph_{{\hat I}(\De)}) (x)\biggr)^{4}
\ \ .
\ee

Let now $\De$ be of type II.2.2a, and $a$ be any integer such that 
$0\le a \le k$, where $k$ is the length of the graph $\Gg=(l_{1},...,l_{k})$.
Now consider $B_{a}(\De)$ the block containing $\De$ in the partition
$\Pi_{\Ga}((l_{1},...,l_{a}))$. Since glued cubes must belong to the same 
block, on $\tilde\De \subset \De$, the field
$\ph_{B_{a}(\De)} = 
\sum\limits_{i\ge i(\De)\atop pr_{i}(\De) \in B_{a}}\ph_{i}$, 
that appears in the expression (\ref{rewint})
of the interaction $\cI$ must have 
the form:
\be
\ph_{B_{a}(\De)} = \ph_{I_{0}} + \ph_{I_{\nu_{1}}}+ ... + \ph_{I_{\nu_{r}}}
\ \ ,
\ee
with $1\le \nu_{1}<...<\nu_{r}\le \mu(\De)$, and $r\ge 0$.
Now for any $x\in \tilde \De$ we have by H\"older's inequality
\be
\big| \ph_{I_{0}}(x)\big| = \big| 1\cdot \ph_{B_{a}(\De)}(x) +
a_{\nu_{1}}^{-1} (-a_{\nu_{1}} \ph_{I_{\nu_{1}}}(x)) + ... + 
a_{\nu_{r}}^{-1} (-a_{\nu_{r}} \ph_{I_{\nu_{r}}}(x))|
\ee
\be
\le \bigl[1 + a_{\nu_{1}}^{-4/3}  + ... +    a_{\nu_{r}}^{-4/3}\bigr]^{3/4}
\bigl[ \ph^{4}_{B_{a}(\De)}(x) + a_{\nu_{1}}^{4}
\ph^{4}_{I_{\nu_{1}}}(x)  + ... +   a_{\nu_{r}}^{4}
\ph^{4}_{I_{\nu_{r}}}(x)\bigr]^{1/4}
\ \ .
\ee
Therefore when we integrate over $x$ we get:
\[
\int_{\tilde\De}\ph^{4}_{I_{0}}(x) \le
\bigl(1 + a_{\nu_{1}}^{-4/3}  + ... +    a_{\nu_{r}}^{-4/3}\bigr)^{3}
\]
\be
\times\bigl[ \int_{\tilde\De}\ph^{4}_{B_{a}(\De)}(x) + a_{\nu_{1}}^{-4/3}
a_{\nu_{1}}^{16/3} \int_{\tilde\De}\ph^{4}_{I_{\nu_{1}}}(x) + ... 
+   a_{\nu_{r}}^{-4/3} a_{\nu_{r}}^{16/3}
\int_{\tilde\De}\ph^{4}_{I_{\nu_{r}}}(x)\bigr]
\ \ ,
\ee
and thus 
\[
\int_{\tilde\De}\ph^{4}_{I_{0}}(x) \le 
\bigl(1 + a_{\nu_{1}}^{-4/3}  + ... +    a_{\nu_{r}}^{-4/3}\bigr)^{3}
\]
\be
\times\bigl[ \int_{\tilde\De}\ph^{4}_{B_{a}(\De)}(x) + \bigl(a_{\nu_{1}}^{-4/3}
+ ... +   a_{\nu_{r}}^{-4/3} \bigr) \max\limits_{1\le \nu\le\mu(\De)}
\lp a_{\nu}^{16/3} \int_{\tilde\De}\ph^{4}_{I_{\nu}}(x)\rp  \bigr]
\ \ .
\label{boundnu}
\ee
Now, by definition of the $a_{\nu}'s$, 
\be
\bigl(a_{\nu_{1}}^{-4/3}
+ ... +   a_{\nu_{r}}^{-4/3} \bigr) \le \sum_{\nu =1}^{\mu}
M^{-\frac{1}{10}(i_{\nu}-i(\tilde\De))} \le \sum_{i=0}^{+\infty}
M^{-\frac{i}{10}}
\ee
\be
=(1-M^{-\frac{1}{10}})^{-1} \eqdef S
\ \ .
\ee
Finally if we arrange for having
\be
K_{13}\eqdef{K_{9}\over (1+s)^{3}}-K_{10}S>0
\ \ ,
\ee
we conclude from (\ref{boundnu}), and the fact that in the present case 
$\int_{\tilde\De}\ph_{I_{0}}(x) > K_{9} g^{-(1+\ep_{1})}$ and 
$\max_{1\le \nu \le \mu (\De)} \lp a_{\nu}(\De)^{16/3}\int_{\tilde\De}
\ph^{4}_{I_{\nu}}\rp\le K_{10}g^{-(1+\ep_{1})}$
hold in the domain of integration, that so does
$\int_{\tilde\De} \ph^{4}_{B_{a}(\De)} \ge g^{-(1+\ep_{1})}K_{13}$.
This has to be true for any $a$, $0\le a\le k$. As a result the part
of the interaction that is integrated in $\tilde \De$ satisfies
\be
\cI_{\tilde\De} = g \sum_{a=0}^{k} (h_{a}-h_{a+1})
\sum_{B\in \Pi_{\Ga}((h_{1},...,h_{a}))}
\int_{\tilde\De} \bigl(\ph^{B}(x)\bigr)^{4} dx
\ee
\be
\ge g\sum_{a=0}^{k} (h_{a}-h_{a+1})
\int_{\tilde\De} \bigl(\ph^{B_{a}(\De)}(x)\bigr)^{4} dx
\ge  g\biggl[\sum_{a=0}^{k} (h_{a}-h_{a+1})\biggr] g^{-(1+\ep_{1})}K_{13}
= K_{13} g^{-\ep_{1}}
\ \ .
\ee
Therefore we can extract from the interaction a factor 
$e^{-\cI_{\tilde\De}} \le e^{-K_{13}g^{-\ep_{1}}}$. Indeed the $\tilde\De$
are disjoint, for this is what summital cubes are all about.

It remains to consider the case of $\De$ of type II.2.2b.$\nu$.
We have in the 
domain of integration on the fields the inequality:
\be
\int_{\tilde\De}\ph^{4}_{I_{\nu}(\De)} > K_{10} g^{-(1+\ep_{1})}\cdot
M^{-\frac{1}{10} [i_{\nu}(\De)-i(\De)]}
\ \ .
\label{flucnu}
\ee
But if $\bar\De$ is the cube
${\rm pr}_{i_{\nu}(\De)}(\De)$ containing $\De$ at scale
$i_{\nu}(\De)$, we have since $\bar\De$ is the top cube of a large field block
and $i(\bar\De)>0$, that $\bar\De$ is necessarily a small field cube. Hence
we also have the condition $\int_{\bar\De} \ph^{4}_{I_{\nu}(\De)}\le 
g^{-(1+\ep_{1})}$. This and (\ref{flucnu}) entail the existence of a large 
fluctuation of $\ph_{I_{\nu}(\De)}$ at some intermediate scale between 
$i(\tilde\De)$ and $i(\bar\De)$. Define $\underline\De$ as the unique
cube containing $\De$ at scale $i(\tilde\De)+E[
\frac{1}{5}(i(\bar\De)-i(\tilde\De))]$,
where $E(x)$ means the integral part of $x$.

For any $x$ in $\tilde\De$, we can use Lemma \ref{avrSobo}
with $\De_{1}=\De_{2}=
\underline\De$, to write the decomposition
\be
\ph_{I_{\nu}(\De)}(x)  = \biggl(\ph_{I_{\nu}(\De)}(x) - 
{\rm Fluct}(\underline\De,
\ph_{I_{\nu}(\De)}) (x)\biggr) + {\rm Fluct}(\underline\De,
\ph_{I_{\nu}(\De)}) (x)
\ \ ,
\ee
where following the notations of the lemma
\be
{\rm Fluct}(\underline\De,\ph_{I_{\nu}(\De)}) (x) = 
\int_{\und\De^{2}\times [0,1]^{2}} d {\bf w} 
f_{\und\De, \underline\De}^{\mu,\nu} (x, {\bf w} ) 
\partial_{\mu}\partial_{\nu}\ph_{I_{\nu}(\De)} \bigl(\cX_{\und\De, 
\underline\De} 
(x, {\bf w} ) \bigr)
\ \ .
\ee
The inequality (\ref{avrbd}) provides us with the bound
\be
\bigg|\ph_{I_{\nu}(\De)}(x) - {\rm Fluct}(\underline\De,\ph_{I_{\nu}(\De)}) (x)
\bigg| \le K_{7} \biggl({1\over |\underline\De |} \int_{\underline\De}
dy\ \ph^{4}_{I_{\nu}}(y)\biggr)^{1/4}
\ \ ,
\ee
since here $i(\De_{2}) -i(\De_{1})=0$.

We can now perform a similar reasoning to that for type II.1 cubes. We readily
obtain
\be
\int_{\tilde\De} \ph^{4}_{I_{\nu}(\De)}  
\le \int_{\tilde\De} dx \biggl( \biggl[ 
\ph_{I_{\nu}(\De)}(x) - {\rm Fluct}(\underline\De,\ph_{I_{\nu}(\De)}) (x)
\biggr]+{\rm Fluct}(\underline\De,\ph_{I_{\nu}(\De)}) (x)\biggr)^{4} 
\ee
\be
\le 8 \int_{\tilde\De} dx \biggl( 
\ph_{I_{\nu}(\De)}(x) - {\rm Fluct}(\underline\De,\ph_{I_{\nu}(\De)}) (x)
\biggr)^{4}+8 \int_{\tilde\De} dx
\biggl({\rm Fluct}(\underline\De,\ph_{I_{\nu}(\De)}) (x)\biggr)^{4}
\ee
\be
\le 8|\tilde\De| \cdot K_{7}^{4}\cdot {1 \over |\underline\De |} 
\int_{\underline\De} dy\ \ph^{4}_{I_{\nu}}(y) + 8 \int_{\tilde\De} dx
\biggl({\rm Fluct}(\underline\De,\ph_{I_{\nu}(\De)}) (x)\biggr)^{4}
\ee
\be
\le 8 \cdot M^{-4(i(\underline\De) -i(\tilde\De))} K_{7}^{4} \int_{\bar\De}
dy\ \ph_{I_{\nu}}^{4}(y) + 8 \int_{\tilde\De}dx 
 \biggl({\rm Fluct}(\underline\De,\ph_{I_{\nu}(\De)}) (x)\biggr)^{4}
\ \ .
\label{flnucomp}
\ee
Indeed $i(\underline\De)\le i(\tilde\De) + \frac{1}{5}
(i(\bar\De)-i(\tilde\De))
\le i(\bar\De)$ entails $\underline\De \subset \bar \De$, from which we deduced
the last inequality. Now from
$\int_{\til\De}\ph_{I_{\nu}(\De)}^4\le g^{-(1+\ep_1)}$,
(\ref{flucnu}) and (\ref{flnucomp}), we infer that
\[
\int_{\tilde\De}dx 
 \biggl({\rm Fluct}(\underline\De,\ph_{I_{\nu}(\De)}) (x)\biggr)^{4}
\]
\be
\ge {g^{-(1+\ep_{1})}\over 8} \biggl[K_{10} 
M^{-\frac{1}{10}(i(\bar\De)-i(\tilde\De))} - 8 K_{7}^{4} 
M^{-4(i(\underline\De)-i(\tilde\De))} \biggr]
\ee
\be
= {g^{-(1+\ep_{1})}\over 8}M^{-\frac{1}{10}(i(\bar\De)-i(\tilde\De))} 
\biggl[K_{10}
 - 8 K_{7}^{4} 
M^{-4(i(\underline\De)-i(\tilde\De))+\frac{1}{10}
(i(\bar\De)-i(\tilde\De))} \biggr]\ \ .
\ee
But $i(\underline\De)-i(\tilde\De)=E[{1\over 5}(i(\bar\De)-i(\tilde\De))]\ge
\frac{1}{5}(i(\bar\De)-i(\tilde\De))-1$.
Besides, since $\De$ is a type II.2.2 cube, we have $j_0(\De)\ge D+1+i(\De)$,
so that
$i(\bar\De)-i(\tilde\De))\ge
i_1(\De)-i(\tilde\De))\ge D+2$.
Therefore we have
\be
M^{-4(i(\underline\De)-i(\tilde\De))+ \frac{1}{10}
(i(\bar\De)-i(\tilde\De))} \le 
M^{4-\frac{7}{10}(i(\bar\De)-i(\tilde\De))}\le M^{4-\frac{7}{10}(D+2)} 
\ \ .
\ee
As a result
\be
\int_{\tilde\De}dx 
 \biggl({\rm Fluct}(\underline\De,\ph_{I_{\nu}(\De)}) (x)\biggr)^{4}
\ge g^{-(1+\ep_{1})}M^{-\frac{1}{10}(i(\bar\De)-i(\tilde\De))} K_{14}
\ \ ,
\ee
where $K_{14}\eqdef{1\over 8}\bigl[K_{10}-8K_{7}^{4}M^{4
-\frac{7}{10}(D+2)}\bigr]$.
Again we shall see that the constants can be fixed so that $K_{14}>0$. We
then
introduce in the integral the majoration
\be
1 \le g^{1+\ep_{1}}M^{(\frac{1}{10}(i(\bar\De)-i(\tilde\De))} K^{-1}_{14}
\cdot \int_{\tilde\De}dx 
 \biggl({\rm Fluct}(\underline\De,\ph_{I_{\nu}(\De)}) (x)\biggr)^{4}
\ee
that is valid in the domain of integration.

The discussion of all possible cases is now complete and we can write a 
bound on a generic term ${\GC}$ of (\ref{contsmf}). We have:
\[
|{\GC}|\le
\int d\mu_{C_{[T_{{\Gg},\Ga}({\bf h})]}} (\ph) 
\prod_{\De\ {\rm isolated}} g^{\ep_5}\times\prod_{\De \ {\rm type \ I}}
\biggl(K_{8}^{-1} g^{1+\ep_{1}}  \int_{\tilde\De}
\ph^{4}_{\beta(\De)}  \biggr) 
\]
\[
\times
\prod_{\De \ {\rm type \ II.1}} \biggl( K^{-1}_{11}g^{1+\ep_{1}} 
\int_{\tilde\De} dx 
 \biggl({\rm Fluct}(\hat\De,\ph_{\hat I(\De)}) (x)\biggr)^{4}\biggr)
\]
\[
\times\prod_{\De \ {\rm type \ II.2.1}} \biggl( K^{-1}_{12}g^{1+\ep_{1}} 
\int_{\tilde\De} dx 
 \biggl({\rm Fluct}(\hat\De,\ph_{\hat I(\De)}) (x)\biggr)^{4}\biggr)
\times
\prod_{\De \ {\rm type \ II.2.2a}} \biggl( e^{-K_{13}g^{-\ep_{1}}}  \biggr)
\]
\be
\times\prod_{\De \ {\rm type \ II.2.2b.}\nu}\biggl(  K^{-1}_{14}
g^{1+\ep_{1}}M^{(1/10)(i(\bar\De)-i(\tilde\De))}
\cdot \int_{\tilde\De}dx 
 \biggl({\rm Fluct}(\underline\De,\ph_{I_{\nu}(\De)}) (x)\biggr)^{4}\biggr)
\ \ .
\ee
In the right hand side we can get the spatial integrations on $x$ in
$\tilde \De$, as well as the integrals over the parameters
$\bf w$, out from the functional integral, that is performed first.
The integrand for this Gaussian
functional integral is now a polynomial expression in
the fields $\ph_{i}$ and their double gradients $\partial_{\mu}\partial_{\nu}
\ph_{i}$. We bound this functional integral with Lemma \ref{Gauss}.
Once integrated, every field $\ph_{i}$ gives a factor $K_{4}M^{-i}$, and 
every double gradient $\partial_{\mu}\partial_{\nu}\ph_{i}$ gives  
a much better factor $K_{4}M^{-3i}$. Besides we get a product of local 
factorials $\prod_{\De\in Y} (n(\De)!)^{1/2}$ to deal with. We recall that
$n(\De)$ denotes the number of fields $\ph_{i}(x)$ as well as double
gradients $\partial_{\mu}\partial_{\nu}\ph_{i}(x)$ located in $\De$, that is
with $\De(x,i)=\De$, that are integrated with respect to 
$d\mu_{C_{[T_{{\Gg},\Ga}({\bf h})]}}$.

One has also to use the $L^{1}$ bound (\ref{L1bd})
on the densities defining the
fluctuation fields. Remark that after the bound (\ref{Gaussbd})
on the Gaussian 
integration, nothing depends on the parameters $\bf w$ except the 
corresponding densities $f_{\De_{1},\De_{2}}^{\mu, \nu}(x, {\bf w})$.
Therefore there is no problem of factorization in using these $L^{1}$
bounds. Finally remark that spatial integration in $\tilde\De$ costs a
volume factor $M^{4i(\tilde\De)}$. Collecting all these factors we obtain
\[
|{\GC}|\le
\prod_{\De\ {\rm isolated}} g^{\ep_5}\times\prod_{\De \ {\rm type \ I}}
\bigl(K_{8}^{-1} K_4^4 M^{-4(\beta(\De)-i(\til\De))} g^{1+\ep_{1}}  \bigr)
\times\prod_{\De \ {\rm type \ II.2.2a}} 
\bigl( e^{-K_{13}g^{-\ep_{1}}}  \bigr)
\]
\[
\times
\prod_{\De \ {\rm type \ II.1}} \biggl( K^{-1}_{11}
K_4^4 g^{1+\ep_{1}} 
M^{4i(\tilde\De)} \sum_{\al_{1},...,\al_{4}=i_{0}(\De)+1}^{j_{0}(\De)}
\bigl(M^{2i(\hat\De)-3\al_{1}}\bigr)...
\bigl(M^{2i(\hat\De)-3\al_{4}}\bigr)\biggr)
\]
\[
\times
\prod_{\De \ {\rm type \ II.2.1}} \biggl( K^{-1}_{12}
K_4^4 g^{1+\ep_{1}}
M^{4i(\tilde\De)} \sum_{\al_{1},...,\al_{4}=i_{0}(\De)+1}^{j_{0}(\De)}
\bigl(M^{2i(\hat\De)-3\al_{1}}\bigr)...
\bigl(M^{2i(\hat\De)-3\al_{4}}\bigr)\biggr)
\]
\[
\times\prod_{\De \ {\rm type \ II.2.2b.}\nu}\biggl(  K^{-1}_{14}
K_4^4 g^{1+\ep_{1}}M^{(1/10)(i(\bar\De)-i(\tilde\De))}
\cdot M^{4i(\tilde\De)}
\]
\[  
.\sum_{\al_{1},...,\al_{4}=i_{\nu}(\De)+1}^{j_{\nu}(\De)}
\bigl(M^{2i(\underline\De)-3\al_{1}}\bigr)...
\bigl(M^{2i(\underline\De)-3\al_{4}}\bigr)\biggr)
\]
\be
\times\prod_{\De\in Y} (n(\De)!)^{1/2}
\ \ .
\label{bill}
\ee
The indices $\al$ are the slice labels of the individual fields that have 
been integrated and that come in bunches of four. We used the fact that
$\hat I  (\De ) =\{i_{0}(\De)+1 ,...,j_{0}(\De)\}$ and $I_{\nu}(\De)=
\{i_{\nu}(\De),...,j_{\nu}(\De)\}$. Needless to say that in (\ref{bill}) the 
numbers $n(\De)$ depend on the previous choice of the $\al$'s. Note that for a
type II.1  cube,
since $i(\hat\De) -i(\tilde\De) \le 2$, we can bound the factor
belonging to $\De$  in (\ref{bill}) by
\be
K_{11}^{-1} \cdot K_4^4\cdot M^{16}\cdot 
\sum_{\al_{1},...,\al_{4}=i_{0}(\De)+1}^{j_{0}(\De)}
\bigl(M^{-3(\al_{1}-i(\tilde\De))}\bigr)...
\bigl(M^{-3(\al_{4}-i(\tilde\De))}\bigr)
\ \ .
\label{collect1}
\ee
The analog remark applies to
type II.2.1 cubes. For a type II.2.2b.$\nu$ cube $\De$,
in the corresponding factor in (\ref{bill}) we can isolate the contribution of 
one of the four fields located say at slice $\al \in \{ i_{\nu}(\De),...,
j_{\nu}(\De)\}$.   It gives
\be
M^{(\frac{1}{40}
(i_{\nu}(\De)-i(\tilde\De))+ i(\tilde\De) +2 i(\underline\De)-3\al}
\le M^{-3(\al -i(\tilde\De)) + \half (i_{\nu}(\De)-i(\tilde\De))}
\ \ ,
\label{collect2}
\ee
since $i(\underline\De)\le i(\tilde\De)+\frac{1}{5}(i_{\nu}(\De)-i(\tilde\De))$
by definition.
 
Now remark that $\al \ge i_{\nu}(\De)$. As a result we bound the right hand 
side of (\ref{collect2}) by $M^{-(5/2)(\al-i(\De))}$. Therefore, the factor
corresponding to $\De$ is bounded by 
\be
K_{13}^{-1}K_4^4g^{-(1+\ep_1)}
\sum_{\al_{1},...,\al_{4}=i_{\nu}(\De)}^{j_{\nu}(\De)}
M^{-\frac{5}{2}(\al_{1}-i(\De))}...M^{-\frac{5}{2}(\al_{4}-i(\De))}
\ \ .
\label{collect3} 
\ee
To get rid of the local factorials, we use Lemma \ref{displ}.
Indeed for every pair of boxes $\De_{1},\De_{2}$ in $\cE\eqdef\cD^{(N)}$,
define the function
\be
G(\De_{1},\De_{2})\eqdef
\left\{
\begin{array}{ll}
(1-M^{-\half}) M^{-\frac{9}{2}(i(\De_{1})-i(\De_{2}))} &
{\rm if}
\ \ \De_{1}\supset \De_{2}\\
0 & {else.}
\end{array}
\right.
\label{defGdis}
\ee
For any $\De_{1}\in \cD^{(N)}$
we have, 
\be
\sum_{\De_{2}\in \cD^{(N)}} G(\De_{1},\De_{2}) = (1-M^{\half}) \sum_{0\le i\le
i(\De_{1})} M^{-\frac{9}{2}(i-i(\De_{1}))} \#(\{\De_{2}|i(\De_{2})=i,
\De_{2}\subset\De_{1}\})
\ee
\be
= (1-M^\half) \sum_{0\le i\le i(\De_{1})} M^{-\half(i-i(\De_{1}))} \le 1
\ \ .
\ee
Now we take the contracted fields to form the set ${\cO}$ in the context
of Lemma \ref{displ}.
We define the map $S_1$ as the one that to a field associates
its location in phase space, i.e. the cube of the corresponding scale $\al$
under the concerned $\tilde\De$. We define now the ``location of the field 
after displacement'' or its image by $S_{2}$ as the corresponding cube $\De$.

Note that $\#({\cO})$ is bounded by $4\# (Y_{S})$.
Besides, $\prod_{\De\in \cE} n_{1}(\De)! = \prod_{\De\in Y}
n(\De)!$, where we used the notations of Lemma \ref{displ}.
Now remark that
\be
\prod_{\De\in \cE} n_{2}(\De)! =
\prod\limits_{
\De {\rm\ type\ I \ or\ type \ II.1}
\atop
{{\rm or \  type \  II.2.1}\atop{{\rm or \  type \  II.2.2b}.\nu}}
}
4! \le 24 ^{\# (Y_{S})}
\ \ .
\ee
Indeed a $\De$ produces exactly 4 contracted fields if of type
I and 4 double gradients if of type II.1,
II.2.1 or II.2.2b.$\nu$. Lemma \ref{displ}
entails 
\be
\prod_{\De\in Y} (n(\De)!)^{1/2} \le e^{2\#(Y_{S})} \cdot 24^{\half\#(Y_{S})}
\cdot \prod_{{\rm field}} \biggl( (1-M^\half)^{-1/2}
M^{\frac{9}{4}(\al-i(\De))}\biggr)
\ \ .
\label{dsploc}
\ee
Remark that for a double gradient field, considering (\ref{collect1}),
(\ref{collect2}) and (\ref{collect3}), we see that we can extract at least
$M^{-\frac{9}{4}(\al-i(\De))}$ to beat the corresponding factor in
(\ref{dsploc}).
However, for a field at scale $\beta(\De)$ coming from a type I cube
$\De$, the available factor $M^{-(\beta(\De)-i(\til\De))}$
in (\ref{bill}) is not sufficient.
We have a loss of $M^{\frac{5}{4}(\beta(\De)-i(\til\De))}\le
M^{\frac{5}{4}D}$, per such field, to account for.

As a result, (\ref{bill}) can be improved to yield
\[
|{\GC}|\le
\prod_{\De\ {\rm isolated}} g^{\ep_5}
\times\prod_{\De \ {\rm type \ I}}
\bigl(K_{8}^{-1} K_4^4g^{1+\ep_{1}} M^{5D} \bigr)
\times\prod_{\De \ {\rm type \ II.2.2a}} 
\bigl( e^{-K_{13}g^{-\ep_{1}}}  \bigr)
\]
\[
\times\prod_{\De \ {\rm type \ II.1}} \biggl( K^{-1}_{11}K_4^4g^{1+\ep_{1}} 
M^{16}. \sum_{\al_{1},...,\al_{4}=i_{0}(\De)+1}^{j_{0}(\De)}
M^{-\frac{3}{4}(\al_{1}-i(\De))}...
M^{-\frac{3}{4}(\al_{4}-i(\De))}\biggr)
\]
\[
\prod_{\De \ {\rm type \ II.2.1}} \biggl( K^{-1}_{12}
K_4^4 g^{1+\ep_{1}} 
M^{16}. \sum_{\al_{1},...,\al_{4}=i_{0}(\De)+1}^{j_{0}(\De)}
M^{-\frac{3}{4}(\al_{1}-i(\De))}...
M^{-\frac{3}{4}(\al_{4}-i(\De))}\biggr)
\]
\[
\prod_{\De \ {\rm type \ II.2.2b.}\nu}\biggl(  K^{-1}_{14}
K_4^4 g^{1+\ep_{1}}.
\sum_{\al_{1},...,\al_{4}=i_{\nu}(\De)+1}^{j_{\nu}(\De)}
M^{-\frac{1}{4}(\al_{1}-i(\De)}...
M^{-\frac{1}{4}(\al_{4}-i(\De)}\bigr)\biggr) 
\]
\be
\times\biggr[e^{2}\sqrt{24}(1-M^{-\half})^{-2}\biggl]^{\# (Y_{S})}
\ \ .
\ee
To sum over the $\al$'s in the case of a type II.1
or type II.2.1 cube, we use the rough bound
\be
\sum_{\al=i_{0}(\De)+1}^{j_{0}(\De)} M^{-\frac{3}{4}(\al-i(\De))} \le 
\sum_{i=0}^{\infty} M^{-\frac{1}{4}i} = (1-M^{-\frac{1}{4}})^{-1}
\ \ .
\ee
In the case of a type II.2.2b.$\nu$ cube, we use 
\be
\sum_{\al=i_{\nu}(\De)}^{j_{\nu}(\De)} M^{-\frac{1}{4}(\al-i_{\nu}(\De))} \le 
 (1-M^{-\frac{1}{4}})^{-1}
\ \ ,
\ee
and we keep the factor $M^{-\frac{1}{4}(i_{\nu}(\De)-i(\De))}$ per field,
in order to sum over the possible cases. Finally:
\[
|{\GC}|\le
\prod_{\De \ {\rm isolated}\atop{\rm and\ summital}} 
g^{\ep_5}\times\prod_{\De \ {\rm type \ I}}
\bigl(K_{8}^{-1} K_4^4 g^{1+\ep_{1}} M^{5D} \bigr)
\times\prod_{\De \ {\rm type \ II.2.2a}} 
\bigl( e^{-K_{13}g^{-\ep_{1}}}  \bigr)
\]
\[
\times\prod_{\De \ {\rm type \ II.1}} \biggl( K^{-1}_{11}K_4^4 g^{1+\ep_{1}} 
M^{16} (1-M^{-\frac{1}{4}})^{-4}\biggr)
\]
\[
\times\prod_{\De \ {\rm type \ II.2.1}} \biggl( K^{-1}_{12}K_4^4 g^{1+\ep_{1}} 
M^{16} (1-M^{-\frac{1}{4}})^{-4}\biggr)
\]
\[
\times\prod_{\De \ {\rm type \ II.2.2b.}\nu}\biggl(  K^{-1}_{14}K_4^4
g^{1+\ep_{1}}(1-M^{-\frac{1}{4}})^{-4} M^{-(i_{\nu}(\De)-i(\De))} \biggr)
\]
\be
\times\biggr[e^{2}\sqrt{24}(1-M^{-\half})^{-2}\biggl]^{\# (Y_{S})}
\ \ .
\label{bill2}
\ee
Now to bound the left hand side of (\ref{smfact}),
one has to sum over the different 
possibilities for non isolated cubes. The sum over the first decomposition
(\ref{fstdec})
of the domain of integration costs a factor $2^{D+1}$ per summital
non isolated cube.
The dichotomy type II.1/type II.2
costs again a factor 2 per type II cube, and the dichotomy
between type II.2.2a and type II.2.2b an other 2 per type II.2.2 cube. Finally
the sum over $\nu$ for a type II.2.2b cube is done using the factor
$M^{-(i_{\nu}(\De)-i(\De))}$ in (\ref{bill2}) and its cost is bounded by 
$\sum_{i}M^{-i}=(1-M^{-1})^{-1}$. Therefore the left hand side of (\ref{bill2})
is bounded by $V(g)^{\# (Y_{S})}$ where
\[
V(g) \eqdef
e^{2}\sqrt{24}(1-M^{-1/2})^{-2} \cdot \max \Bigl[g^{\ep_5},
2^{D+1}K_{8}^{-1}K_4^4 M^{5D} g^{1+\ep_{1}},
\]
\[
2^{D+2} K^{-1}_{11}K_4^4 g^{1+\ep_{1}}M^{16} (1-M^{-\frac{1}{4}})^{-4},
2^{D+2} K^{-1}_{12}K_4^4 g^{1+\ep_{1}}M^{16} (1-M^{-\frac{1}{4}})^{-4},
\]
\be
2^{D+3} e^{-K_{13}g^{-\ep_{1}}},
2^{D+3} K^{-1}_{14} K_4^4
g^{1+\ep_{1}}(1-M^{-\frac{1}{4}})^{-4}(1-M^{-1})^{-1}\Bigr)
\ \ .
\ee
Since $\lim_{g\to 0}V(g)=0$ and since by Lemma (\ref{roof})
$\# (Y_{S})\le \#(Y) \le {M^{4}\over M^{4}-1}\# (Y_{S})$, if we take
$U(g) \eqdef
\max ( V(g) , V(g) ^{{M^{4}\over M^{4}-1}})$, we get the desired bound.

The only thing to check now to complete the proof is the consistency in fixing
$K_{8}, K_{9}, K_{10}$ and $D$.
We see that the constraints that we need to satisfy
are:
\be
K_{11}=\frac{1}{8}[\half-64 M^{16}K_{7}^{4}(K_{8}+K_{9})>0
\ \ ,
\label{k11}
\ee
\be
K_{12} = \frac{1}{8}[\half-8M^{16}K_{7}^{4}D^4K_8]>0
\ \ ,
\label{k12}
\ee
\be
K_{13} = \frac{K_9}{(1+S)^3}-K_{10}S>0
\ \ ,
\label{k13}
\ee
and
\be
K_{14}=\frac{1}{8}[K_{10}-8K_7^4 M^{4-\frac{7}{10}(D+2)}]>0
\ \ .
\label{k14}
\ee
The first step is to choose $K_9$ small enough such that
\be
\half-64 M^{16}K_{7}^{4}K_{9}>0
\ \ ,
\ee
then pick $K_{10}$ small enough so that (\ref{k13}) is verified.
One has then to take the integer constant $D$ large enough for (\ref{k14})
to be fulfilled. Finally, we choose $K_8$ small enough so that
(\ref{k11}) and (\ref{k12}) are satisfied. 
\Endproof

\subsection{Rearrangement of the expansion}

Before computing the functional derivatives, once we have chosen a large
field region $\Ga$, it is essential to group together the
contributions of the graphs $\Gg$, according to the following notions.

First we recall that a dressed large field block is a
connected component of $Y$ with respect to gluing by $\Ga$,
i.e. an element of $\Pi_\Ga(\emptyset)$.
Next we say that two links $l$ and $l'$ are \und{form-equivalent}
if they are equal 2-links or are both 4-links and satisfy
the following property.
We require that $l^\pi={l'}^\pi$, where we posed $\pi\eqdef
\Pi_\Ga(\emptyset)$ and used the notation of Section 2.2.3 for the reduced
link with respect to a partition $\pi$.
This means that for any dressed large field block $X\in\pi$,
\be
l^\pi(X)\eqdef\sum_{\De\in X} l(\De)=\sum_{\De\in X} l'(\De)
\eqdef {l'}^\pi(X)
\ \ .
\ee
We now define an equivalence relation among graphs in $\cG$.
We say that two ordered graphs $\Gg=(l_1,\ldots,l_k)$
and $\Gg'=(l'_1,\ldots,l'_k)$ with the same length are
{\em form-equivalent}, if for any $a$, $1\le a\le k$,
$l_a$ and $l'_a$ are form-equivalent.
An equivalence class of graphs is called a \und{form}.
The partition $\Pi_\Ga(\Gg)$ into 4-VI components of a graph
$\Gg$ depends only on the form of $\Gg$ we denote by $<\Gg>$.
Therefore if $\Gg$ and $\Gg'$ are form-equivalent, $\Gg$ is allowed
if and only if $\Gg'$ is allowed too.

Hereupon we will use the natural and consistent notations
$\Pi_\Ga(<\Gg>)\eqdef\Pi_\Ga(\Gg)$
and $T_{<\Gg>,\Ga}({\bf h})\eqdef T_{\Gg,\Ga}({\bf h})$.
The notion of allowedness naturally extends to forms.
It is easy to see that for two allowed form-equivalent graphs
$\Gg$ and $\Gg'$, for a fixed value of the fields and the $h$
parameters, the exponentiated interactions are the same.
We can therefore factorize the sum of functional integrals appearing
in (\ref{activ})
that correspond to the graphs in a certain allowed form $<\Gg>$.
Indeed consider ${\rm Ver}(l)$ the sum of vertices associated
to 4-links $l'$ that are form equivalent to a 4-link $l$.
We suppose that the reduced
link $l^\pi$ has support $\{X_1,\ldots,X_\al\}$, with $1\le\al\le 4$
and $X_1,\ldots,X_\al$ distinct dressed large field blocks.
We can write from (\ref{kernel}) and (\ref{activ}),
\be
{\rm Ver}(l)=\sum_{{l'\ {\rm 4-link\ in\ }Y}\atop{{l'}^\pi=l^\pi}}
\sum_{{(\De_1,\ldots,\De_4)\in Y^4}
\atop{l[\De_1,\ldots,\De_4]=l'}}
(-g)
\int_{\De_1\cap\ldots\cap\De_4}
dx
\ \ph_{i(\De_1)}(x)\ldots\ph_{i(\De_4)}(x)
\ \ .
\ee
We define for any $x\in\La$ and any $B\subset Y$,
\be
\ph^B(x)\eqdef
\sum_{i=0}^N
\bbbone_{\{\De(x,i)\in B\}}
\ph_i(x)
\ee
from the collection of fields $(\ph_Y)$. Thus we have
\be
{\rm Ver}(l)=
\sum_{{(\De_1,\ldots,\De_4)\in Y^4}
\atop{l[\De_1,\ldots,\De_4]^\pi=l^\pi}}
(-g)
\int_\La dx
\ \ph^{\{\De_1\}}(x)\ldots\ph^{\{\De_4\}}(x)
\ \ .
\ee
Here $l[\De_1,\ldots,\De_4]^\pi=l^\pi$ means that among
the possibly repeated cubes $(\De_1,\ldots,\De_4)$,
$l^\pi(X_\beta)$ of them must belong to $X_\beta$, for every $\beta$,
$1\le \beta\le \al$.
The sum over which subset of indices in $\{1,2,3,4\}$
label the cubes that belong to $X_\beta$, for each $\beta$
gives a combinatorial factor
\be
\frac{4!}{\prod_{\beta=1}^\al l^\pi(X_\beta)!}
\ \ .
\ee
The remaining sum over the positions of the boxes is done thanks to
the identity
\be
\sum_{\De\in X}
\ph^{\{\De\}}(x)=\ph^X(x)
\ \ ,
\ee
for every $x\in\La$.
Finally if we pose $m_\beta\eqdef l^\pi(X_\beta)$, for each $\beta$,
$1\le \beta\le \al$, in order that $m_1+\ldots+m_\al=4$, we have
\be
{\rm Ver}(l)=
(-g)
\frac{4!}{m_1!\ldots m_\al!}
\int_\La dx
\lp\ph^{X_1}(x)\rp^{m_1}\ldots\lp\ph^{X_\al}(x)\rp^{m_\al}
\ \ .
\label{verl}
\ee
The above cited factorization allows us to rewrite the activity (\ref{activ})
of a polymer $Y$ in the somewhat more glamorous form
\[
\cA(Y,(\De_s^{\rm ext},\ze_s^{\rm ext})_{s\in\cS})=
\sum_{\Ga}
\ \sum_{{<\Gg>\ {\rm allowed\ form}}\atop{Y\ {\rm 4-VI}}}
\ \int_{1>h_1>\dots>h_k>0}dh_1\ldots dh_k
\]
\[
\int d\mu_{C[T_{<\Gg>,\Ga}({\bf h})]}(\ph)
\prod_{{1\le j\le k}\atop{l_j\in\cL_2}}
\lp
\int_{\De^j_1}dx^j_1\int_{\De^j_2}dx^j_2
\ C(x_1^j,i(\De^j_1);x_2^j,i(\De^j_2))
\frac{\de}{\de\ph_{i(\De^j_1)}(x_1^j)}
\frac{\de}{\de\ph_{i(\De^j_2)}(x_2^j)}
\rp
\]
\[
\times\prod_{{1\le j\le k}\atop{l_j\in\cL_4}}
\lp
(-g)
\frac{4!}{m_{j,1}!\ldots m_{j,\al_j}!}
\int_\La dx
\lp\ph^{X_{j,1}}(x)\rp^{m_{j,1}}\ldots\lp\ph^{X_{j,\al_j}}(x)
\rp^{m_{j,\al_j}}
\rp
\]
\be
\times\prod_{s\in\cS}\ph_{i(\De_s^{\rm ext})}(\ze_s^{\rm ext})
\times
\ch_\Ga((\ph))
\times\exp
\lp
-g\sum_{a=0}^k(h_a-h_{a+1})
\sum_{B\in\Pi_\Ga((l_1,\ldots,l_a))}
\int_\La dx\lp\ph^B(x)\rp^4
\rp
\ \ .
\label{activ2}
\ee
Here $\Gg=(l_1,\ldots,l_k)$ is a choice of representative of the form
$<\Gg>$; $\De_1^j$ and $\De_2^j$ are a choice of cubes such that
$l[\De_1^j,\De_2^j]=l_j$, for any $j$, $1\le j\le k$ with
$l_j\in\cL_2$.
For the 4-links $l$ we have used the same notation as in (\ref{verl})
but with an additional index $j$ to distinguish between the vertices.

Before we proceed, we first pose to describe more thoroughly the
structure of dressed large field blocks.
The links $\{\De_1,\De_2\}$, with $\De_1$ just on top of $\De_2$,
naturally endow
$\cD^{(N)}$ with a tree structure.
Once we choose a large field region $\Ga$ in $Y$ we select
from the above mentioned links those for which $\De_2\in\Ga\subset Y$.
These selected links form a subforest of the previous tree,
whose connected components in $Y$ are the elements of $\pi$.
We therefore see that if $\De_1$ and $\De_2$ are two elements of
$X\in\pi$, then every cube $\De\in\cD^{(N)}$ satisfying $\De_1
\subset\De\subset\De_2$ must belong to $X$.
Indeed $\De$ is on the unique path in the forest of selected links
that connects $\De_1$ and $\De_2$.
Besides, there exists a largest cube ${\rm bg}(X)\in X$ that contain
all the cubes in $X$. Briefly,$X$ has to be a stack of cubes lying on
${\rm bg}(X)$, and at the top of the heap the cubes must be small field
ones except possibly if they are in the highest layer
$\cD_0^{(N)}$.

We now introduce an order relation among dressed large field blocks.
If $X$ and $X'$ are in $\pi$, we say that $X$ is above $X'$
if there exist $\De\in X$ and $\De'\in X'$ auch that $\De\subset\De'$.
Remark that this is equivalent to ${\rm bg}(X)\subset
{\rm bg}(X')$. This order relation is not total but, if $X_1,\ldots,X_\al$
are the blocks making the support of $l^\pi$, for some 4-link $l$
whose contribution ${\rm Ver}(l)$ does not vanish, then
$X_1,\ldots,X_\al$ must be pairwise comparable for this order
relation. Therefore in the formula (\ref{activ2}) we suppose
$X_{j,1},\ldots,X_{j,\al_j}$ have been labeled such that $X_{j,1}$
is above $X_{j,\beta}$ for every $\beta$, $2\le \beta\le\al_j$.
$X_{j,1}$ is the \und{top block} of the vertex.

If $B$ is a subset of $\cD^{(N)}$ and $\De\in\cD^{(N)}$, we define
the {\em shadow} of $B$ in $\De$ as
\be
{\rm sh}(\De,B)\eqdef
\cup_{{\til{\De}\in B\backslash\{\De\}}
\atop{\til{\De}\subset\De}}\til{\De}
\ \ .
\ee
Therefore if $\De\in B$, it is easy to see that $\De$ is the disjoint union
of the nonempty sets $\til{\De}\backslash{\rm sh}(\til{\De},B)$,
where $\til{\De}$ ranges through
\be
S(\De,B)\eqdef
\{\til{\De}\in B|\til{\De}\subset\De,\til{\De}\backslash{\rm sh}
(\til{\De},B)\neq\emptyset\}
\ \ .
\ee
We now expand in (\ref{activ2}), each vertex $l_j$ as
\be
(-g)
\frac{4!}{m_{j,1}!\ldots m_{j,\al_j}!}
\sum_{\til{\De}_j\in S({\rm bg}(X_{j,1}),X_{j,1})}
\int_{\til{\De}_j\backslash{\rm sh}(\til{\De}_j,X_{j,1})} dx
\lp\ph^{X_{j,1}}(x)\rp^{m_{j,1}}\ldots\lp\ph^{X_{j,\al_j}}(x)
\rp^{m_{j,\al_j}}
\ \ .
\label{decver}
\ee
We simply decomposed the domain of integration and used the fact that the
integrand vanishes out of ${\rm bg}(X_{j,1})$.
The fields like $\ph^{X_{j,1}}(x)$ are called the \und{high
momentum fields}, whereas $\ph^{X_{j,2}}(x), \ldots, \ph^{X_{j,\al_j}}(x)$
are called the \und{low momentum fields} of the considered
explicit vertex.

After making these decompositions, we finally
compute the functional derivatives.
We avoid excessive formalization by denoting a generic derivation
procedure by $\GP$. This contains for every operator
$\de/\de\ph_i(x)$ in (\ref{activ2}) the information about weather
if derives an explicit vertex, a large field condition $\ch_\Ga$,
an external source $\ph_{i(\De^{\rm ext})}(\ze^{\rm ext})$,
or the exponentiated interaction.
This contains also the information about which particular field appearing
in some monomial expression does the contraction $\de/\de\ph_i(x)$
hook to.
We perform the sum over $\GP$ through several steps.
First we sum over the choices $\Gc$, for every operator $\de/\de\ph$,
between deriving an external source $\ph_{i(\De_s^{\rm ext})}
(\ze_s^{\rm ext})$ or not.
Then we sum over the derivation procedures $\GP_{\rm ext}$, concerning
the $\de/\de\ph$ operators that act on the external sources.
Next we introduce a decomposition for the domain of integration
of the $\de/\de\ph$ operators that do not act on the sources,
writing
\be
\int_{\De_1^j}dx_1^j=\sum_{\til{\De}_1^j\in S(\De_1^j,Y)}
\int_{\til{\De}_1^j\backslash{\rm sh}(\til{\De}_1^j,Y)}dx_1^j
\label{decder1}
\ee
and
\be
\int_{\De_2^j}dx_2^j=\sum_{\til{\De}_2^j\in S(\De_2^j,Y)}
\int_{\til{\De}_2^j\backslash{\rm sh}(\til{\De}_2^j,Y)}dx_2^j
\ \ .
\label{decder2}
\ee
Finally we sum over the derivation procedures $\GP_{\rm int}$ of the
$\de/\de\ph$ operators that do not act on the sources.
Therefore the ultimate version of the sum we have to bound in
(\ref{polybound}) looks like
\be
\sum_{Y}\sum_\Ga\sum_{<\Gg>}
\sum_{{(\til{\De})\ {\rm of}}\atop{\rm vertices}}
\sum_{\Gc}
\sum_{\GP_{\rm ext}}
\sum_{{(\til{\De})\ {\rm of}}\atop{\rm derivations}}
\sum_{\GP_{\rm int}}
\int d{\bf h}
\ \ \GC
\ \ .
\label{lotsums}
\ee
The sum over
the $\til{\De}$'s is over 
the decompositions (\ref{decver}), (\ref{decder1})
and (\ref{decder2}) of the spatial integrations involved.
Note also that $\GC$ denotes a generic individual
contribution consisting in a functional integral with a few
spatial integrations depending on all the previous data. 
How do we bound such a contribution is the purpose of the next section.

\subsection{The bound on individual contributions. The domination
of low momentum fields.}

The main ingredient is the bound on the low momentum fields of explicit
vertices.
The high momentum fields are simply bounded with the Gaussian
measure. The low-momentum fields, however must be ``dominated'' [FMRS, R1].
One already had a flavor of this method in Proposition \ref{propsmf}.
The idea is
that since low-momentum fields are of much lower frequency (i.e. higher
index) than the 
localization cube in which they are integrated, they are statistically
almost constant. Therefore we comp5are them with their average in some cube,
whose scale is intermediate between that of the field and that of the 
localization cube. On the other hand, the bound on the average
is of the H\"older type and uses the small field conditions. Hence
it stems indirectly from the positivity of the interaction itself. Finally
the bound on the fluctuation part is Gaussian, since the involved double
gradients are well behaved concerning local factorials. The tricky issue
is the tuning of the intermediate cube of averaging. The frequency $i$
of this cube is the one appearing in the ``effective'' power counting
factor $M^{-i}$ that we gain for a low momentum field after the domination
procedure. If this averaging scale is too high ($i$ too small), the power
counting factor is bad and cannot pay all the combinatorial sums we have
to perform. If it is too low ($i$ too big), we risk the fluctuation term 
behaving as badly as would have
the initial low momentum field, if bounded by the Gaussian
measure only.
  
To bound an individual contribution ${\GC}$, first we bound everything by
its modulus. Then we bound every propagator coming from a 2-link 
$l_{j}=l[\De^{j}_{1},\De^{j}_{2}]$, using (\ref{bdprop}), by:
\be
\big| C( x_{1}^{j}, i(\De_{1}^{j}); x_{2}^{j}, i(\De_{2}^{j})) \big|
\le K_{3} (r) M^{-2i(l_{j})} \bigl( 1+ M^{-i(l_{j})}  d_{2}(l_{j}) \bigr)^{-r}
\ \ .
\label{bpropd}
\ee
The notation is that of Proposition \ref{propbd};
$r$ is some large enough fixed integer.
We also take out of the functional integral the spatial integrations of the 
explicit vertices. From (\ref{bpropd})
the line factors $M^{-i(l_{j})}$ are affected
to the corresponding contracted fields, as we explain below.

\subsubsection{The explicit vertices}
First we consider the case of an {\it explicit vertex} $l_{j}$. After
the functional derivations and taking the modulus, it has the form
\be
g\cdot {4! \over m_{1}! ... m_{\al}!}
\int_{\tilde \De \backslash {\rm sh}(\tilde\De,X_{1})} 
dx\ \big| \ph^{X_{1}}(x)\big|^{m'_{1}} ... \big| \ph^{X_{\al}}(x)
\big|^{m'_{\al}} M^{-i_{1}}...M^{-i_{\ga}}
\ \ ,
\label{vertexder}
\ee
where the $m'$ integers are the exponents left after derivation, and
$M^{-i_{1}},...,\ M^{-i_{n}}$ are the scaling factors of the derived fields
collected from (\ref{bpropd}).
We have $0\le m'_{\beta}\le m_{\beta}$ for any $\beta$,
$1\le \beta \le \al$, and $m'_{1}+...+ m'_{\al} +\ga =4$. 

From now on, we call {\it field} any one of the four factors appearing
in (\ref{vertexder}),
including the factors $M^{-i}$. We treat separately each field,
generically denoted by the symbol $\ph$ that we use to label the
associated structures, like the averaging cube ${\underline\De}_{\ph}$ for
instance, which is defined independently for each field.

We find it simpler in the sequel to define ${\und\De}_\vph$
for any low momentum field, even those that have been derived.
However the only ${\und\De}_\vph$ that will be used for averaging
are those of the remaining fields.

First consider a low momentum field $| \ph^{X_{\vph}}(x) |$. We denote by
$\tilde\De_{\vph}$ the domain of integration of the 
corresponding vertex $l_{\vph}$, and by $\bar \De_{\vph}$ the smallest cube of
$X_{\vph}$ containing $\tilde\De_{\vph}$.
Remark that $\bar\De_{\vph}$ cannot contain
any other cube of $X_{\vph}$. Indeed, by definition, the cube just above
$\bar\De_{\vph}$ containing $\tilde\De_{\vph}$ is not in $X_{\vph}$,
therefore
$\bar\De_{\vph}$ has to be a small field cube and glues to nothing above it.
Note that in this argument, we used the fact $\tilde \De_{\vph}$ is in the
top block of the vertex we denote by $X_{1,\vph}$, which is distinct from 
$X_{\vph}$, and therefore the inclusion $\tilde\De_{\vph}\subset\bar\De_{\vph}$
is strict. The averaging cube $\bar \De_{\vph}$ is chosen such that
$\tilde\De_{\vph}\subset \underline\De_{\vph}\subset \bar \De_{\vph}$; the rule
will be precised later. 

Now we write:
\be
\big|\ph _{X_{\ph}}(x)\big| \le \big|\ph _{X_{\ph}}(x)- {\rm Fluct} 
\bigl(\underline\De_{\ph} ,\ph^{X_{\ph}}\bigr)(x)    \big| + \big|{\rm Fluct} 
\bigl(\underline\De_{\ph} ,\ph^{X_{\ph}}\bigr)(x)\big|
\ \ ,
\ee 
where using the notation of Lemma \ref{avrSobo} with 
$\De_{1}=\De_{2}=\underline\De_{\ph}$, we posed
\be
{\rm Fluct} 
\bigl(\underline\De_{\ph} ,\ph^{X_{\ph}}\bigr)(x)\eqdef 
\int_{\underline\De_{\ph}^{2}\times[0,1]^{2}} d{\bf w}
\ f^{\mu,\nu}_{\underline\De_{\ph},\underline\De_{\ph}}(x,{\bf w} )
\partial_{\mu}\partial_{\nu} \ph^{X_{\ph}}
({\cal X}_{\underline\De_{\ph},\underline\De_{\ph}} (x,{\bf w} )
\ \ .
\label{flucexpli}
\ee
Still by Lemma \ref{avrSobo} we have the estimate
\be
\big|\ph _{X_{\ph}}(x)- {\rm Fluct} 
\bigl(\underline\De_{\ph} ,\ph^{X_{\ph}}\bigr)(x)    \big|\le
K_{7} \biggl({1\over |\underline\De_{\ph}|}\int _{\underline\De_{\ph}}
\bigl(\ph^{X_{\ph}}(y)\bigr)^{4}\biggr)^{1/4}
\ \ .
\ee
Note that we used the fact that $\ph^{X_{\ph}}$ is smooth in 
$\underline\De_{\ph}$, which is true since by our previous remark on
$\bar\De_{\ph}$, the bunch of frequencies making up $\ph^{X_{\ph}}(\xi)$
does not vary when $\xi$ ranges through $\bar\De_{\ph}$. Since $\bar\De_{\ph}$
is a small field cube, thanks to the function $\ch_{\Ga}(\ph)$, 
even modified
by the functional derivatives, we have in the domain with non zero
integrand, with respect to the fields, the condition
\be
\int _{\underline\De_{\ph}}
\bigl(\ph^{X_{\ph}}(y)\bigr)^{4} \le g^{-(1+\ep_{1})}
\ \ .
\ee
This entails, since $\underline\De_{\ph} \subset \bar\De_{\ph}$,
\be
\big|\ph _{X_{\ph}}(x)- {\rm Fluct} 
\bigl(\underline\De_{\ph} ,\ph^{X_{\ph}}\bigr)(x)    \big|\le K_{7} 
M^{-i(\underline\De_{\ph})} \cdot g^{-\frac{1+\ep_{1}}{4}}
\ \ .
\ee
The fluctuation part is bounded with the Gaussian measure. In fact we write
\[
\big| {\rm Fluct} 
\bigl(\underline\De_{\ph} ,\ph^{X_{\ph}}\bigr)(x)\big| \le \sum_{j\in I_{\ph}}
\sum_{\mu,\nu=1}^{4} 
\int_{\underline\De_{\ph}^{2}\times[0,1]^{2}} d{\bf w}
\]
\be
\big|f^{\mu,\nu}_{\underline\De_{\ph},\underline\De_{\ph}}(x,{\bf w} )
\big|\ \big|
\partial_{\mu}\partial_{\nu} \ph^{X_{\ph}}
({\cal X}_{\underline\De_{\ph},\underline\De_{\ph}} (x,{\bf w})\big|
\ \ ,
\ee
where $I_{\ph}$ is the set of scales making up $\ph^{X_{\ph}}$. The sums over
$j$, $\mu$ and $\nu$, the integration over $\bf w$ and the
functions $f^{\mu\nu}_{\underline\De_{\ph},\underline\De_{\ph}}$
are taken out of the functional integral.
We will later bound the double gradient by a bound like (\ref{Gaussbd}) where
the $x$ and $\bf w$ and $\mu$, $\nu$ dependences disappear. This enables
the factorization of the expressions
$\sum_{\mu,\nu=1}^{4}\int_{\underline\De_{\ph}^{2}
\times[0,1]^{2}} d{\bf w} 
\big|f^{\mu\nu}_{\underline\De_{\ph},\underline\De_{\ph}}(x,{\bf w} )
\big|$ that are bounded by 16$M^{2i(\underline\De_{\ph})}$ from (\ref{L1bd}).

We consider now the case of a high momentum field
$\ph^{X_{\ph}}$. With the same notations, we simply bound it by 
$\sum_{j\in I_{\ph}} \big|\ph_{j}(x)\big|$, and take the sum over $j$
out of the functional integral.

\subsubsection{External sources}
We consider now an external source
$\ph_{i(\De^{\rm ext}_{s})}(\ze_{s}^{\rm ext})$.
If no derivation contracts to it,
we keep $\ph_{i(\De^{\rm ext}_{s})}(\ze_{s}^{\rm ext})$ within the integral,
to be
controlled by the Gaussian bound (\ref{Gaussbd}). If it is derived, it has
to be by a 2-link hooked to $\De^{\rm ext}_{s}$, and we simply collect 
from (\ref{bpropd}) the corresponding factor $M^{-i(\De^{\rm ext}_{s})}$.

\subsubsection{Large field condition vertices}
We now point our attention to the large field condition $\ch_{\Ga}$ defined
in (\ref{lfcond}). The action of the functional derivatives will produce
new vertices of the form 
\be
\int_{\De} dx\ \big| \ph_{I_{\Ga}(\De)} (x) \big| ^{\al} g^{1+\ep_{1}} \cdot
M^{-i_{1}}\ldots M^{-i_{\ga}}
\ \ ,
\label{lfvertex}
\ee
with $0\le \al\le 3$, $\al + \ga =4$,
the $M^{-i_{\beta}}$, $1\le \beta \le \ga$,
being the scaling factors of the derived fields. These derivations are
produced by extremities $\De_{\beta}$ of 2-links with 
$i(\De_{\beta})\in I_{\Ga}(\De)$ and
$\De \subset \De_{\beta}$. Now remember that
any derivative $\ch^{(n)}$ of the function (\ref{smooth})
used to define the large field
conditions has support in $[\half,1]$. As a result, regardless of the small or
large field nature of $\De$ we have always at hand the constraint
\be
\int_{\De} dx\ \ph_{\Ga(\De)}^{4}  g^{1+\ep_{1}} \le 1
\ \ .
\ee
Therefore using H\"older's inequality, we bound (\ref{lfvertex}) by
\be
|\De|^{1-{\al\over 4}} \cdot \biggl(\int_{\De}\ph_{\Ga(\De)}^{4}\biggr)^{\al/4}
\cdot g^{1+\ep_{1}}\cdot M^{-i_{1}} ... M^{-i_{\ga}} \le 
M^{-(i_{1}-i(\De))}... M^{-(i_{\ga}-i(\De))}\cdot g^{(1+\ep_{1}){\ga \over 4}}
\ \ ,
\ee
with $1\le\ga\le 4$. Remark that, in fact, the domain of spatial
integration  of this vertex may be 
shrank by the choice of $\til\De$'s
for the involved derivation operators,
but this does not alter this bound. Besides, 
one can show by explicit computation that the derivatives of the function
$\ch$ satisfy for any $t\in \RR$
\be
|\ch^{(n)}(t) | \le K_{15}\cdot (n!)^{2}
\ \ ,
\label{derchi}
\ee
for some constant $K_{15}\ge 1$. 

Note that for each small field cube $\De$, we can introduce without changing
the functional integral a factor $\bbbone _{\{\int_{\De} 
\ph^{4}_{I_{\Ga}(\De)}\le g^{-(1+\ep_{1})} \}}$.
We introduce also a factor  $\bbbone _{\{\int_{\De} 
\ph^{4}_{I_{\Ga}(\De)}\ge \half g^{-(1+\ep_{1})} \}}$
per large field cube $\De$.
This prepares
the functional integral for applying Proposition \ref{propsmf}.

\subsubsection{Interaction vertices}
Finally we have to consider the bound on vertices that are derived from the
interaction. Such a vertex has the following form
\[
{\cal V} \eqdef g \int_{\cap_{1\le\beta\le\ga}
(\tilde\De_{\ga}\backslash {\rm sh}(\tilde\De_{\ga},Y))}
dx\ \sum_{a=0}^{k}(h_{a}-h_{a+1})
\]
\be
.\sum_{B\in \Pi_{\Ga}((l_{1},...,l_{a}))}
\bbbone_{\{\De_{1},...,\De_{\ga} \in B\}}\ |\ph^{B}(x)|^{4-\ga} \cdot
M^{-i_{1}}...M^{-i_{\ga}}
\ \ ,
\ee
where $\ga$, $1\le \ga\le 4$, is the number of derivations
${\de\over \de\ph}$ that acted on this vertex. They are labeled by the
subscript $\beta$, $1\le \beta\le\ga$.
$\De_{\beta}$ is the extremity of the 2-link
that produced the derivation with label $\beta$. $\tilde\De_{\beta}$ denotes
the chosen element of $S(\De_{\beta},Y)$ to define the domain of integration;
finally $i_{\beta}$ is shorthand for
$i(\De_{\beta})$. Remark that $\cV$ gives a non trivial contribution
only if $\cap_{1\le\beta\le\ga}(\tilde\De_{\ga}\backslash
{\rm sh}(\tilde\De_{\ga},Y))\ne 
\emptyset$, and this enforces $\tilde\De_{1}=...= \tilde\De_{\ga}\eqdef
\tilde\De_\cV$.

Note that if there is a $B$ in $ \Pi_{\Ga}((l_{1},...,l_{a}))$ such that
$\De_{1},...,\De_{\ga}\in B$, then $B$ is unique; therefore we denote
it by $B_{a}$. Now we have by H\"older's inequality
\[
{\cal V} \le g M^{-i_{1}}...M^{-i_{\ga}}\sum_{a=0}^{k}(h_{a}-h_{a+1})
\bbbone_{\{\exists B_{a}\in \Pi_{\Ga}((l_{1},...,l_{a}))|\De_{1},
...,\De_{\ga}\in B_{a}\}}
\]
\be 
|\tilde\De_{\cV}\backslash{\rm sh}(\tilde\De_{\cV},Y))|^{\frac{\ga}{4}}
\biggl(\int_{\tilde\De_{\cV}\backslash
{\rm sh}(\tilde\De_{\cV},Y))}dx
\ \ph^{B_{a}}(x)^{4}\biggr)^{{4-\ga \over 4}}
\ee
\[
\le g M^{-(i_{1}-i(\til\De_\cV))}...M^{-(i_{\ga}-i(\til\De_\cV))}
\sum_{a=0}^{k}(h_{a}-h_{a+1})
\]
\be
.\biggl(\bbbone_{\{\exists B_{a}\in \Pi_{\Ga}((l_{1},...,l_{a}))|\De_{1},
...,\De_{\ga}\in B_{a}\}}\int_{\tilde\De_{\cV}-sh(\tilde\De_{\cV},Y))}dx
\bigl(\ph^{B_{a}}(x)\bigr)^{4}\biggr)^{{4-\ga \over 4}}
\ \ .
\ee

Now since the $h_a-h_{a+1}$ are positive and add up to 1, by concavity
of the function $t\mapsto t^{\frac{4-\ga}{4}}$, we have
\be
\cV\le g.M^{-(i_1-i(\til{\De}_\cV))}\ldots
M^{-(i_\ga-i(\til{\De}_\cV))}
\lp
\sum_{a=0}^k
(h_a-h_{a+1})
\bbbone_{\{\exists B_a\}}
\int_{\til{\De}_\cV\backslash{\rm sh}(\til{\De}_\cV,Y)}dx
\lp\ph^{B_a}(x)\rp^4
\rp^{\frac{4-\ga}{4}}
\ee
\be
\le
g^{\frac{\ga}{4}}
.M^{-(i_1-i(\til{\De}_\cV))}\ldots
M^{-(i_\ga-i(\til{\De}_\cV))}
\cY_{\til\De_\cV}^{\frac{4-\ga}{4}}
\ \ ,
\ee
with
\be
\cY_{\til\De}\eqdef
g\int_{\til{\De}\backslash{\rm sh}(\til{\De},Y)}dx
\lp
\sum_{a=0}^k (h_a-h_{a+1})
\sum_{B\in\Pi_\Ga((l_1,\ldots,l_a))}
\lp\ph^{B}(x)\rp^4
\rp
\ \ ,
\ee
for any $\til{\De}\in\cD^{(N)}$.

Now $\La$ being the unique cube in the lowest layer $\cD_N^{(N)}$,
we can write the interaction in the exponential as
\be
\cI=\sum_{\til{\De}\in S(\La,Y)}\cY_{\til{\De}}
\ \ .
\ee
We now use half of if to bound the product of
$\cY_{\De_\cV}^{\frac{4-\ga}{4}}$ generated by the derived vertices.
Indeed thanks again to the inequality $y^\nu e^{-y}\le \nu!$ for $y\ge 0$,
we have
\be
\prod_{{\cV\ {\rm derived}}\atop{\rm vertex}}
\cY_{\til\De_\cV}^{\frac{4-\ga_\cV}{4}}
\times\exp(-\frac{1}{2}\cI)
\le
2^{-\frac{U}{4}}.\prod_{\til{\De}\in S(\La,Y)}
\lp{\rm mult}_1(\til{\De})!\rp^{\frac{1}{4}}
\ \ ,
\ee
where $U$ is the number of underived fields in derived vertices i.e.
\be
U\eqdef
\sum_{{\cV\ {\rm derived}}\atop{\rm vertex}}
(4-\ga_\cV)
\ \ .
\ee
Likewise, ${\rm mult}_1(\til{\De})$ counts
the number of underived fields localized
in $\til{\De}\backslash{\rm sh}(\til{\De},Y)$
i.e.
\be
{\rm mult}_1(\til{\De})\eqdef
\sum_{{\cV\ {\rm derived}}\atop{\rm vertex}\ |\ \til{\De}_\cV=\til{\De}}
(4-\ga_\cV)
\ \ .
\ee

\subsubsection{The bound on the functional integral}
Now that we have explained the bound on field monomials appearing in the
functional integral, we explain the bound on the latter.
We take out of it all the bounding numerical constants including
powers of $g$, of ${\rm mult_1}(\til{\De})!$, scaling factors
$M^{-i}$, bounds like (\ref{bpropd}) on the propagators,
bounds as in (\ref{derchi}) on $\ch^{(n)}$.
The sums over scales $j$ of fields, and the spatial
integrations are taken out of the functional integral, so that what
remains in the end is an expression of the form
\[
\GC'=\int d\mu_{C[T_{<\Gg>,\Ga}({\bf h})]}
\ \prod |\ph_j(x)|\times\prod|\ph_{i(\De_s^{\rm ext})}(\ze_s^{\rm ext})|
\times\prod |\partial_\mu\partial_\nu\ph_j(x)|
\]
\be
\times\prod \bbbone_{\{\int_\De \Ph_{I_\Ga(\De)}^4\ge\frac{1}{2}\}}
\times\prod\bbbone_{\{\int_\De \Ph_{I_\Ga(\De)}^4\le 1\}}
\times
\exp(-\frac{1}{2}\cI)
\ \ .
\label{defcprim}
\ee
The product over $|\ph_j(x)|$ is on the remaining
high momentum fields, that one over $|\partial_\mu\partial_\nu\ph_j(x)|$
is on the double gradients generated by the fluctuation terms in
(\ref{flucexpli}).
Thanks to the Cauchy-Schwarz inequality we now bound $\GC'$
by $(I_1.I_2)^\half$ where
\be
I_1\eqdef
\int d\mu_{C[T_{<\Gg>,\Ga}({\bf h})]}
\ \prod (\ph_j(x))^2\times
\prod (\partial_\mu\partial_\nu\ph_j(x))^2
\ee
and
\be
I_2\eqdef
\int d\mu_{C[T_{<\Gg>,\Ga}({\bf h})]}
\ \prod \bbbone_{\{\int_\De \Ph_{I_\Ga(\De)}^4\ge\frac{1}{2}\}}
\times\prod\bbbone_{\{\int_\De \Ph_{I_\Ga(\De)}^4\le 1\}}
\times
\exp(-\cI)
\ \ .
\ee

The first integral is bounded by Lemma \ref{Gauss}. The second is bounded
thanks to Proposition \ref{propsmf} by
\be
I_2\le
U(g)^{\half\#(Y)}.
\prod_{\De{\rm\ isolated\ in\ }Y}
g^{-\frac{\ep_5}{2}
\ \ .}
\ee

\subsubsection{The choice of averaging cubes}
We are left with the task of explaining the rule for choosing the
averaging cubes ${\und\De}_\vph$, for low momentum fields. The method consists
in two operations.

In the first we define a family $({\und\De}_\vph^0)_\vph$ of cubes indexed by
low momentum fields, by letting ${\und\De}_\vph^0$ be the unique cube
containing ${\til\De}_\vph$ at scale
\be
i({\und\De}_\vph^0)\eqdef
i({\til\De}_\vph)+E[\frac{1}{4}(i({\Br\De}_\vph)-i({\til\De}_\vph))]
\ \ .
\ee
Note that necessarily
${\til\De}_\vph\subset {\und\De}_\vph^0
\subset {\Br\De}_\vph$.

The second operation is inductive and goes from small to large scales.
Following the notations of
Section 3.2, we let $i$ be any scale, $0\le i< i_{\rm max}(Y)$,
we then construct a family $({\und\De}_\vph^i)_\vph$ of cubes in $\Br Y$, as
follows.
Suppose $i\ge 1$, and $({\und\De}_\vph^{i-1})_\vph$ has
been constructed. We consider the ordinary graph $F$ made by the 2-links
of $<\Gg>$. Since $<\Gg>$ is allowed, $F$ must be a forest.
Besides, $\Pi_\Ga(\Gg)=\{Y\}$. As a consequence, for any component $A$
in $\cR_i(Y,F)$, there must be at least one gluing link
or five explicit vertices crossing the lower boundary of $A$.
This is to say, their support intersects
$Y_{>i}\eqdef
Y\cap(\cup_{j>i}\cD_j^{(N)})$, as well as $Y\cap {\rm pr}_i^{-1}(A)$,
the set of cubes in $Y$ contained in some cube of $A$.
Indeed if that was not true, we would be able to disconnect
$Y\cup{\rm  pr}_i^{-1}(A)$ from the (nonempty since $i<i_{\rm max}(Y)$)
rest of $Y$, by removing at most four 4-links from $\Gg$,
contradicting the four vertex irreducibility of $Y$ with
respect to $\Gg$ and $\Ga$.
Suppose there is no gluing link crossing the lower boundary of $A$,
then there are at least five explicit vertices which do.
It is not difficult to see that
we can pick five low momentum fields $\vph$
(one in each of the above mentioned
vertices) such that ${\til\De}_\vph\in Y\cap {\rm pr}_i^{-1}(A)$
and $i({\Br\De}_\vph)>i$.
If some of these five chosen fields verify
$i({\und\De}_\vph^{i+1})\le i$, we define for such a field
${\und\De}_\vph^{i}$ to be the unique cube containing ${\til\De}_\vph$
at scale $i+1$.
For all the other low momentum fields $\vph$ with
${\til\De}_\vph\in Y\cap {\rm pr}_i^{-1}(A)$ we keep
${\und\De}_\vph^{i}\eqdef{\und\De}_\vph^{i-1}$.
We do the same for all the components $A$ in $\cR_i(Y,F)$, and
this completes the definition of $({\und\De}_\vph^{i})_\vph$.

Finally we define $({\und\De}_\vph)_\vph$ to be
$({\und\De}_\vph^{i_{\rm max}(Y)})_\vph$.
It is easy to check that the cubes ${\und\De}_\vph$ belong to $\Br Y$,
and for any $\De\in{\Br Y}$, the number of fields $\vph$ such that
${\und\De}_\vph=\De$ and ${\und\De}_\vph\neq{\und\De}_\vph^o$
is at most $5M^4$, i.e. five times the maximal number of components
$A$ having an element just above $\De$.

\subsubsection{The small factors per cube}
Section 3.6.1 shows that, in the bound for an
explicit vertex, we obtain at least
a factor $g^{\frac{1-3\ep_1}{4}}$ ($g$ is supposed in the interval
$]0,1[$). If now we choose $\ep_1\eqdef \frac{1}{6}$, we
obtain $g^{\frac{1}{8}}$. From a large field condition vertex, we can extract
$g^{\frac{1+\ep_1}{4}}=g^{\frac{7}{24}}$. Finally from an interaction vertex we
can extract at least a $g^{1/4}$.
We can thus say that from a vertex of any kind we obtain at least
a factor as small as
$g^{\frac{1}{8}}$. Now if we chose $\ep_5=\frac{1}{16}$, we collect
\be
\prod_{{\rm vertices\ of}\atop{\rm any\ kind}}g^{\frac{1}{8}}
\le
\prod_{{\rm isolated\ cubes\ of}\ Y}g^{\frac{\ep_5}{2}}
\ \ .
\label{distsmf}
\ee
Indeed, any isolated cube $\De$ of $Y$ forms a dressed large field
block in the support of
an explicit vertex, or is an extremity of a 2-link.
In the first case we can say that one fourth of the $g^{\frac{1}{8}}$
factor of such an explicit vertex is attributed to this cube.
If we cannot find such an explicit vertex, having chosen an order to perform
the functional derivatives, we consider the first computed $\de/
\de\ph_i(x)$ operator coming form a 2-link hooked to $\De$.
This derivation has to derive a new vertex from the exponentiated
interaction or from a large field condition. We can decide that
the $g^{1/8}$ factor of this vertex is attributed to $\De$.
In any case we get better that $g^{\frac{1}{32}}=g^{\frac{\ep_5}{2}}$.
It is easy to see that we have not attributed in this way the same
factor twice; and this proves (\ref{distsmf}).

\subsubsection{The handling of Gaussian factorials}
For any cube $\De$ of $Y$, we define ${\rm mult}_2(\De)$
to be the number of fields $\ph_j(x)$ or $\partial_\mu\partial_\nu
\ph_j(x)$ or $\ph_{i(\De_s^{\rm ext})}(\ze_s^{\rm ext})$
in the integral $\GC'$ of equation (\ref{defcprim}),
such that $\De(x,j)=\De$ or $\De_s^{\rm ext}=\De$,
i.e. located in $\De$. Note that every initial field
appearing in (\ref{defcprim}) is now counted twice after the Cauchy-Schwarz
argument.
The Gaussian bound of Lemma \ref{Gauss} applied to $I_1$
yields a product of local
factorials
$\prod_{\De\in Y}((2.{\rm mult}_2(\De))!)^\half$ to deal with.

Now we define ${\rm mult}_3(\De)$ to count the high momentum fields,
of the form $\ph_j(x)$, located in $\De$, and ${\rm mult}_4(\De)$ to count
the low momentum fields, of the form $\partial_\mu\partial_\nu
\ph_j(x)$, located in $\De$. We recall that $E_\De$ counts
the external sources in $\De$, and $\cS$ labels them.
By the elementary inequality
\be
(p_1+\ldots+p_m)!
\le m^{p_1+\ldots+p_m}p_1!\ldots p_m!
\ \ ,
\label{nommulti}
\ee
and since the number
of fields in $I_1$ is at most $2.4.k+2.\#(\cS)$, we have
\[
\prod_{\De\in Y}((2.{\rm mult}_2(\De))!)^\half
\le
3^{4k+\#(\cS)}\times
\prod_{\De\in Y}((2.{\rm mult}_3(\De))!)^\half
\]
\be
\times\prod_{\De\in Y}((2.{\rm mult}_4(\De))!)^\half
\times\prod_{\De\in Y}((2.E_\De)!)^\half
\ \ .
\ee
We now use the elementary multinomial inequality (\ref{multinom}) to write
\be
\prod_{\De\in Y}((2.{\rm mult}_3(\De))!)^\half
\le
\prod_{X\in \pi}((2.{\rm mult}_5(X))!)^\half
\ \ ,
\ee
where ${\rm mult}_5(X)\eqdef\sum_{\De\in X}{\rm mult}_3(\De)$, for any
$X\in\pi$.
We make another distinction by defining
${\rm mult}_6(\De)$ to count the low momentum fields
$\vph$ located in $\De$ and such that ${\und\De}_\vph\neq
{\und\De}_\vph^0$. We define also ${\rm mult}_7(\De)$ to count the other
low momentum fields located in $\De$.
By the same line of argument and since the total number of occurrences
of low momentum
fields in $I_1$
is no greater than $2.3.k$, we
have by (\ref{nommulti})
\be
\prod_{\De\in Y}((2.{\rm mult}_4(\De))!)^\half
\le
2^{3k}\times
\prod_{\De\in Y}((2.{\rm mult}_6(\De))!)^\half
\times
\prod_{\De\in Y}((2.{\rm mult}_7(\De))!)^\half
\ \ .
\ee
We now use the lemma of displacement of local factorials to bound
the last two products.

First we consider the fields $\vph$ such that ${\und\De}_\vph=
{\und\De}_\vph^0$. Remark that the spatial integration on the corresponding
vertex that has been taken out of the functional integral $\GC'$,
produces a factor $|\til\De_\vph\backslash
{\rm sh}(\til\De_\vph,X_{1,\vph})|\le M^{-4i(\til\De_\vph)}$.
This factor can be distributed equally to the four fields
composing the vertex. Now if we take the share of a low momentum
field $\vph$ inside the square root $I_1^\half$, we have such a factor for
each of the two copies of the field.
But the Gaussian bound (\ref{Gaussbd}) produces a constant $K_4$ and a scaling 
factor $M^{-3j_\vph}$ for each of these copies,
since it is of double gradient type.
Recall that we have also $L^1$ bounds on the smearing functions
$f_{\und\De_\vph,\und\De_\vph}^{\mu,\nu}$
of these double gradients, that produce a factor $16M^{2i(\und\De_\vph)}$.
Now the total factor attributed to a copy of a field $\vph$ is
a vertical exponential decay, of varying strength weather we are
between the double gradient scale $j_\vph$ and the averaging scale
$i(\und\De_\vph)$, or
between the latter and the localization scale $i(\til\De_\vph)$.
Namely it is
\be
16K_4
M^{-3(j_\vph-i(\Br\De_\vph))-3(i(\Br\De_\vph)-i(\und\De_\vph))
-(i(\und\De_\vph)-i(\til\De_\vph))}
\ \ .
\label{recoil}
\ee
Now in the present case
$i(\und\De_\vph)=
i({\til\De}_\vph)+E[\frac{1}{4}(i({\Br\De}_\vph)-i({\til\De}_\vph))]$
entails
$3(i(\Br\De_\vph)-i(\und\De_\vph))
\ge\frac{9}{4}(i(\Br\De_\vph)-i(\til\De_\vph))$,
as well as
$\frac{1}{10}(i(\und\De_\vph)-i({\til\De}_\vph))\ge
\frac{1}{40}(i(\Br\De_\vph)-i({\til\De}_\vph))-\frac{1}{10}$.
Therefore we readily bound (\ref{recoil}) by
\be
16K_4 M^{-\frac{1}{10}}(1-M^{-\half})^{-\half}
.M^{-\frac{9}{10}(i(\und\De_\vph)-i(\til\De_\vph))}
.M^{-\frac{1}{40}(j_\vph-i(\til\De_\vph))}
.
\sqrt{(1-M^{-\half}) M^{-\frac{9}{2}(j_\vph-i(\til\De_\vph))}}
\ \ .
\ee
We recognize in the square root the function $G$ of (\ref{defGdis})
that allows
to compute the local factorials of the presently considered
fields as if they were
located in $\til\De_\vph$, at the cost of an extra factor
$e^\half$.

Now we consider fields $\vph$ such that ${\und\De}_\vph\neq
{\und\De}_\vph^0$. By construction we must have
\be
i(\und\De_\vph)\ge
i({\til\De}_\vph)+E(\frac{1}{4}(i({\Br\De}_\vph)-i({\til\De}_\vph)))
\ \ .
\ee
Since we deal again with double gradients we have still a factor
(\ref{recoil}), per copy of $\vph$, that is grossly bounded by
\be
16K_4 M^{-\frac{1}{10}}(1-M^{-\half})^{-\half}
.M^{-\frac{9}{10}(i(\und\De_\vph)-i(\til\De_\vph))}
.M^{-\frac{1}{40}(j_\vph-i(\til\De_\vph))}
.\sqrt{(1-M^{-\half}) M^{-\frac{9}{2}(j_\vph-i(\und\De_\vph))}}
\ \ .
\ee
Now recall our previous remark that for a given
$\und\De_\vph$
there can be at most
$5M^4$ low momentum fields $\vph$ such that ${\und\De}_\vph\neq
{\und\De}_\vph^0$.
Therefore the product of local factorials we get after displacement by
Lemma \ref{displ} is bounded by
\be
\prod_{{\De\in{\Br Y}}\atop{\exists\vph,{\und\De}_\vph=\De}}
((10M^4)!)^\half
\le
((10M^4)!)^{\frac{3}{2}k}
\ \ .
\ee

\subsubsection{The bound on an individual contribution $\GC$}
We now have all the elements to write such a bound and it is
\[
|\GC|\le
U(g)^{\half\#(Y)}
\times\prod_{l\in F}\lp K_3(r)(1+M^{-i(l)}d_2(l))^{-r}\rp
\times\prod_{s\in\cS}\lp K_4
M^{-i(\De_s^{\rm ext})}\rp
\]
\[
\times\prod_{\De\in Y}((2.E_\De)!)^{\frac{1}{4}}
\times\lp 2(1-M^{-\frac{1}{40}})^{-1}\rp^{4k}\times
(4!)^k\times
K_4^{3k}
\]
\[
\times
\prod_{\vph}
\lp
\max[K_7,16 K_4 e^\half (1-M^{-\half})^{-\half}]
.M^{-\frac{1}{10}}
.M^{-\frac{9}{10}(i(\und\De_\vph)-i(\til\De_\vph))}
.M^{-\frac{1}{40}(i(\Br\De_\vph)-i(\til\De_\vph))}
\rp
\]
\[
\times
\prod_{\de}
\lp
M^{-\frac{3}{40}(i(\De_\de)-i(\til\De_\de))}
\rp
\times K_{15}^{\#(Y)}
\times 3^{2k+\half \#(\cS)}\times 2^{\frac{3}{2}k}
\times\prod_{\De\in Y}({\rm mult}_8(\De)!)^2
\]
\be
\times
\prod_{\til\De\in S(\La,Y)}({\rm mult}_1(\til\De)!)^\half
\times
\prod_{X\in\pi}(2.{\rm mult}_5(X)!)^{\frac{1}{4}}
\times
\prod_{\De\in Y}(2.{\rm mult}_9(\De)!)^{\frac{1}{4}}
\times((10M^4)!)^{\frac{3k}{4}}
\ \ .
\label{indbd}
\ee

In this inequality, $F$ is the forest of 2-links of $\Gg$; $s\in\cS$
labels the sources; $\vph$ ranges through all the low momentum fields,
derivated or not, for which $\und\De_\vph$ has always been
defined; and $\de$ labels the functional differential
operators $\de/\de\ph_i(x)$, attached to the extremities of the 2-links,
that did not contract to the external sources.
$\De_\de$ is the concerned extremity, whereas $\til\De_\de$
is the corresponding chosen integration cube in $S(\De_\de,Y)$
of equations (\ref{decder1}) and (\ref{decder2}). For every $\De\in Y$,
${\rm mult}_8(\De)$ counts how many times the associated large field
condition $\ch_\De$ of equation (\ref{lfcond}), is derived.
Finally, ${\rm mult}_9(\De)$ counts the number of low momentum fields
$\vph$ such that $\til\De_\vph=\De$.

Note that the factor $\lp 2(1-M^{-\frac{1}{40}})^{-1}\rp^{4k}$
comes from the choice paid by a factor 2,
for each low momentum field, between the
fluctuation or the average term, in addition to the choice
of scale $j_\vph$, which is summed thanks to the factor
$M^{-\frac{1}{40}(j_\vph-i(\Br\De_\vph))}$. In the case of a high
momentum field we sum also on the scale thanks only to a mere fraction
$M^{-\frac{1}{40}(j-\til\De)}$ of the available vertical decay
$M^{-(j-i(\til\De))}$.
Finally the number of high or low momentum fields is bounded by $4k$.

The $(4!)^k$ factor bounds the symmetry factors of the explicit
vertices. The $K_4^{3k}$ term bounds
the constants coming
with the scaling factor of the high momentum fields, when integrated with
the Gaussian bound.

Remark that for a derivated field in an explicit vertex, the volume of
integration of the latter is shrunk to $\til\De_\de\backslash
{\rm sh}(\De_\de,Y)$, where $\de$ is the involved derivation.
However we use this improvement, only when considering the fourth power of
this volume that is attributed to the mentioned derivated field.
In any case, for a field derivated by $\de$, we collect
a factor $M^{-((i(\De_\de)-i(\til\De_\de))}$. This is more than enough
to pay the $M^{-\frac{3}{40}(i(\De_\de)-i(\til\De_\de))}$ in case the
field is in an interaction vertex, or a large field condition vertex,
or a source, or a high momentum field.
In case it is a low momentum field $\vph$, we need a fraction
$37/40$ of the initial decay
to pay for $M^{-\frac{9}{10}(i(\und\De_\vph)-i(\til\De_\vph))}$
and $M^{-\frac{1}{40}(i(\Br\De_\vph)-i(\til\De_\vph))}$,
and we cannot count on more than
$M^{-\frac{3}{40}(i(\De_\de)-i(\til\De_\de))}$ to be allocated
to the product over the $\de$'s.

After justifying (\ref{indbd}),
we can improve it by writing a nicer bound where the Gaussian factorials have
undergone a first treatment. By the inequality (\ref{nommulti})
with $m=2$, we have
\be
\prod_{X\in\pi}(2.{\rm mult}_5(X)!)^{\frac{1}{4}}
\times\prod_{\De\in Y}(2.{\rm mult}_9(\De)!)^{\frac{1}{4}}
\le
2^{2k}\times
\prod_{X\in\pi}({\rm mult}_5(X)!)^{\frac{1}{2}}
\times\prod_{\De\in Y}({\rm mult}_9(\De)!)^{\frac{1}{2}}
\ \ ,
\ee
since $\sum_{X\in\pi}{\rm mult}_5(X)+\sum_{\De\in Y}{\rm mult}_9(\De)$
is bounded by the total number of fields in explicit vertices,
i.e. by $4k$.
Now if we define for any $X\in\pi$, ${\rm mult}_{10}(X)$ to count
the number of low momentum fields $\vph$ with $\til\De_\vph\in X$,
by the inequality (\ref{multinom}), we have
\be
\prod_{\De\in Y}{\rm mult}_9(\De)!
\le
\prod_{X\in\pi}{\rm mult}_{10}(X)!
\ \ .
\ee
Now let $\GV(X)$ denote the number of explicit vertices in the form
$<\Gg>$ whose top block is $X$. This implies, for any $X\in\pi$,
${\rm mult}_{5}(X)+{\rm mult}_{10}(X)=4\GV(X)$.
Thus by (\ref{multinom}) and (\ref{nommulti}), we have
\be
\prod_{X\in\pi}({\rm mult}_5(X)!)^{\frac{1}{2}}
\times\prod_{X \in \pi}({\rm mult}_{10}(X)!)^{\frac{1}{2}}
\le\prod_{X \in \pi}[(4\GV(X))!]^\half\le
2^{4k}\times\prod_{X\in\pi}(\GV(X)!)^2
\ \ .
\ee
If ${\rm mult}_{11}(\til\De)$ counts by definition the derivations
$\de$ such that $\til\De_\de=\til\De$, we have
${\rm mult}_1(\til\De)\le 3.{\rm mult}_{11}(\til\De)$,
and therefore $\prod_{\til\De\in S(\La,Y)}({\rm multi}_{1}(\til\De)!)^\half
\le 3^{3k}.\prod{\til\De\in S(\La,Y)}({\rm multi}_{11}
(\til\De)!)^{\frac{3}{2}}$, since the total number of $\de$'s is bounded
by $2k$.
We can now use Lemma \ref{displ} to bound the local factorials
in ${\rm mult}_{8}(\De)$. Indeed, if $\de$ acts on the large field
condition of $\De$, we must have $\til\De_\de\subset\De\subset\De_\de$,
and therefore
\be
\prod_{\De\in Y}
({\rm mult}_{8}(\De)!)^2
\times\prod_{\de}
\lp
M^{-\frac{1}{40}(i(\De_\de)-i(\til\De_\de))}
\rp
\le
\lp
\prod_{\De\in Y}
({\rm mult}_{8}(\De)!)\times
\prod_{{\de\ {\rm acting}}\atop{{\rm on\ }\ch}}
M^{-\frac{1}{80}(i(\De_\de)-i(\hat\De_\de))}
\rp^2
\ \ ,
\ee
where $\hat\De_\de$ denotes the cube that labels the large field condition
on which $\de$ acts. 
Now define ${\rm mult}_{12}(\De)$ to be the number of 2-links
hooked to $\De$. It must be no less than the
number of internal derivations $\de$ such that $\De_\de=\De$.
By Lemma \ref{displ}, with $\cO$ equal to the
set of $\de$'s acting on $\ch$, $S_1(\de)\eqdef\hat\De_\de$ and
$S_2(\de)\eqdef\De_\de$, we deduce
\be
\prod_{\De\in Y}
({\rm mult}_{8}(\De)!)^2
\times\prod_{\de}
\lp
M^{\frac{1}{40}(i(\De_\de)-i(\til\De_\de))}
\rp
\le
\prod_{\De\in Y}
({\rm mult}_{12}(\De)!)^2
\times
\lp
e^2.(1-M^{-\frac{1}{80}})^{-2}
\rp^{2k}
\ \ .
\ee

Finally, using Lemma \ref{fivefor} to bound the exponents $k$ by $6\#(Y)$, and
posing
\[
K_{16}\eqdef 
2^{87}.3^{27}.e^{24}.K_4^{18}.K_{15}.
(\max[K_7,16 K_4 e^\half (1-M^{-\half})^{-\half}].M^{-\frac{1}{10}})^{18}
\]
\be
.\lp
1-M^{-\frac{1}{40}}
\rp^{-24}
.\lp
1-M^{-\frac{1}{80}}
\rp^{-24}
.((10M^4)!)^{\frac{9}{2}}
\ \ ,
\ee
the results of this section boil down to
\[
|\GC|\le
U(g)^{\half\#(Y)}\times K_{16}^{\#(Y)}\times\prod_{\De\in Y}
(E_\De!)^\half\times
\prod_{s\in\cS}
\lp
2^{\frac{1}{2}}.3^\half.K_4.M^{-i(\De_s^{\rm ext})}
\rp
\]
\[
\times
\prod_{l\in F}\lp K_3(r)(1+M^{-i(l)}d_2(l))^{-r}\rp
\times
\prod_{\vph}
\lp
M^{-\frac{9}{10}(i(\und\De_\vph)-i(\til\De_\vph))}
.M^{-\frac{1}{40}(i(\Br\De_\vph)-i(\til\De_\vph))}
\rp
\]
\be
\times
\prod_{\de}
\lp
M^{-\frac{1}{20}(i(\De_\de)-i(\til\De_\de))}
\rp
\times
\prod_{\De\in Y}
({\rm mult}_{12}(\De)!)^{2}
\times
\prod_{\til\De\in S(\La,Y)}
({\rm mult}_{11}(\til\De)!)^{\frac{3}{2}}
\times
\prod_{X\in\pi}
(\GV(X)!)^2
\ \ .
\ee

\subsection{The sum over the procedures of internal derivations}

We note that the previous bound do not depend on $\GP_{\rm int}$.
Therefore we only have to bound the number of such derivation procedures.
We suppose an order has been chosen for the action of the derivation
operators generically denoted by $\de$.
For each $\de$ we have a tree of possibilities. Each node of the tree
means a new discussion of possible cases. The number we have to bound
is the number of procedures as filtered by the successive discussions,
i.e. the number of leafs in this tree.
At each node we bound the sum over the subordinate procedures, by the number of
cases at this stage times the supremum of the analog sums for
the chosen case to be discussed at the next stage.
This is the standard method of combinatoric factors, see [GJ2, R1].
Now considering a derivation operator $\de$.
By a factor 2, we decide weather it derives a new vertex either from
the interaction or a large field condition or the product
of explicit vertices, or weather it derives an already derived
vertex.
In the first case we choose by a factor 3 between
the cited possibilities.

If $\de$ derives a new vertex from the interaction, we say we have
a type I derivation and we only
have to choose among the fields of this vertex the one
to contract, by a factor 4.

If $\de$ derives an explicit vertex $l_j$, we say that we have
a type II derivation. First, we have to find its localization
cube $\til\De_j$. The latter has to verify $\til\De_\de\subset
\til\De_j\subset\De_\de$, therefore there are at most $1+i(\De_\de)-
i(\til\De_\de)$ possibilities. Since $\til\De_j$ is in
$S({\rm bg}(X_{j,1}),X_{j,1})$ and thus in $X_\de\eqdef X_{j,1}$, the top
large field block of $l_j$,
we have to pay a factor $\GV(X_\de)$ to find the concerned vertex.
Finally we pay a factor 4, to find the field to contract in the vertex.

If $\de$ derives the large field condition $\ch_\De$ of some cube
$\De$, we say we have a type III derivation
and we must have again $\til\De_\de\subset\De\subset\De_\de$.
As a result we have to pay a factor $1+i(\De_\de)-
i(\til\De_\de)$, then a factor 4 to find the field in the vertex.

Finally if $\de$ derives an already derived vertex, we say we have a
type IV derivation. Now we have to find
the previous derivation $\de'$ that first derived the vertex on
which $\de$ acts.
In order to yield a non zero contribution, we must have
the match $\til\De_{\de'}=\til\De_\de$. As a result we have to
pay the factor ${\rm mult}_{11}(\til\De_\de)$ to find $\de'$, then a factor
3 to choose the field to contract in the vertex, since one has
already been derived.

To summarize the previous considerations, we write the bound
\be
\sum_{\GP_{\rm int}} 1
\le
\max_{\GP_{\rm int}}
\ {\rm Comb}(\GP_{\rm int})
\ \ ,
\ee
where
\[
{\rm Comb}(\GP_{\rm int})
\eqdef
\lp
\prod_{{\de\ {\rm of\ type\ I}}\atop{{\rm in}\ \GP_{\rm int}}}
24
\rp
\times
\lp
\prod_{{\de\ {\rm of\ type\ II}}\atop{{\rm in}\ \GP_{\rm int}}}
\lp 24.(1+i(\De_\de)-i(\til\De_\de)).\GV(X_\de)\rp
\rp
\]
\be
\times
\lp
\prod_{{\de\ {\rm of\ type\ III}}\atop{{\rm in}\ \GP_{\rm int}}}
\lp 24.(1+i(\De_\de)-i(\til\De_\de))\rp
\rp
\times
\lp
\prod_{{\de\ {\rm of\ type\ IV}}\atop{{\rm in}\ \GP_{\rm int}}}
\lp 6.{\rm mult}_{11}(\til\De_\de)\rp
\rp
\ \ .
\label{combder}
\ee
Since the number of occurrences of a factor $\GV(X)$ in (\ref{combder})
is the number of derivations that contract to an explicit
vertex with
top block $X$, it is bounded by $4\GV(X)$.
Similarly, the number of occurrences of a factor
${\rm mult}_{11}(\til\De)$ is at most the number of
$\de$'s such that $\til\De_\de=\til\De$, i.e. ${\rm mult}_{11}(\til\De)$.
It is now  easy to derive
\be
\sum_{\GP_{\rm int}} 1
\le
24^{2k}\times
\prod_{\de}
(1+i(\De_\de)-i(\til\De_\de))
\times
\prod_{{X\in\pi}\atop{\GV(X)\neq 0}}\GV(X)^{4\GV(X)}
\times
\prod_{{\til\De\in S(Y,\La)}\atop{{\rm mult}_{11}(\til\De)\neq 0}}
{\rm mult}_{11}(\til\De)^{{\rm mult}_{11}(\til\De)}
\ \ ,
\ee
or
\be
\sum_{\GP_{\rm int}} 1
\le
24^{2k}.e^{6k}\times
\prod_{\de}
(1+i(\De_\de)-i(\til\De_\de))
\times
\prod_{X\in\pi}
(\GV(X)!)^4
\times
\prod_{\til\De\in S(Y,\La)}
({\rm mult}_{11}(\til\De))!
\ \ .
\ee
Note that using the inequality $u.e^{-u}\le 1$, for $u\ge 0$, we
readily derive
\be
\prod_\de(1+i(\De_\de)-i(\til\De_\de))\le
\prod_\de
\lp
\frac{60.M^{\frac{1}{60}}}{\log M}.
M^{\frac{1}{60}(i(\De_\de)-i(\til\De_\de))}\rp
\ \ .
\ee
Next, we can use Lemma \ref{displ} to get rid of the local factorials
$({\rm mult}_{11}(\til\De)!)^{\frac{5}{2}}$,
that appear after the bound on the sum over $\GP_{\rm int}$.
Indeed, we take $\cO$ to be the set of internal
derivations $\de$, $\cE\eqdef Y$, 
$S_1(\de)\eqdef \til\De_\de$, and $S_2(\de)\eqdef \De_\de$, for
every $\de$.
Finally, we use the function
\be
G(\De_1,\De_2)\eqdef
\left\{
\begin{array}{ll}
(1-M^{-\frac{1}{150}}).M^{-\frac{1}{150}(i(\De_2)-i(\De_1))} &
{\rm if}\ \ \De_1\subset\De_2\\
0 & {\rm else}.
\end{array}
\right.
\ee
Now, Lemma \ref{displ} yields
\be
\prod_{\til\De\in S(\La,Y)}({\rm mult}_{11}(\til\De)!)^{\frac{5}{2}})
\le
\prod_\de
\lp
e^{\frac{5}{2}}(1-M^{-\frac{1}{150}})^{-\frac{5}{2}}.
M^{\frac{1}{60}(i(\De_\de)-i(\til\De_\de))}\rp
\times
\prod_{\De\in Y}
({\rm mult}_{12}(\De)!)^{\frac{5}{2}}
\ \ .
\ee
Besides remark that nothing in the present
bounds depend on the $h$ parameters and therefore the
integral $\int_{1>h_1>\ldots>h_k>0}dh$ is simply
bounded by the powerful factor $\frac{1}{k!}$.
As a result
\[
\sum_{\GP_{\rm int}}\int d{\bf h}|\GC|\le
\frac{1}{k!}.U(g)^{\half\#(Y)}.K_{17}^{\#(Y)}.\prod_{\De\in Y}
(E_\De!)^\half
\]
\[
\times\prod_{s\in\cS}(2^\half.3^\half.K_4.M^{-i(\De_s^{\rm ext})})
\times
\prod_{l\in F}\lp K_3(r)(1+M^{-i(l)}d_2(l))^{-r}\rp
\]
\[
\times
\prod_{\vph}
\lp
M^{-\frac{9}{10}(i(\und\De_\vph)-i(\til\De_\vph))}
.M^{-\frac{1}{40}(i(\Br\De_\vph)-i(\til\De_\vph))}
\rp
\]
\be
\times
\prod_{\de}
\lp
M^{-\frac{1}{60}(i(\De_\de)-i(\til\De_\de))}
\rp
\times
\prod_{\De\in Y}
({\rm mult}_{12}(\De)!)^{\frac{9}{2}}
\times
\prod_{X\in\pi}
(\GV(X)!)^6
\ \ ,
\ee
where
\be
K_{17}\eqdef
K_{16}.24^{12}.e^{36}.
\lp
\frac{60.M^{\frac{1}{60}}}{\log M}.e^{\frac{5}{2}}.
(1-M^{-\frac{1}{150}})^{-\frac{5}{2}}\rp^{12}
\ \ .
\ee

\subsection{The final bound}
We now describe the bounds on the remaining sums in (\ref{lotsums}).
We first bound the sum over the localization cubes $\til\De_\de$
of the internal derivations $\de$ thanks to Lemma \ref{rainlem}
that gives
\be
\sum_{{(\til\De)\ {\rm of}}\atop{\rm derivations}}
\prod_\de\lp M^{-\frac{1}{60}(i(\De_\de)-i(\til\De_\de))}\rp
\le
e^{\#(Y)}.(1-M^{-\frac{1}{120}})^{-24\#(Y)}.
\prod_{\De\in Y}{\rm mult}_{12}(\De)!
\ \ .
\ee
Indeed, we only have to apply (\ref{rainres}) with
$K_6=e$, $\ep_4=\frac{1}{60}$, and noting that
$n_1+\ldots+n_p$ is bounded by the number of $\de$'s thus by $2k\le 12\#(Y)$.

To sum over the procedures $\GP_{\rm ext}$ of external derivations, we
have to sum for each such derivation operator $\de/\de\ph_i(x)$
located in $\De=\De(x,i)$, over the external source in $\De$
to contract. There are $E_\De$ possibilities, hence we have
\be
\sum_{\GP_{\rm ext}} 1
\le
\prod_{\De\in Y}\lp (E_\De)^{{\rm mult}_{12}(\De)}\rp
\le
\prod_{\De\in Y}
\lp e^{E_\De}.{\rm mult}_{12}(\De)!\rp
\ \ .
\ee
The bound over the distinction $\Gc$ between external and
internal derivations is given by
\be
\sum_{\Gc}1\le 2^{2k}\le 2^{12\#(Y)}
\ \ .
\ee
Therefore at this stage we have
\[
\sum_{\Gc}
\sum_{\GP_{\rm ext}}
\sum_{{(\til\De)\ {\rm of}}\atop{\rm derivations}}
\sum_{\GP_{\rm int}}
\int d{\bf h}|\GC|\le
\frac{1}{k!}
U(g)^{\half\#(Y)}.\lp 2^{12}.e.(1-M^{-\frac{1}{120}})^{-24}.K_{17}
\rp^{\#(Y)}
\]
\[
\prod_{\De\in\cD^{(N)}}(E_\De !)^\half\times\prod_{s\in\cS}
M^{-i(\De_s^{\rm ext})}
\times
C^{\#(\cS)}\times
\prod_{l\in F}\lp K_3(r)(1+M^{-i(l)}d_2(l))^{-r}\rp
\]
\be
\times
\prod_{\vph}
\lp
M^{-\frac{9}{10}(i(\und\De_\vph)-i(\til\De_\vph))}
.M^{-\frac{1}{40}(i(\Br\De_\vph)-i(\til\De_\vph))}
\rp
\times
\prod_{\De\in Y}
({\rm mult}_{12}(\De)!)^{\frac{13}{2}}
\times
\prod_{X\in\pi}
(\GV(X)!)^6
\ \ ,
\label{intermbd}
\ee
where $C\eqdef e.2^\half.3^\half.K_4$ is the constant appearing
in (\ref{polybound}).

The next step is to get rid of the local factorials thanks to volume effects.
First concerning the multiplicities $\GV(X)$.
Remark that there exists a numerical constant $K_{18}\ge 1$,
such that for any integer $p\ge 1$,
\be
6p\log p-\frac{(\log M)p^2}{4000}
\le
K_{18}
\ \ .
\ee
We then claim that
\begin{lem}
\be
\prod_{X\in\pi}(\GV(X)!)^6
\times
\prod_\vph
M^{-\frac{1}{80}(i(\Br\De_\vph)-i(\til\De_\vph))}
\le
K_{18}^{\#(Y)}
\ee
\end{lem}

\noindent{\bf Proof:\ \ }
We must show that, for any $X\in\pi$,
\be
(\GV(X)!)^6
\times
\prod_{\vph\ {\rm produced\ by\ }X}
M^{-\frac{1}{80}(i(\Br\De_\vph)-i(\til\De_\vph))}
\le
K_{18}
\ \ ,
\label{volumver}
\ee
where produced means that $\vph$ belongs to an explicit vertex
whose top block is $X$.
Let us choose a representative $\Gg$ of the given form $<\Gg>$ satisfying the
following requirements. If $l$ is any of its 4-links, and $X_l$
is the top block of $l^\pi$, then $supp\ l\cap X_l=\{{\rm bg}(X_l)\}$.
Furthermore, we ask that for any other block $X$ in $supp\ l^\pi$,
$supp\ l\cap X$ be reduced to $\{\Br\De\}$, where $\Br\De$ is
the smallest box in $X$ containing ${\rm bg}(X_l)$. Such a choice is
always possible.

Now we consider a fixed $X\in\pi$,
and we extract the subsequence $\Gg_X$ of $\Gg$ made by the 4-links
whose top block is $X$. As a subsequence of an allowed graph,
$\Gg_X$ is allowed in $Y$, with respect to the large field
region $\Ga$. It is easy to see that it must also be allowed
with respect to the empty large field region, in the subset $W_X$
of $Y$ made by the union of the supports of the links in $\Gg_X$.
Now from the proof of Lemma \ref{fivefor} we deduce that the length $\GV(X)$
of $\Gg_X$ must satisfy $\GV(X)\le 5\#(W_X)-5$. As a result
if $W'_X$ denotes the set $W_X\backslash\{{\rm bg}(X)\}$, we
have
\be
\#(W'_X)\ge\frac{\GV(X)}{5}
\ \ .
\ee
But it is easy to check that
\be
\sum_{{\vph\ {\rm produced}}\atop{{\rm by}\ X}}
\frac{1}{80}(i(\Br\De_\vph)-i(\til\De_\vph))
\ge
\sum_{\De\in W'_X}
\frac{1}{80}(i(\De)-i({\rm bg}(X)))
\ge
\sum_{j=1}^{\#(W'_X)}\frac{j}{80}
\ge
\frac{(\#(W'_X))^2}{160}
\ \ ,
\ee
and thus
\be
\sum_{\vph\ {\rm produced\ by\ }X}
\frac{1}{80}(i(\Br\De_\vph)-i(\til\De_\vph))
\ge
\frac{(\GV(X))^2}{4000}
\ \ ,
\ee
from which (\ref{volumver}) follows.
\Endproof

Concerning the local factorials in ${\rm mult}_{12}(\De)$, we use
the following classical volume argument.
\begin{lem}
There exists a constant $K_{19}\ge 1$ such that for
any $\De\in Y$, we have
\be
({\rm mult}_{12}(\De)!)^{\frac{13}{2}}
.\prod_{{l\in F}\ |\ {\De\in l}}
(1+M^{-i(l)}d_2(l))^{-52}
\le K_{19}^{{\rm mult}_{12}(\De)}
\ \ .
\label{volumprop}
\ee
\end{lem}

\noindent{\bf Proof:\ \ }
Remark that since $F$ is a forest, the cubes $\De'$ forming
the extremities of the links $l=\{\De,\De'\}$ appearing in (\ref{volumprop}),
must be distinct. Remark also that the cubes $\De'$ are of the
same scale as $\De$.
If $d\ge 0$ is some number, 
one can easily see by a volume argument
that the number of cubes $\De'$ in $\cD^{(N)}_{i(\De)}$
such that $d_2(\De,\De')\le d.M^{i(\De)}$ is at
most $\half\pi^2(d+4)^4$.
Thus if ${\rm mult}_{12}(\De)\ge 256\pi^2$, we take
\be
d\eqdef
E\lp
\lp\frac{{\rm mult}_{12}(\De)}{\pi^2}\rp^{\frac{1}{4}}-4
\rp
\ge 0
\ \ ,
\ee
and have
$
\half\pi^2(d+4)^4\le\half {\rm mult}_{12}(\De)
$,
and therefore at least half of the cubes $\De'$
verify $d_2(\De,\De')>d.M^{i(\De)}$.
As a consequence
\be
({\rm mult}_{12}(\De)!)^{\frac{13}{2}}.
\prod_{{l\in F}\ |\ {\De\in l}}
(1+M^{-i(l)}d_2(l))^{-52}
\le
\lp{\rm mult}_{12}(\De)^{\frac{13}{2}}
(1+d)^{-26}
\rp^{{\rm mult}_{12}(\De)}
\ \ .
\ee
On the other hand,
\be
1+d\ge 
\lp\frac{{\rm mult}_{12}(\De)}{\pi^2}\rp^{\frac{1}{4}}-4
\ \ ,
\ee
and thus if we choose $K_{19}$ large enough so that
$K_{19}\ge (256\pi^2)^{\frac{13}{2}}$
and for any $p\ge 256\pi^2$ we have
\be
p^{\frac{13}{2}}.
\lp\lp\frac{p}{\pi^2}\rp^{\frac{1}{4}}-4
\rp^{-26}
\le
K_{19}
\ \ ,
\ee
then we easily arrive at (\ref{volumprop}).
\Endproof

Now the bound on the sum over the forms $<\Gg>$ and the $(\til\De)$ of the
vertices is done by the same move. The method parallels that of
the sum over $\GP_{\rm int}$.
First we chose the value of $k$, for which by the Lemma \ref{fivefor}
there are less than $6\#(Y)$ possibilities.
Next for each index $j$, $1\le j\le k$, we decide weather
the link $l_a$ of $\Gg$ is a 2-link or a 4-link.
This costs a factor $2^k\le 2^{6\#(Y)}$.
If $l_j$ is chosen to be a 4-link, we pay a huge factor
$\#(Y)$ to find the localization cube $\til\De_j$ of the vertex.
We use here the notations of Section 3.5.
Once we know $\til\De_j$, we will also know the top block $X_{j,1}$ of
the vertex, it is the unique $X\in\pi$ such that $\til\De_j\in X$.
Next we choose by a factor 3 the multiplicity
$l_j^\pi(X_{j,1})$ i.e. the number of high momentum fields.
We choose by a factor 3 the number $\al_j$ of blocks in the support
of $l_j^\pi$. Suppose we canonically label these disjoint blocks such that
$X_{j,\beta}$ is above $X_{j,\beta+1}$ for any $\beta$,
$1\le\beta\le \al_j-1$. We boldly bound by $3^3$ the number of choices
of multiplicities $m_{j,\beta}$ for the blocks
$X_{j,\beta}$, $2\le\beta\le\al_j$.
It remains to sum over the location of these blocks.
It is enough to know for each $X_{j,\beta}$ the smallest of its cubes
$\Br\De_{j,\beta}$
containing $\til\De_j$. But this cube has to be the $\Br\De_\vph$
of at least one low momentum field located in $X_{j,\beta}$.
Therefore we can use the corresponding factor
$M^{-\frac{1}{80}(i(\Br\De_\vph)-i(\til\De_\vph))}$
especially spared for that purpose in (\ref{intermbd}).

If $l_j$ is a 2-link, and if some total ordering was
chosen on $\cD^{(N)}$, we sum over the smallest cube
of $supp\ l_j$, with a factor $\#(Y)$. Next we sum on the second
cube thanks for instance to a decay $(1-M^{-i(l_j)}d_2(l_j))^{-5}$ that
has to be extracted
from (\ref{intermbd}).

We summarize these considerations by the following bound
\[
\sum_{<\Gg>}\sum_{{(\til\De)\ {\rm of}}\atop{\rm derivations}}
\frac{1}{k!}
\prod_\vph
M^{-\frac{1}{80}(i(\Br\De_\vph)-i(\til\De_\vph))}
\prod_{l\in F}
(1-M^{-i(l)}d_2(l))^{-5}
\]
\be
\le
\sum_{0\le k\le 6\#(Y)-6}
\frac{\#(Y)^k}{k!}
\lp 2^6.\max[3^{30}.(1-M^{-\frac{1}{80}})^{-18},K_{5}^6]
\rp^{\#(Y)}
\ \ ,
\ee
where $K_5$ is
the constant appearing in (\ref{defK5}).
Now $\frac{\#(Y)^k}{k!}$ is simply bounded by $e^{\#(Y)}$,
besides the sum over $\Ga \in Y$, costs a factor $2^{\#(Y)}$.
It remains to perform the sum over $Y$ using Proposition \ref{PropY}.
By construction of the averaging cubes $\und\De_\vph$,
for any $i$, $0\le i\le i_{\rm max}(Y)-1$, and for any component
$A \in\cR_i(Y,F)$, there are two possibilities.
Either there is a gluing link crossing the
lower boundary of $A$, or there are at least five low momentum fields
$\vph$
such that $i(\und\De_\vph)>i\ge i(\til\De_\vph)$, and
${\rm pr}_i(\til\De_\vph)\in A$.
As result
using the notations of Proposition \ref{PropY} with $\ep_2=\half$, 
we have
\be
M^{-\frac{9}{2}\#(Y)}
\prod_{l\in F}(1-M^{-i(l)}d_2(l))^{-\frac{9}{2}}
\prod_\vph
M^{-\frac{9}{10}(i(\und\De_\vph)-i(\til\De_\vph))}
\le
\cT_{\ep_2}(Y)
\ \ .
\ee

Therefore we need to choose $r$ at the beginning so as to fulfill
$r\ge 52+5+\frac{9}{2}$.
Now if we define the new constant
\[
K_{20}\eqdef 2^{19}.e^2.(1-M^{-\frac{1}{120}})^{-24}.K_{17}.K_3(r)
\]
\be
.K_{18}.K_{19}^2.M^{\frac{9}{2}}.\max[3^{30}(1-M^{-\frac{1}{80}})^{-18},
K_{5}^6].K_1(4,M,\half)
\ \ ,
\ee
we have by applying Proposition \ref{PropY},
and reverting to the notations of Theorem 2
\[
\sum_{{{Y|Y\subset\De^{(N)}}\atop{\De_{\rm ext}\in Y}}
\atop{\{\De_s^{\rm ext}|s\in\cS\}\subset Y}}
|\cA(Y,(\De_s^{\rm ext},\ze_s^{\rm ext})_{s\in\cS})|.K^{\#(Y)}
\]
\be
\le
\lp\sum_{Q\ge 1}6Q.(K.K_{20}.U(g)^\half)^Q
\rp
.C^{\#(\cS)}.\prod_{\De\in\cD^{(N)}}(E_\De!)^{1/2}
.\prod_{s\in\cS}M^{-i(\De_s^{\rm ext})}
\ \ .
\label{lastf}
\ee
Now since by Proposition \ref{propsmf} $\lim_{g\rightarrow 0^+}U(g)=0$,
for any $\et$ and $K$, we can find a $g_0$,
$0<g_0<1$, such that $0<g\le g_0$
implies that the first factor in the right side of (\ref{lastf}),
is smaller than $\et$.
One can easily check that this result is uniform in the IR cut-off $N$,
the volume cut-off $\La$, and the collection of external sources.
This proves (\ref{polybound}), at last.
\Endproof

\noindent{\Large\bf Aknowledgements}

We thank Jacques Magnen for illuminating discussions,
and for the trick in Lemma \ref{avrSobo}.

\end{document}